\documentclass[useAMS,usenatbib]{mn2e}

\usepackage{graphicx,natbib, bm, amssymb, amsmath, longtable, subfigure, multirow}

\title[Stokes tomography of radio pulsar magnetospheres. I. Linear polarization]{Stokes tomography of radio pulsar magnetospheres. I. Linear polarization}

\author[C. T. Y. Chung et al.]{C. T. Y.~Chung $^1$\thanks{E-mail: c.chung4@pgrad.unimelb.edu.au} and A.~Melatos $^1$ \\                                           
 $^1$ School of Physics, University of Melbourne, Parkville, VIC               
 3010, Australia}

\begin{document}

\date{ }

\pagerange{\pageref{firstpage}--\pageref{lastpage}} \pubyear{2010}

\maketitle

\label{firstpage}

\begin{abstract}
Polarimetric studies of pulsar radio emission traditionally concentrate on how the Stokes vector $(I, Q, U, V)$ varies with pulse longitude, with special emphasis on the polarization angle (PA) swing of the linearly polarized component. The interpretation of the PA swing in terms of the rotating vector model is limited by the assumption of an axisymmetric magnetic field and the degeneracy of the output with respect to the orientation and magnetic geometry of the pulsar; different combinations of the latter two properties can produce similar PA swings.
This paper introduces Stokes phase portraits as a supplementary diagnostic tool with which the orientation and magnetic geometry can be inferred more accurately. The Stokes phase portraits feature unique patterns in the $I$-$Q$, $I$-$U$, and $Q$-$U$ planes, whose shapes depend sensitively on the magnetic geometry, inclination angle, beam and polarization patterns, and emission altitude. We construct look-up tables of Stokes phase portraits and PA swings for pure and current-modified dipole fields, filled core and hollow cone beams, and two empirical linear polarization models, $L/I = \cos \theta_0$ and $L/I = \sin \theta_0$, where $\theta_0$ is the colatitude of the emission point. We compare our look-up tables to the measured phase portraits of 24 pulsars in the European Pulsar Network online database. We find evidence in 60\% of the objects that the radio emission region may depart significantly from low altitudes, even when the PA swing is S-shaped and/or the pulse-width-period relation is well satisfied. On the other hand, the data are explained adequately if the emission altitude exceeds $\sim$ 10\% of the light cylinder radius. We conclude that Stokes phase portraits should be analysed concurrently with the PA swing and pulse profiles in future when interpreting radio pulsar polarization data.
\end{abstract}


\section{Introduction}
\label{intro}
Polarimetric studies of pulsar radio emission probe the geometry of the pulsar magnetosphere. Although the phenomenology of the emission is often complex, there are some general trends in the observed pulse shapes and polarization angle (PA) swings that have been explained qualitatively with simple magnetospheric models.
 In many objects, the data are approximately consistent with \textit{low-altitude} emission from open \textit{dipolar} field lines centred on the magnetic axis \citep[e.g.][]{rankin83, lyne88}, although in some objects, like the Crab, the emission is thought to emerge as a fan beam from higher altitudes in the outer magnetosphere \citep[e.g.][]{cheng77, kaspi00}.

Three lines of evidence favour an approximately dipolar magnetic field in the radio emission region. First, pulses narrow with increasing frequency. According to the radius-to-frequency mapping, the frequency of the radiation is related to the local particle density, which decreases with altitude, as the dipole field lines diverge \citep{ruderman75, cordes78}. The width of the pulse profile is commonly used to infer the emission height, with high-frequency pulses originating closer to the surface \citep[e.g.][]{manchester96, vonhoensbroech97, karas07, weltevrede08}.
Second, in many objects, the pulse width $W$ is inversely related to the spin period $P$ of the pulsar according to $W \propto P^{-1/2}$ \citep{rankin93}. This relation can be explained geometrically assuming an emission cone tangent to the last open field line of a dipole \citep[e.g.][]{gangadhara01}. Under this assumption, the measured pulse widths can also be related to the cone half-opening angle and the angle between the magnetic and rotation axes \citep{gil84}.
Third, the plane of linear polarization traces out a characteristic S-shaped swing over one pulse period. The swing has been modelled successfully by the rotating vector model assuming a dipole field (or, more generally, an axisymmetric field). The model relates the gradient of the S-shape at the inflection point to the inclination angle of the observer and the angle between the magnetic and rotation axes \citep{radha69, lynman88, hibschman01}. The phase lag between the centroid of the pulse profile and the inflection point has also been used to estimate the emission altitude \citep{blaskiewicz91}.

Unfortunately, experience shows that there are limitations in relying only on the pulse profile and PA swing. For example, the three trends mentioned above fail to fix the geometry uniquely. The PA swing looks remarkably similar for many different orientations of the observer and the magnetic axis. Moreover, observations show that many pulsars do not display the clean, S-shaped swing expected from the rotating vector model. Instead, the S-shape is often distorted, suggesting nondipolar configurations \citep{karas05, jonkar08, han09}. In particular, many millisecond pulsars have flat PA profiles \citep{ stairs99, ord04}, highly distorted profiles \citep[e.g. PSR J0437$-$4715;][]{navarro97} or seemingly random PAs \citep[e.g. giant pulses from PSR J1824$-$2452A;][]{knight06}. Multi-frequency observations show that the PA swing varies with frequency for many pulsars \citep{johnston08}, indicating that the magnetic geometry changes appreciably with altitude. Several mechanisms can distort a dipole field, e.g. a current flowing along the field lines \citep{hibschman01, dyks08}, or rotational sweepback near the light cylinder \citep{hibschman01, dyks04, dyks08}.

One diagnostic tool which has been used sparingly but holds considerable promise is the \textit{Stokes phase portrait}, that is, the pattern traced out by the four Stokes parameters $I$, $Q$, $U$ and $V$ when plotted against each other over one pulse period. PA swings describe how the quantity $U/Q$ varies with time, but less effort has been devoted to studying how $Q$ and $U$ (say) depend on each other. We show here that Stokes phase portraits harbour a great deal of extra information regarding the magnetospheric geometry. The analysis of polarization data on the $Q$-$U$ plane is common practice when calculating rotation measure transfer functions for polarized emission from radio galaxies \citep[e.g.][]{burn66, brentjens05, haverkorn06}, and in optical and UV pulsar data analysis \citep[e.g.][]{smith88, graham96, slowikowska09}.

In this paper, we calculate theoretically how the Stokes parameters vary with pulse longitude for several magnetic field configurations and beam patterns. We apply the theoretical results to several radio pulsars to determine the observer's inclination angle, $i$, the angle between the magnetic and rotation axes, $\alpha$, the emission altitude, and the approximate geometry of the magnetosphere (e.g. the ratio of poloidal to toroidal field). We call this reconstruction technique \textit{Stokes tomography}. In Section \ref{sec:stokes}, we define the model and the algorithm used to generate the Stokes phase portraits. Look-up tables of phase portraits, PA swings, and pulse profiles for a pure dipole field at low emission altitudes are presented in Section \ref{sec:dipole}. The look-up tables can be compared directly against observational data for any pulsar to determine the emission geometry. In Section \ref{sec:dipoleaberration}, we present look-up tables for a pure dipole at 10\% of the light cylinder radius, where relativistic aberration effects are important. It is shown that many pulsars that are categorised as pure dipoles on the strength of their PA swing possess Stokes phase portraits that are inconsistent with a pure dipole at low emission altitudes. In Section \ref{sec:toroidal}, look-up tables are presented for a current-modified dipole field with a radially increasing toroidal component, as in standard magnetospheric models \citep{hibschman01}. Future applications of Stokes tomography are canvassed briefly in Section \ref{sec:conclusion}. Companion papers will generalize the approach to circularly polarized pulsar radio emission and realistic (e.g. force-free) magnetic fields.

\section{Stokes tomography}
\label{sec:stokes}
\subsection{Radiation field}
\label{sec:radiation}
In the magnetosphere, radiation from highly relativistic particles tied to magnetic field lines is narrowly beamed along the particle velocity vector. The observed emission point $\mathbf{x}_0(t)$ at any time $t$ is therefore located where the magnetic vector $\mathbf{B [x}_0(t),(t)]$ points along the observer's line-of-sight vector $\mathbf{w}$. The unit tangent vector to the magnetic field at $\mathbf{x}_0(t)$ is defined as $\mathbf{t} = \mathbf{B}[\mathbf{x}_0(t), t]/|\mathbf{B[x}_0(t), t]|$. For non-relativistic particles, to find the emission point at time $t$, we must therefore solve the equation $\mathbf{t} = \mathbf{w}$. In general, however, when considering emission altitudes of $r \gtrsim 0.1 r_\text{LC}$, where $r_\text{LC} = c/\Omega$ is the light cylinder radius, we must account for relativistic aberration. Aberration shifts the emission point and electric field vector by different amounts of order $r/r_\text{LC}$. The emission point $\mathbf{x}_0 (t)$ at time $t$ satisfies the equation \citep{blaskiewicz91}
\begin{equation}
\label{eqn:tangent}
\mathbf{w} = \frac{\mathbf{t}  + \mathbf{\Omega} \times \mathbf{x}_0/c}{\lvert \mathbf{t}  + \mathbf{\Omega} \times \mathbf{x}_0/c \rvert},
\end{equation}
where $\mathbf{\Omega}$ is the angular velocity vector.
Note that $\mathbf{B}$ varies with $t$ intrinsically, as the star rotates in the observer's frame, not just through $\mathbf{x}_0(t)$. The normal to the field at $\mathbf{x}_0$ is defined as $\mathbf{n} = \bm{\kappa}/|\kappa|$, where $\bm{\kappa} = (\mathbf{t} \cdot \nabla)\mathbf{t}$ is evaluated at $\mathbf{x}_0$. The binormal is defined as $\mathbf{b} = \mathbf{t} \times \mathbf{n}$. 

The amplitude and direction of the complex electric field vector $\mathbf{E}$ at the emission point are determined by the radiation physics, which is not understood in detail. A popular assumption is that $\mathbf{E}$ at time $t$ points along the instantaneous acceleration vector of the particle, $\mathbf{a}$. To first order in $r/r_\text{LC}$, the acceleration is given by \citep{dyks08}
\begin{equation}
\label{eqn:acceleration}
\mathbf{a} = c \frac{\partial \mathbf{t}}{\partial t} + c (\mathbf{v} \cdot \nabla) \mathbf{t} + c \mathbf{\Omega} \times \mathbf{t} + \mathbf{\Omega} \times (\mathbf{\Omega} \times \mathbf{x}_0),
\end{equation}
in terms of the instantaneous velocity vector $\mathbf{v} = c \mathbf{t} + \mathbf{\Omega} \times \mathbf{x}_0$, assuming no precession ($\partial \mathbf{\Omega} / \partial t = 0$). From left to right, the four terms on the right-hand side of (\ref{eqn:acceleration}) describe: (i) the change in field aligned velocity $c \mathbf{t}$ as the magnetic dipole moment (and hence $\mathbf{t}$) rotates infinitesimally in the inertial frame; (ii) the change in field-aligned velocity $c \mathbf{t}$ as the charge moves infinitesimally along $\mathbf{v} = c \mathbf{t} + \mathbf{\Omega} \times \mathbf{x}_0$ and hence $\mathbf{t}$ changes; (iii) the change in corotation velocity $\mathbf{\Omega} \times \mathbf{x}_0$ as the charge moves infinitesimally along the field $\mathbf{t}$; and (iv) the change in corotation velocity $\mathbf{\Omega} \times \mathbf{x}_0$ as the charge corotates infinitesimally with the magnetosphere along $\mathbf{\Omega} \times \mathbf{x}_0$ \citep{dyks08, hibschman01}.

The associated Stokes parameters ($I, Q, U, V$) describing the polarization state are defined as
\begin{eqnarray}
I &=& \lvert E_x \rvert^2 + \lvert E_y \rvert^2\\
\label{Qeqn} Q &=&\lvert E_x \rvert^2 - \lvert E_y \rvert^2\\
\label{Ueqn} U &=& 2 \text{Re} (E_x E_y^*)\\
V &=& 2 \text{Im} (E_x E_y^*),
\end{eqnarray}
where $I$ is the total intensity, $L = (Q^2 + U^2)^{1/2}$ is the linearly polarized component, and $V$ is the circularly polarized component.
The $x$- and $y$- components are measured with respect to an orthonormal basis $(\hat{\mathbf{x}}, \hat{\mathbf{y}})$ which is fixed in the plane of the sky. In this paper, we choose $\hat{\mathbf{x}} = \mathbf{\Omega}_\text{p}/\lvert \mathbf{\Omega}_\text{p}\rvert$ and $\mathbf{\hat{y}} = \hat{\mathbf{x}} \times \mathbf{w}$, where $\mathbf{\Omega}_\text{p} = \mathbf{\Omega - (\Omega \cdot w) w}$ is the projection of the angular velocity $\mathbf{\Omega}$ onto the plane of the sky.
Then the polarization angle, $\psi$, measured counter-clockwise between $\hat{\mathbf{x}}$ and the linearly polarized part of $\mathbf{E}$ is given by 
\begin{equation}
\label{paeqn}
\psi = \frac{1}{2} \tan^{-1} \frac{U}{Q}.
\end{equation}

\subsubsection{Orientation of $\hat{\mathbf{x}}$ and $\hat{\mathbf{y}}$}
\label{sec:orientationxy}
In the observational data, the direction of $\mathbf{\Omega}$ is usually not known. The polarization basis in which the observed Stokes parameters are expressed may therefore differ from the canonical basis. This affects the Stokes phase portraits in the following way. 

Consider two polarization bases, ($\hat{\mathbf{x}}$, $\hat{\mathbf{y}}$) and ($\hat{\mathbf{x}}'$, $\hat{\mathbf{y}}'$), where ($\hat{\mathbf{x}}'$, $\hat{\mathbf{y}}'$) is rotated clockwise by an angle $\beta$ with respect to ($\hat{\mathbf{x}}$, $\hat{\mathbf{y}}$). In the canonical (unprimed) basis, the electric field vector transforms as 
\begin{eqnarray}
E_x &=& \cos \beta E_x' + \sin \beta E_y',\\
E_y &=& -\sin \beta E_x' + \cos \beta E_y',
\end{eqnarray}
and the Stokes parameters transform as
\begin{eqnarray}
I &=& I'\\
\label{eq:qdash} Q &=& Q' \cos 2\beta + U' \sin 2\beta\\
\label{eq:udash} U &=& -Q' \sin 2\beta + U' \cos 2\beta.
\end{eqnarray}

Hence, in the $Q$-$U$ phase portrait, the pattern rotates counter-clockwise through an angle $2 \beta$ with respect to the $Q'$-$U'$ phase portrait, but its shape remains invariant. In contrast, in the $I$-$Q$ and $I$-$U$ phase portraits, both the shape and orientation of the pattern change. When fitting our model to data, in cases where $\beta$ is unknown, we first work to match the shape of the pattern in the $Q$-$U$ data, without worrying about the orientation. Then, armed with a good fit to the $Q$-$U$ shape, we rotate the basis to match the orientation as well and deduce $\beta$. Finally, knowing $\beta$, we match the patterns in the $I$-$Q$ and $I$-$U$ data.  Note that the PAs in the two bases are related by $\psi = \psi' - \beta$, as expected.

\subsection{Numerical algorithm}
We now describe how to construct ($I, Q, U, V$) as functions of pulse longitude (and hence each other) for a prescribed magnetic geometry.

\begin{figure}
\includegraphics[scale=0.6]{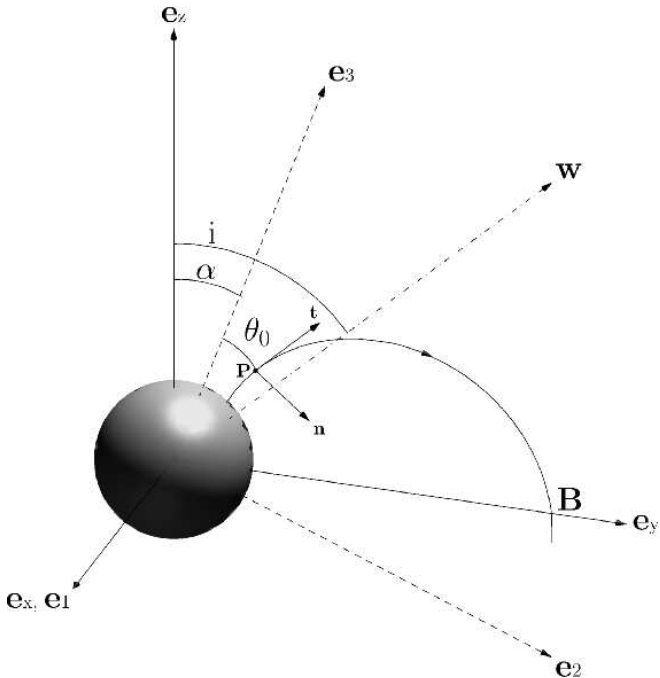}
\caption{Orientation of the inertial axes $\mathbf{e}_x, \mathbf{e}_y, \mathbf{e}_z$ (solid lines) and pulsar body axes $\mathbf{e}_1, \mathbf{e}_2, \mathbf{e}_3$ (dashed lines) at an instant where $\mathbf{e}_2$ and $\mathbf{e}_3$ lie in the $\mathbf{e}_y$-$\mathbf{e}_z$ plane. The line of sight vector $\mathbf{w}$ (dashed-dotted line) is defined to always lie in the $\mathbf{e}_y$-$\mathbf{e}_z$ plane at an angle $i$ to $\mathbf{e}_z$. $\mathbf{e}_3$ is tilted with respect to the rotation axis $\mathbf{e}_z$ by an angle $\alpha$. An example of a dipolar field line ($\mathbf{B}$) and its corresponding emission point $\mathbf{x}_0(t)$ (labelled $P$) are drawn. At $P$, the tangent vector is $\mathbf{t = w}$ and the normal vector is $\mathbf{n}$. The angle between $\mathbf{t}$ and $\mathbf{e}_3$ is $\theta_0$. The shading on the sphere represents the beam pattern $I(r, \theta, \phi)$; $I$ is a maximum where the shading is brightest.}
\label{axes1}
\end{figure}

Let us define two reference frames, as depicted in Figure \ref{axes1}: the inertial frame in which the observer is at rest, with Cartesian axes ($\mathbf{e}_x, \mathbf{e}_y, \mathbf{e}_z$), and the body frame of the rotating pulsar, with Cartesian axes ($\mathbf{e}_1, \mathbf{e}_2, \mathbf{e}_3$). The line-of-sight vector $\mathbf{w}$ is chosen to lie in the $\mathbf{e}_y$-$\mathbf{e}_z$ plane, making an angle $i$ with $\mathbf{e}_z$. The rotation and magnetic axes are chosen to lie along $\mathbf{e}_z$ and $\mathbf{e}_3$ respectively. The orientation of the body axes with respect to the inertial axes is computed as a function of time by solving Euler's equations of motion for a rigid body; precession can be included in general, although we do not examine it in this paper. 

We define a spherical polar grid $(r, \theta, \phi)$ in the body frame covering the region $x_\text{min} \leq r/r_\text{LC} \leq x_\text{max}$, $0 \leq \theta \leq \pi$, $0 \leq \phi \leq 2\pi$, with $64 \times 256 \times 64$ grid cells, where the line $\theta = 0$ lies along $\mathbf{e}_3$. Most of our analysis is done in the middle to outer magnetosphere, with $x_\text{min} = 0.001$ and $x_\text{max} = 0.38$. Nonetheless, other choices are possible and easy to implement. Given a magnetic field $\mathbf{B}(r, \theta, \phi)$, we compute $\mathbf{t, n,}$ and $\mathbf{b}$ at each grid point using second-order central finite differencing and ghost cells at the edges. 

In order to calculate ($I, Q, U, V$) versus $t$, we search the grid at a fixed altitude $r_0$ to find the emission point satisfying (\ref{eqn:tangent}) (the point $P$ in Figure \ref{axes1}). At $P$, $(\mathbf{t} + \mathbf{\Omega} \times \mathbf{x}_0/c) \cdot \mathbf{w}$ is a maximum; note that the location of $P$ changes with time in both the body frame and the inertial frame.\footnote{In general, we must consider $\pm \mathbf{t}$ in (\ref{eqn:tangent}) to capture emission from particles flowing out from both north and south poles. In this paper, however, we restrict out attention to emission from one pole only, to avoid cluttering the look-up tables in Sections \ref{sec:dipole} and \ref{sec:toroidal}. We therefore do not capture ``interpulse" emission with our model. The 24 objects studied in Section \ref{sec:rmapping} exhibit either zero or weak interpulse emission in existing data.} To fine-tune the location, we interpolate $\mathbf{t}$ linearly between the four surrounding points and re-compute (\ref{eqn:tangent}) for each interpolated point. In general, there are two points $\mathbf{x}_0(t)$ satisfying (\ref{eqn:tangent}) on opposite sides of the pulsar, but the observer only sees the one facing him.  

We stipulate the shape of the beam pattern $I(r, \theta, \phi)$ and the degree of linear polarization $L(r, \theta, \phi)$, and assume that $\mathbf{E}$ is constant and lies along $\mathbf{a}$, i.e.
\begin{equation}
\mathbf{E} = L \mathbf{a}/ I.
\end{equation}
In Figure \ref{axes1}, $I(r, \theta, \phi)$ is represented by the shading on the surface of the sphere. It peaks along $\mathbf{e}_3$ (i.e. $\theta = 0$), i.e. Figure \ref{axes1} depicts a centre-filled beam. Other beam patterns, e.g. a hollow cone, are also considered in Sections \ref{sec:dipole} and \ref{sec:toroidal}.
Note that $\mathbf{E}$ is purely real in this paper because we assume $V = 0$, i.e. zero circular polarization. Examples of $I$ and $L$ are given in Sections \ref{sec:beamshape} and \ref{sec:linear}, drawn either from a phenomenological theory or directly from data.

\subsection{Relativistic aberration}
\label{sec:aberration}
Relativistic aberration acts to shift the pulse centroid relative to the inflection point of the PA swing, such that they are separated in phase by $\approx 4 r/r_\text{LC}$\,radians to leading order \citep{blaskiewicz91, hibschman01, dyks08}. 
Hence, aberration introduces an altitude dependence in the case of a pure dipole, which is absent in the non-aberrated case. At large emission altitudes ($\gtrsim 0.1 r_\text{LC}$), the phase shift can significantly alter the shapes of the Stokes phase portraits. This is investigated further in Section \ref{sec:dipoleaberration}.
Additionally, if the magnetic field sweeps back rotationally, or develops a helical, current-modified component, the emission point at an instant in time varies as a function of altitude. The effects of aberration combined with a current-modified field are investigated in Section \ref{sec:toroidal}.
According to the radius-to-frequency mapping, observing emission from different altitudes is analogous to making observations at several different frequencies.

 Equation (\ref{eqn:tangent}) ignores cross-field particle motions (e.g. curvature and gradient drifts) in the outer magnetosphere, where the cyclotron cooling time exceeds the flow time, for the sake of simplicity.

\subsection{Beam pattern}
\label{sec:beamshape}
The two most common beam patterns in the literature are the core-and-cone model \citep{rankin83} and the patchy-beam model \citep{lyne88}. In the core-and-cone model, single-peaked profiles are explained by emission filling the polar flux tube, double-peaked profiles are explained by emission from a hollow cone tangential to the polar flux tube, and triple-peaked profiles are explained by a combination of the two. The patchy-beam model was proposed to explain asymmetric profiles. It suggests that emission comes from randomly distributed locations within an emission beam.
In this paper, for simplicity, we consider two beam shapes based on the Rankin model: (i) a filled core beam (single-peaked pulse profile) and (ii) a hollow cone (double-peaked pulse profile). Both beams are centred on $\mathbf{e}_3$ and are cylindrically symmetric about it.

The colatitude of the emission point at time $t$, $\theta_0(t)$, is obtained from $\theta_0 = \cos^{-1} [r_0^{-1} \mathbf{x}_0(t) \cdot \mathbf{e}_3]$ and can range over $0 \leq \theta_0 \leq \pi$. We model the polarized intensity $I(t) = I[\theta_0(t)]$ as a gaussian beam,
\begin{equation}
\label{eq:beam}
I(\theta_0) = (2 \pi \sigma^2)^{-1/2} \text{exp}\left[-(\theta_0 - \rho)^2/(2 \sigma^2)\right] 
\end{equation}
where $\sigma$ defines the beam width, and $\rho$ is the half-opening angle of the emission cone. In general, $\sigma$ and $\rho$ vary from pulsar to pulsar. However, to generate the look-up tables in Sections \ref{sec:dipole}--\ref{sec:toroidal}, we use fixed, representative values: $\rho = 0^\circ$ and $\sigma = 10^\circ$ for a filled core, and $\rho = 25^\circ$ and $\sigma = 10^\circ$ for a hollow cone. For the sake of simplicity, we assume in what follows that only one pole shines (see footnote above); it is trivial to modify this assumption if necessary.

Equation (\ref{eq:beam}) can only be used to model approximately symmetric pulses. To fit asymmetric profiles, more complex models, such as a `broken cone' or `horseshoe', are required, in which the intensity varies with longitude $\phi_0(t)$ as well as $\theta_0(t)$. Such asymmetric intensity maps are inspired by the patchy beam model \citep{lyne88, kramer08}. 

\subsubsection{Tilted axis of symmetry}
\label{sec:displaced}
Although the magnetic-pole model for pulsar radio emission is commonly adopted, it is possible that the beam pattern is centred on another axis that is slightly tilted with respect to the magnetic axis. Fan beams in the outer magnetosphere are an extreme instance of such a geometry \citep{cheng00, watters09}, but less dramatic displacements can occur if there are local, quadrupolar surface fields in the vicinity of the magnetic poles.

\begin{figure}
\includegraphics[scale=0.9]{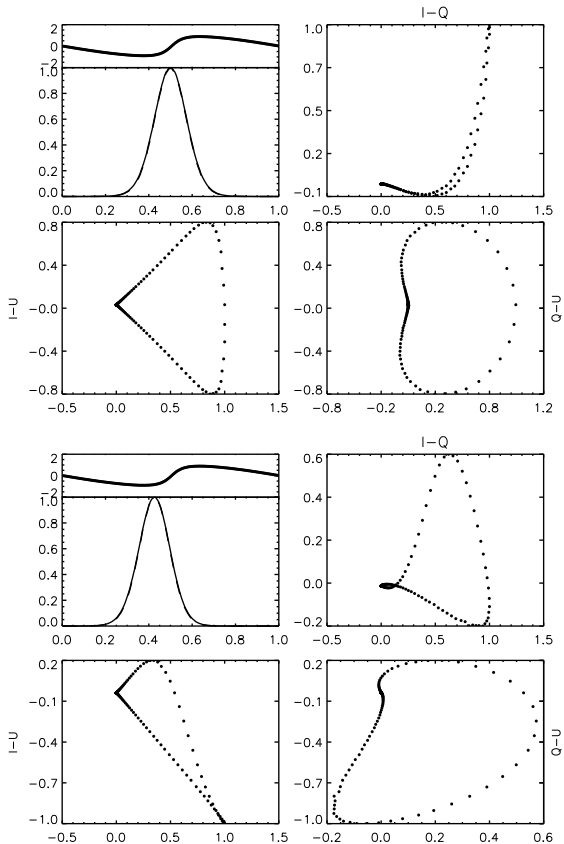}
\caption{Theoretical pulse profiles, PA swings, and Stokes phase portraits for a pure dipole and a beam pattern symmetric about $\mathbf{e}_3$ (top four subpanels), and about $\mathbf{e}_3'$ (bottom four subpanels), an axis tilted by $10^\circ$ with respect to $\mathbf{e}_3$ and $90^\circ$ with respect to $\mathbf{e}_1$, as described in the text. Within each half of the figure, the four subpanels display (clockwise from top left): (a) $I/I_\text{max}$ (bottom) and PA (top; in radians) as functions of pulse longitude, (b) $I$-$Q$, (c) $Q$-$U$, and (d) $I$-$U$. Relativistic aberration is not included in this figure. }
\label{compareoffset}
\end{figure}

To illustrate the above effect, we consider an arbitrary example, where $I$ and $L$ are symmetric about the axis $\mathbf{e}'_3$  in a reference frame $(\mathbf{e}'_1, \mathbf{e}'_2, \mathbf{e}'_3)$, where $\mathbf{e}'_3$ is tilted with respect to $\mathbf{e}_3$ by $10^\circ$, and $\mathbf{e}'_1$ is tilted with respect to $\mathbf{e}_1$ by $90^\circ$.  
Figure \ref{compareoffset} compares the original (top panel) and tilted (bottom panel) pulse profiles, PA swings, and Stokes phase portraits for $(\alpha, i) = (30^\circ, 40^\circ)$. The axis tilt causes the pulse phase to lead the PA swing by $\approx 0.47$ radians. In the bottom panel, the hockey stick in the $I$-$Q$ phase portrait broadens into a balloon, while the shape in the $I$-$U$ plane narrows and tilts. The heart shape in the $Q$-$U$ plane tilts and is no longer symmetric about $U = 0$. 

This effect mimics relativistic aberration, which results in a relative phase shift between the pulse profile and PA swing of $4 r/r_\text{LC}$\,rad. The effects of relativistic aberration on the phase portraits are investigated in detail in Section \ref{sec:dipoleaberration}. We note that neither Stokes tomography nor the rotating vector model can distinguish the two effects. For the sake of clarity, in Sections \ref{sec:dipole}--\ref{sec:toroidal} of this paper, we therefore assume that the beam and polarization patterns are symmetric about $\mathbf{e}_3$, while noting that the Stokes tomography technique can be generalized easily (as in Figure \ref{compareoffset}) to accomodate a tilted axis of symmetry. When modelling individual pulsars in detail, $I$ and $L$ can be fitted iteratively and empirically, and asymmetric beam patterns arise naturally.

\subsection{Linear polarization}
\label{sec:linear}
The uncertainty surrounding the microscopic emission physics in a pulsar makes it difficult to model its polarization properties, although theoretical progress has been made \citep[e.g.][]{gil90, melrose95, mckinnon98, rankin03, xu00}. Much of this work attempts to account for observed orthogonal mode jumps (i.e. $90^\circ$ jumps in the PA swing). \citet{mckinnon98} modelled the $L/I$ distribution statistically, describing $Q$ as a Gaussian random variable. One physical model assumes that emission comes from inverse Compton scattering, and shows that the fractional linear polarization $L/I$ varies with $\theta_0$ \citep{xu00}. 

In the absence of a unique empirical or theoretical rule for how the fractional linear polarization $L/I$ varies across the beam, we investigate two simple cases when constructing the look-up tables in Sections \ref{sec:dipole} and\ref{sec:toroidal}: (i) $L/I \propto \cos \theta_0$ (generally, but not exclusively, suited to single-peaked profiles), and (ii) $L/I \propto \sin \theta_0$ (generally, but not exclusively, suited to double component profiles). The proportionality constants are chosen to be unity in Sections \ref{sec:dipole}--\ref{sec:toroidal} for simplicity, but, in reality, approximately 80\% of normal pulsars have a pulse-averaged $L/I \leq 0.3$ \citep{xilouris98}. 

Although the above models are useful for illustrating the variety of Stokes phase portraits in our look-up tables, they are a rough approximation in any specific object. Model (i) gives $L/I \sim 1$ for small $\theta_0$,  higher than the observed fraction in most pulsars. Moreover, in many objects, there is a phase lag between $I$ and $L$, which we ignore. We do not attempt to model orthogonal mode jumps in this paper.

\subsection{Summary of algorithm}
\label{sec:summary}
\begin{figure*}
\label{flow}
\centering
\subfigure[] {
\label{flow1}
\includegraphics[scale=1]{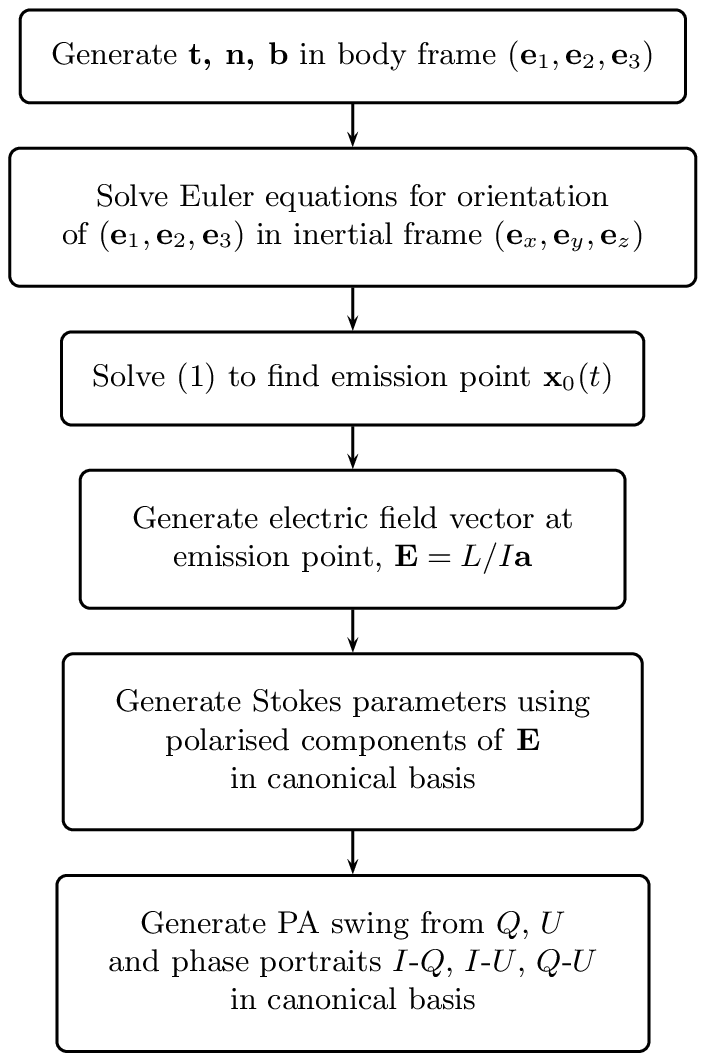}
}
\hspace{1cm}
\subfigure[] {
\label{flow2}
\includegraphics[scale=0.95]{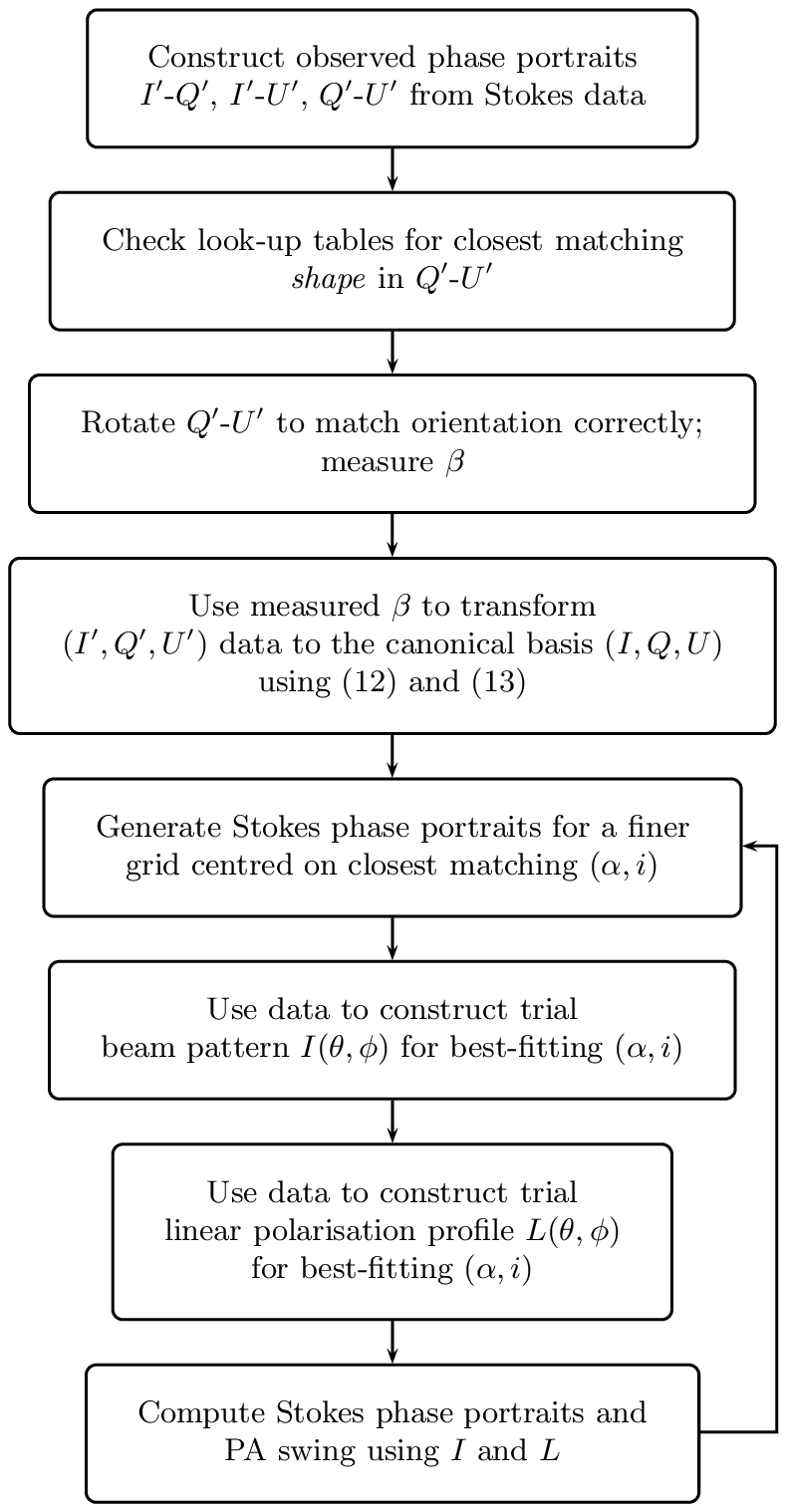}
}
\caption{(a) Recipe to generate look-up tables of Stokes phase portraits, PA swings, and pulse profiles for a given orientation and magnetic geometry. (b) Iterative recipe to extract orientation and magnetic geometry from observational data.}
\end{figure*}

Figure \ref{flow1} summarises, in the form of a flowchart, the steps required to generate the PA swing, pulse profile, and Stokes phase portraits for a pulsar with a given magnetic configuration (as described in Sections \ref{sec:radiation}--\ref{sec:linear} above). Figure \ref{flow2} summarises the steps required to match the theoretical PA swing, pulse profile, and Stokes phase portraits from the look-up tables to observational data and thereby infer the pulsar's orientation and magnetic geometry.  For a nondipolar field, the results depend on altitude, so the process must be repeated using several values of $r$ until a good match is achieved. We note that in principle, if $\beta$ is unknown, one can also repeat the process using several values of $\beta$. However, throughout this paper, we find that once an approximate estimate for $\beta$ is found, it is more useful to explore $\alpha$ and $i$ in detail, as varying $\beta$ only alters the orientation of the $Q$-$U$ pattern, and not its shape.

\section{Dipole field at low emission altitudes}
\label{sec:dipole}
In this section, we calculate Stokes phase portraits, PA swings, and pulse profiles for a pulsar with a purely poloidal, dipolar magnetic field at low emission altitudes, where the effects of relativistic aberration are negligible. We generate a two-dimensional look-up table of phase portraits for orientations $10^\circ \leq i \leq 90^\circ$ and $10^\circ \leq \alpha \leq 90^\circ$. 
For each beam pattern (core emission in Section \ref{sec:dipolecore}, conal emission in Section \ref{sec:dipolecone}) and polarization pattern ($L/I = \cos \theta_0, \sin \theta_0)$, we present three phase portraits ($I$-$Q$, $I$-$U$, $Q$-$U$) for each pair of angles $(\alpha, i)$. Finally, in Section \ref{sec:compdipole}, we analyse a selection of full Stokes data obtained from the European Pulsar Network (EPN)\footnote{Available at: http://www.mpifr-bonn.mpg.de/pulsar/data/} online archive \citep{lorimer98} for objects in \citet{gould98}. We show that a dipole field (and hence low-altitude emission) is inconsistent with the observed phase portraits of many superficially dipole-like pulsars, e.g. objects which obey the pulse-width-period relation or exhibit clean, S-shaped PA swings.
 
\subsection{Filled core emission}
\label{sec:dipolecore}
\subsubsection{$L = I \cos \theta_0$}
\label{sec:dipolelcostheta}
\begin{figure}
\centering
\includegraphics[scale=0.45]{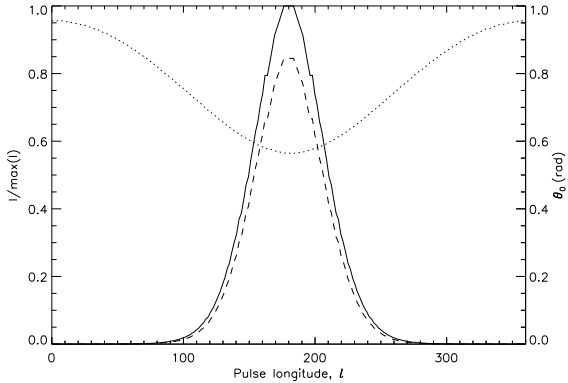}
\caption{Dipole field at a low emission altitude $r \ll r_\text{LC}$. Example of a pulse profile for a filled core beam with linear polarization $L = I \cos \theta_0$ and $(\alpha, i) = (70^\circ, 20^\circ)$. Solid, dashed and dotted curves represent the total polarized intensity $I$, degree of linear polarization $L$, and emission point colatitude $\theta_0$. Pulse longitude $l$ is measured in units of degrees. }
\label{m0w10lcostheta_dprofiles}
\end{figure}

Examples of the $I$ (solid curve) and $L$ (dashed curves) profiles for $(\alpha, i) = (70^\circ, 20^\circ)$ are shown in Figure \ref{m0w10lcostheta_dprofiles}. $I$ is normalised by its peak value. The pulse is single-peaked and maximally polarized at the peak. The dotted curve shows how the magnetic colatitude $\theta_0(t)$ of the emission point varies across one pulse period. We note the following trends. (i) The pulse profile narrows with increasing $i$ and $\alpha$ as the emission point moves further from the magnetic pole. The full-width half-maximum (FWHM)  decreases from $\approx 0.5$ phase units at $(\alpha, i) = (10^\circ, 10^\circ)$ to $\approx 0.15$ at $(\alpha, i) = (90^\circ, 90^\circ)$. (ii) For $\alpha = i$, where $\theta_0 = 0^\circ$ at the pulse peak, we have $L(t) = I(t)$. $L(t)$ decreases as $\lvert \alpha - i \rvert$ increases, e.g. we find $L(t) \approx 0.6 I(t)$ at $(\alpha, i) = (10^\circ, 90^\circ)$. The maximum value of $L/I$ can be calculated in terms of $\alpha$ and $i$ by solving equation (\ref{eq:dipole}) below for $\cos \theta_0$.

\begin{figure}
\centering
\includegraphics[scale=0.5]{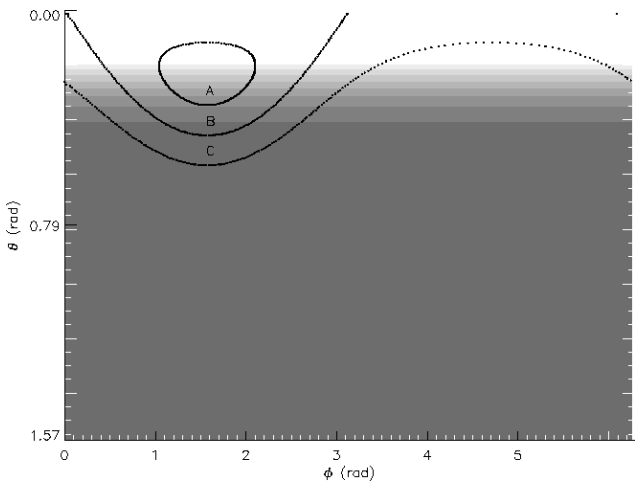}
\caption{Examples of the path traced by the emission point $\hat{\mathbf{x}}_0(t)$ in the body frame (curves), overplotted on the intensity map for a filled core beam (greyscale). The beam is brightest at $\theta = 0$ (corresponding to the $\mathbf{e}_3$ axis). Curves A: $(\alpha, i) = (20^\circ, 10^\circ)$, B: $(\alpha, i) = (20^\circ, 20^\circ)$, and C: $(\alpha, i) = (20^\circ, 30^\circ)$.}
\label{dipoleemission}
\end{figure}

In Figure \ref{dipoleemission}, we plot the path traced by $\mathbf{x}_0(t)$ in the body frame across one pulse period for three cases: (A) $(\alpha, i) = (20^\circ, 10^\circ)$, (B) $(20^\circ, 20^\circ)$, and (C) $(20^\circ, 30^\circ)$. For a dipole field, the emission colatitude $\theta_0$ is related to $\alpha$ and $i$ as follows:
\begin{equation}
\label{eq:dipole}
\cos (i - \alpha) = \frac{3 \cos^2 \theta_0 - 1}{(1 + 3 \cos^2 \theta_0)^{1/2}}.
\end{equation}
The path changes from an undulation at $\alpha < i$ (curve C in Figure \ref{dipoleemission}) to an ellipse at $\alpha > i$ (curve A). The difference in geometry is largely responsible for the difference in polarization properties observed for $\alpha < i$ and $\alpha > i$.

\begin{figure*}
\includegraphics[scale=0.8]{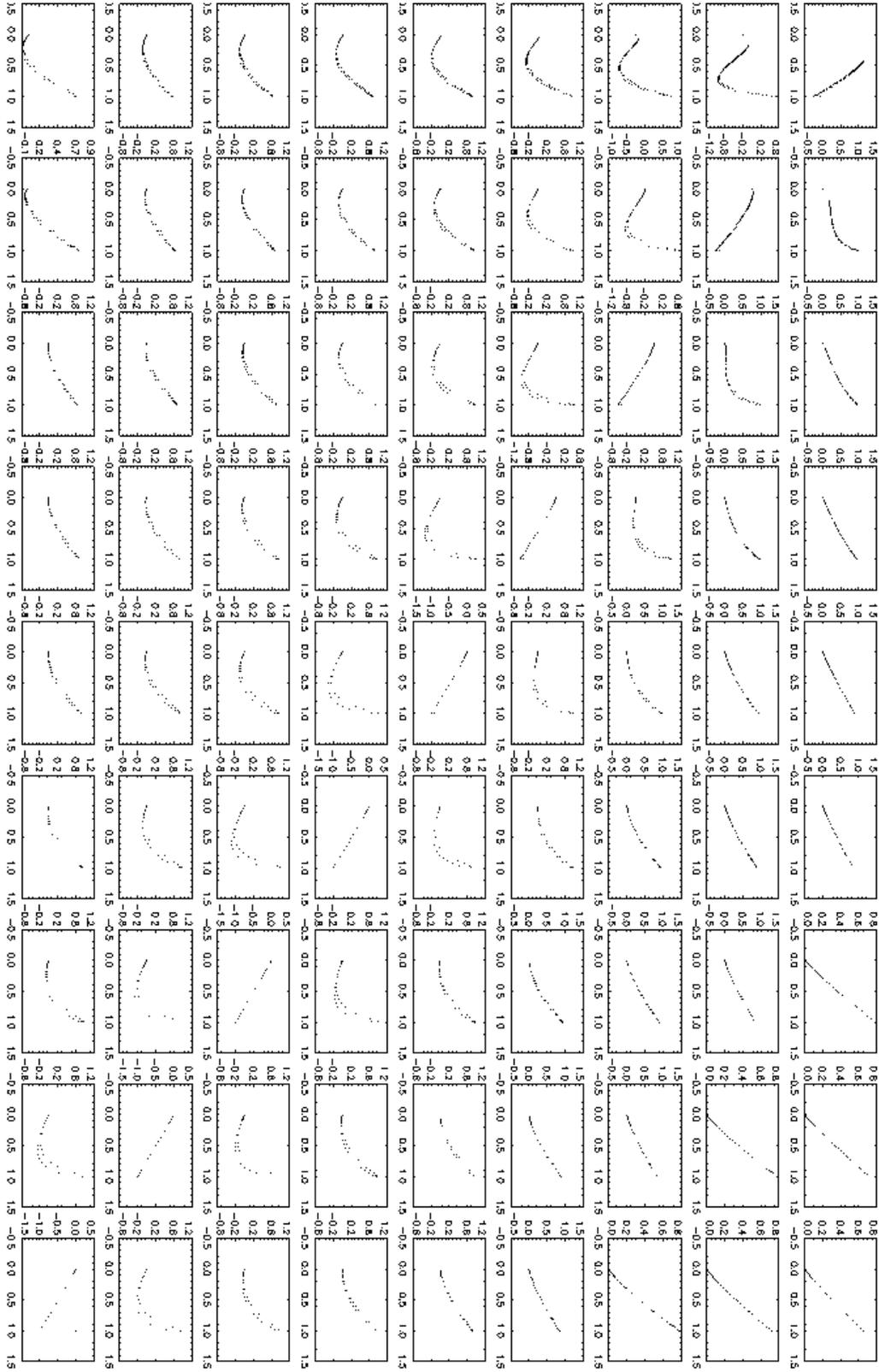}
\caption{Dipole field at a low emission altitude $r \ll r_\text{LC}$. Look-up table of Stokes phase portraits in the $I$-$Q$ plane for a filled core beam with degree of linear polarization $L = I \cos \theta_0$, where $\theta_0$ is the emission point colatitude. The panels are organised in landscape mode, in order of increasing $10^\circ \leq i \leq 90^\circ$ (left--right) and $10^\circ \leq \alpha \leq 90^\circ$ (top--bottom) in intervals of $10^\circ$. $I$ is plotted on the horizontal axis and normalised by its peak value. $Q$ is plotted on the vertical axis.}
\label{m0w10lcostheta_dqvsi}
\end{figure*}

\begin{figure*}
\includegraphics[scale=0.8]{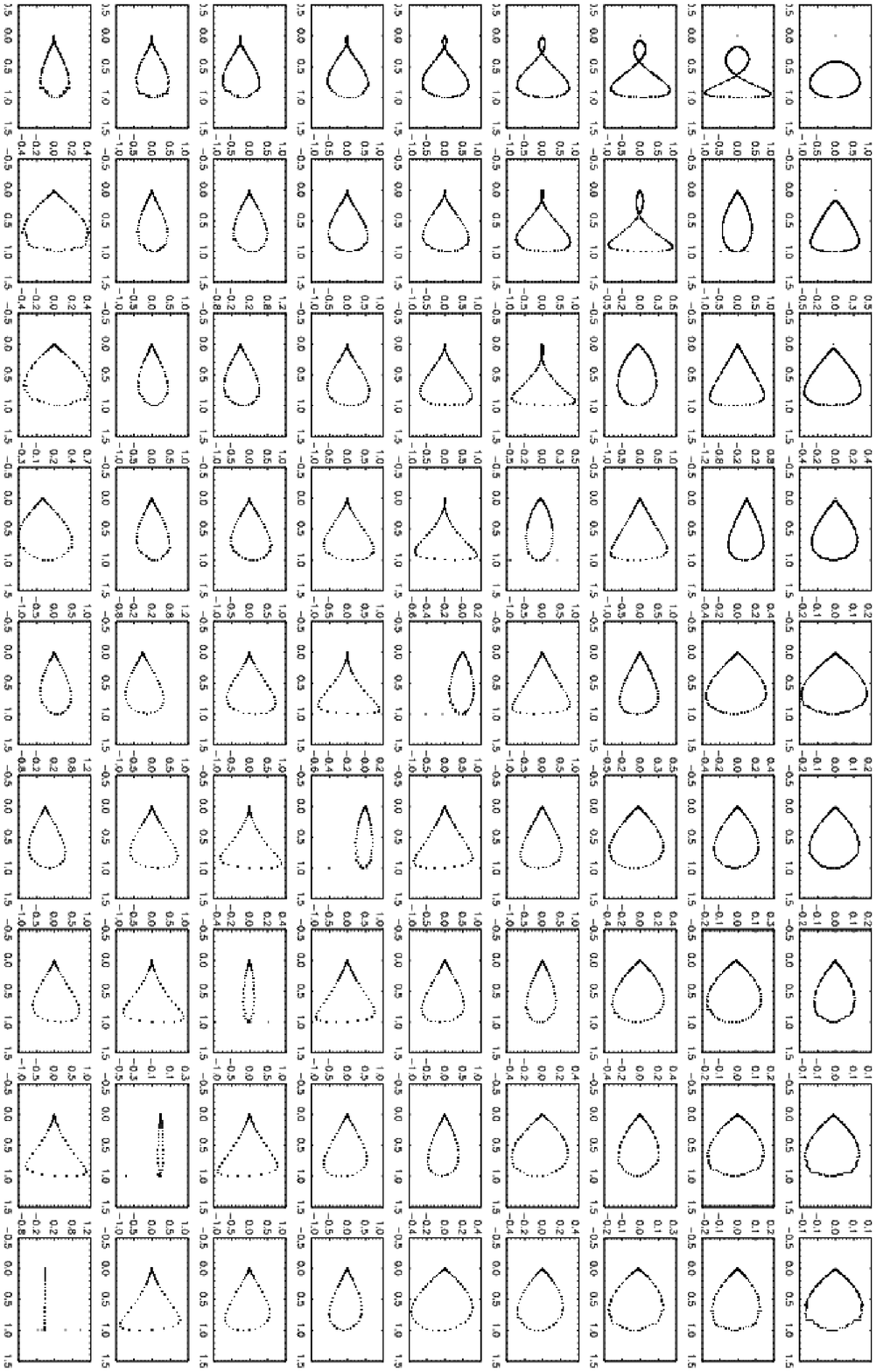}
\caption{Dipole field at $r \ll r_\text{LC}$. Layout as for Figure \ref{m0w10lcostheta_dqvsi}, but for $I$-$U$ ($I$ on the horizontal axis).}
\label{m0w10lcostheta_duvsi}
\end{figure*}

\begin{figure*}
\includegraphics[scale=0.8]{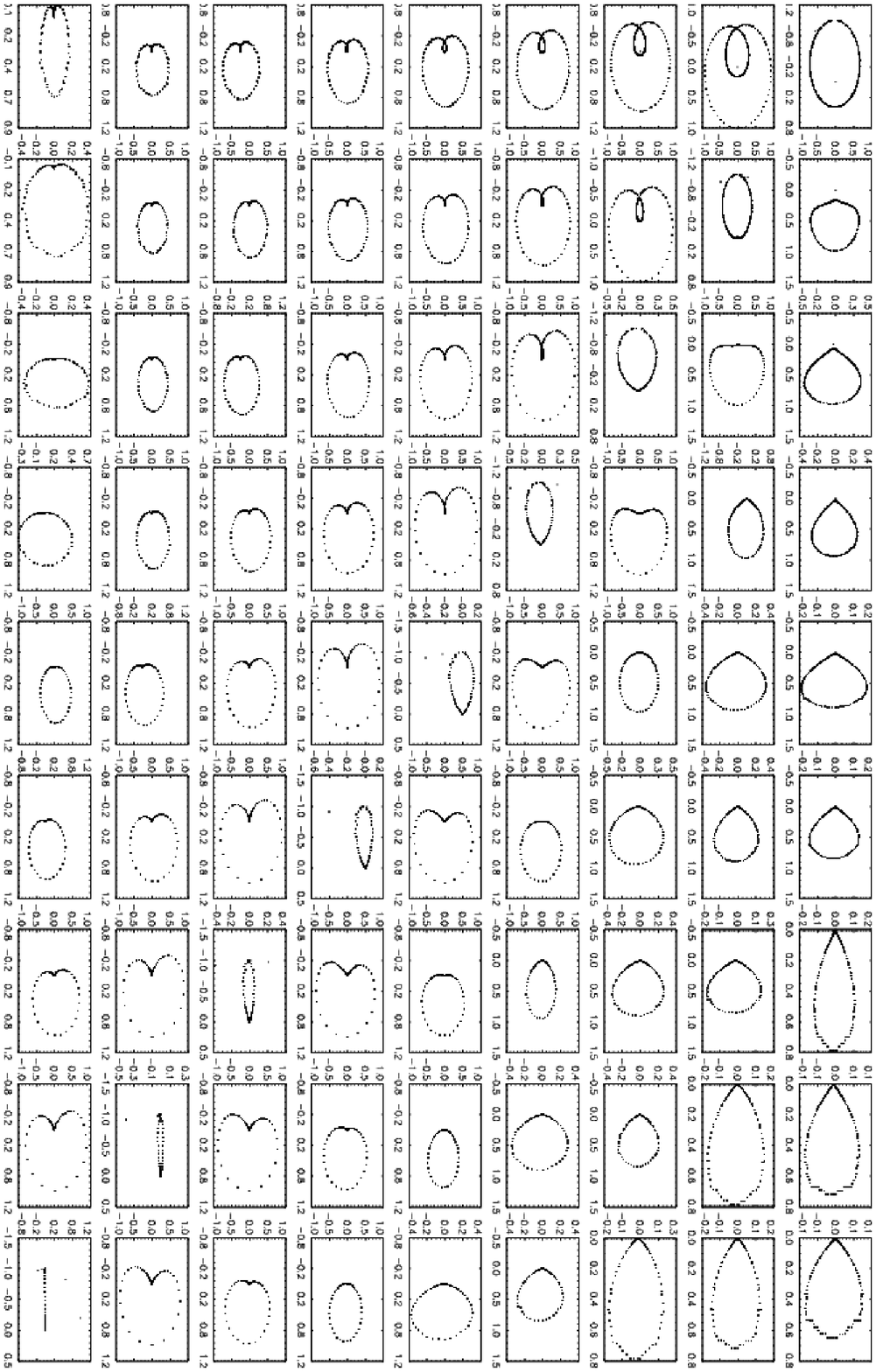}
\caption{Dipole field at $r \ll r_\text{LC}$. Layout as for Figure \ref{m0w10lcostheta_dqvsi}, but for $Q$-$U$ ($Q$ on the horizontal axis).}
\label{m0w10lcostheta_duvsq}
\end{figure*}

The Stokes phase portraits for a filled core with $L/I = \cos \theta_0$ are drawn in Figures \ref{m0w10lcostheta_dqvsi}--\ref{m0w10lcostheta_duvsq}. In each figure, the panels are organised in order of increasing $i$ (left--right) and $\alpha$ (top--bottom) in landscape mode. The panels range over $10^\circ \leq i \leq 90^\circ$ and $10^\circ \leq \alpha \leq 90^\circ$ and are separated by intervals of $10^\circ$. In Figures \ref{m0w10lcostheta_dqvsi} and \ref{m0w10lcostheta_duvsi}, $I$ is plotted on the horizontal axis (normalised by its peak value), and $Q$ and $U$ are plotted on the vertical axis. In Figure \ref{m0w10lcostheta_duvsq}, $Q$ is plotted on the horizontal axis and $U$ is plotted on the vertical axis. The same layout is used in all the look-up tables presented in this paper.

In the $I$-$Q$ plane (Figure \ref{m0w10lcostheta_dqvsi}), we note the following behaviour. (i) For $\alpha \leq 30^\circ$ and $i \geq 50^\circ$, the pulse traces out an approximately straight line segment which evolves into a slender hockey-stick shape as $\alpha$ increases and $i$ decreases. In the body frame, for $\alpha > i$ (below the diagonal in Figure \ref{m0w10lcostheta_dqvsi}), $\hat{\mathbf{x}}_0(t)$ traces out an ellipse in the quadrant $0 \leq \theta \leq \pi/2, 0 \leq \phi \leq \pi$ (see Figure \ref{dipoleemission}). As $\lvert \alpha - i \rvert$ increases, the ellipse moves further from $\theta = 0^\circ$ (and hence away from the beam peak). This reduces the curvature of the hockey stick. For $\alpha < i$ (above the diagonal in Figure \ref{m0w10lcostheta_dqvsi}), $\hat{\mathbf{x}}_0(t)$ traces out a sinusoidal curve across $0 \leq \phi \leq 2 \pi$. Again, the curve moves away from $\theta = 0^\circ$ as $\lvert \alpha - i \rvert$ increases, reducing the curvature of the hockey stick. The left horn of the hockey stick shortens as $\hat{\mathbf{x}}_0(t)$ moves away from $\theta = 0^\circ$.
(ii) For $\alpha = i$, the pattern is a straight line with $dQ/dI < 0$. In the body frame, $\hat{\mathbf{x}}_0(t)$ traces out a circle that crosses $\theta = 0^\circ$ symmetrically. (iii) The shapes are slightly larger in the panels adjacent to the $\alpha = i$ diagonal (i.e. $-0.2 < Q < 1.2$ for $\lvert \alpha - i \rvert \leq 10^\circ$) and decrease to e.g. $0 < Q < 0.6$ at $(\alpha, i) = (10^\circ, 90^\circ)$ at the corner of Figure \ref{m0w10lcostheta_dqvsi}.

In the $I$-$U$ plane  (Figure \ref{m0w10lcostheta_duvsi}), we note the following behaviour. (i) The pattern is symmetric about $U = 0$ and traces a sideways balloon shape for most angles, with the pointy tip at $(U, I) = (0, 0)$. (ii) For $\alpha \leq 50^\circ$ and $i \leq 30^\circ$, the pattern twists into a figure-eight. This twist occurs when the path traced by $\hat{\mathbf{x}}_0(t)$ in the body frame elongates from a circle to an oval. (iii) For $\alpha = i$, the phase portrait is roughly oval, with major and minor axes along the $I$- and $U$-axes respectively. It narrows with increasing $\alpha = i$. (iv) The patterns are broadest adjacent to the $\alpha = i$ diagonal (e.g. $-1 < U < 1$ for $\lvert \alpha - i \rvert < 10^\circ$) and narrow by up to 90\% as $\lvert \alpha - i \rvert$ increases to the corners of Figure \ref{m0w10lcostheta_duvsi}.

In the $Q$-$U$ plane (Figure \ref{m0w10lcostheta_duvsq}), we see a mix of ovals, balloons and heart shapes with the following properties. (i) For $\alpha < i$, above the diagonal, the shapes broaden with increasing $\alpha$, while the reverse is true for $\alpha > i$, below the diagonal. (ii) On the diagonal, the ovals narrow with increasing $\alpha = i$. (iii) In some cases, e.g. $(\alpha, i) = (50^\circ, 40^\circ)$, a cusp forms at the start and end points of the pulse, where $(U, Q) = (0, 0)$. This cusp twists into a secondary oval for $10^\circ \leq \alpha \leq 50^\circ$ and $i \leq 20^\circ$. As in the $I$-$U$ plane, this secondary loop forms when the pattern traced by $\hat{\mathbf{x}}_0(t)$ elongates from a circle to an oval. The size of the loop is proportional to the ellipticity of the oval. (iv) As in Figure \ref{m0w10lcostheta_duvsi}, the patterns in the phase portrait are broadest for $\lvert \alpha - i \rvert < 10^\circ$ directly adjacent to the $\alpha = i$ diagonal, and narrow as $\lvert \alpha - i \rvert$ increases.

\begin{figure*}
\includegraphics[scale=0.8]{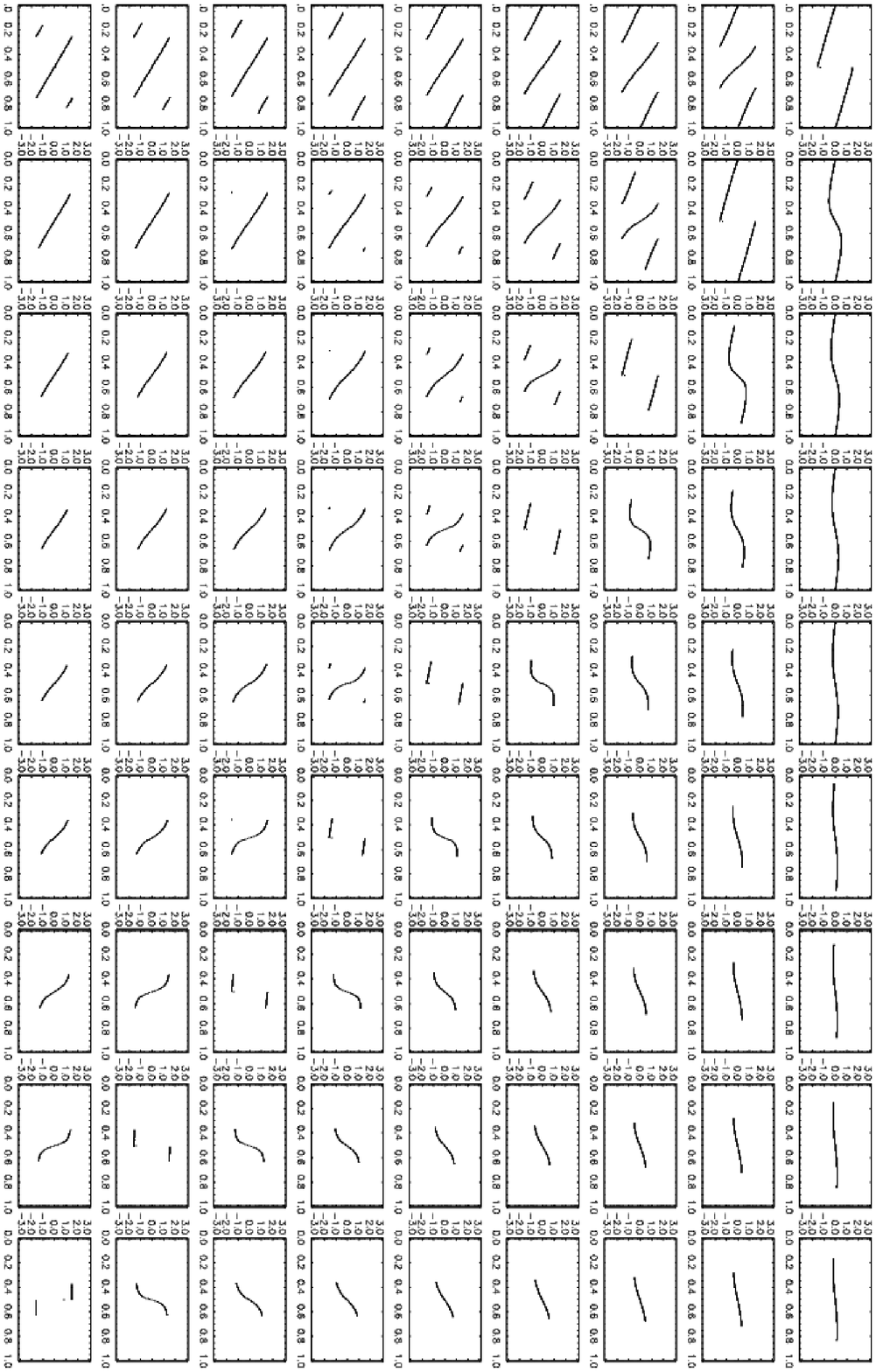}
\caption{Dipole field at $r \ll r_\text{LC}$. Layout as for Figure \ref{m0w10lcostheta_dqvsi}, but for position angle (on the vertical axis in landscape orientation, in units of radians) versus pulse longitude (on the horizontal axis, in units of $2 \pi$ radians).}
\label{m0w10lcostheta_dpa}
\end{figure*}

In Figure \ref{m0w10lcostheta_dpa}, we plot the PA swings corresponding to each panel in Figures \ref{m0w10lcostheta_dqvsi}--\ref{m0w10lcostheta_duvsq}. We plot only the parts of the swing that are illuminated by the pulse, i.e. when $\lvert L \rvert \geq 10^{-2}$. The threshold is arbitrary and should be adjusted for instrumental resolution when modelling actual data. We note the following behaviour. (i) The PA swings are identical to those predicted by the rotating vector model, as expected for a dipole magnetic field, with the S-shape steepening as $\alpha$ and $i$ increase. (ii) Above the $\alpha = i$ diagonal, we find $d(\text{PA})/d l > 0$, where $l$ is the pulse longitude, whereas the opposite is true below the diagonal. This is consistent with the rotating vector model's prediction that $d(\text{PA})/d l = \sin \alpha/\sin (i - \alpha)$ at the inflection point. (iii) Along and below the $\alpha = i$ diagonal, there is phase wrapping in some swings, e.g. clearly visible for $i = 10^\circ$. 

\begin{figure}
\centering
\includegraphics[scale=0.8]{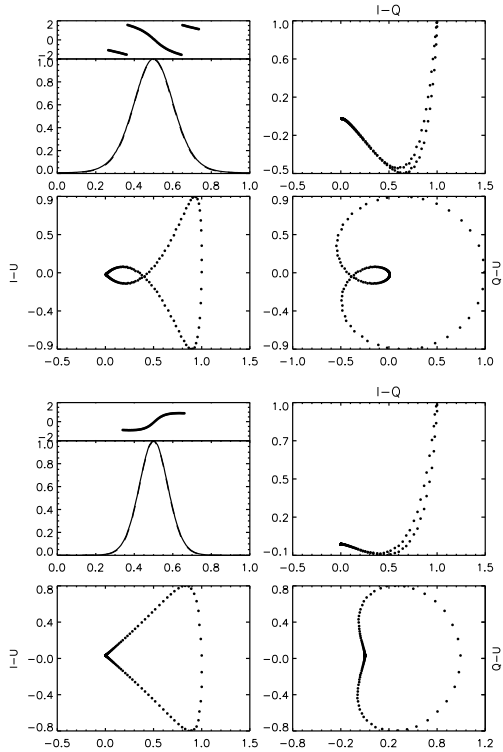}
\caption{Two orientations with equal PA-swing gradients ($\lvert d (\text{PA})/d l \rvert$) at the inflection point, but different Stokes phase portraits. Top four panels: $(\alpha, i) = (30^\circ, 20^\circ)$, bottom four panels: $(\alpha, i) = (30^\circ, 40^\circ)$. Within each half of the figure, the four subpanels display (clockwise from top left): (a) $I/I_\text{max}$ (bottom) and PA (top; in radians) as functions of pulse longitude, (b) $I$-$Q$, (c) $Q$-$U$, and (d) $I$-$U$. }
\label{detailedalpha}
\end{figure}

One limitation of the rotating vector model is that it is difficult to distinguish between orientations with similar $\lvert d (\text{PA})/d l \rvert$. This occurs, for example, at two orientations with equal $\alpha$ and $i - \alpha$. For a pure dipole at $r \ll r_\text{LC}$, the Stokes phase portraits face a similar problem for $\alpha \gtrsim 40^\circ$. Comparing $(\alpha, i) = (50^\circ, 40^\circ)$ and $(50^\circ, 60^\circ)$, the hockey sticks in the $I$-$Q$ plane are look alike. The balloons and heart shapes in the $I$-$U$ and $Q$-$U$ planes are slightly different, although the differences are close to $I = 0$ and may be hidden if the real data is noisy. However, for $\alpha \lesssim 30^\circ$, the Stokes phase portraits clearly discriminate between different orientations with equal $i - \alpha$, in a way that the PA swing does not. For example, let us compare the orientations $(\alpha, i) = (30^\circ, 20^\circ)$ and $(30^\circ, 40^\circ)$ in Figure \ref{detailedalpha}. In the $I$-$Q$ plane, the hockey stick at $(\alpha, i) = (30^\circ, 20^\circ)$ is more curved than at $(\alpha, i) = (30^\circ, 40^\circ)$. In the $I$-$U$ plane, there is a distinct figure-eight instead of a balloon, whereas in the $Q$-$U$ plane, there are two interlocking ovals instead of a heart shape. In this case, the Stokes phase portraits provide an obvious way to differentiate between the two orientations.
We note that in Sections \ref{sec:dipoleaberration} and \ref{sec:toroidal}, the differences in the phase portraits for orientations with equal $\lvert i - \alpha \rvert$ are even more marked. Hence the Stokes phase portraits are useful supplements to the PA swing in many instances.

The Stokes phase portraits produced by $L = I \cos \theta_0$ are extremely similar to those produced by a model with $L \propto I$. For the sake of brevity, we do not investigate the latter model here.

\subsubsection{$L = I \sin \theta_0$}
\label{sec:dipolelsintheta}
\begin{figure}
\includegraphics[scale=0.45]{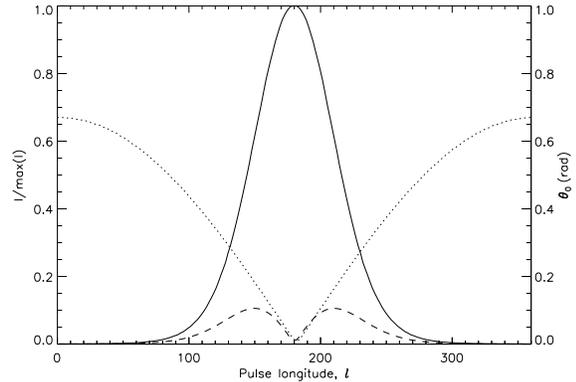}
\caption{Dipole field at $r \ll r_\text{LC}$. Example of a pulse profile for a filled core beam with linear polarization $L = I \sin \theta_0$ and $(\alpha, i) = (30^\circ, 30^\circ)$. Solid, dashed and dotted curves represent the total polarized intensity $I$, degree of linear polarization $L$, and emission point colatitude $\theta_0$. Pulse longitude $l$ is measured in units of degrees.}
\label{m0w10lsintheta_dprofiles}
\end{figure}

Examples of the $I$ (solid curve) and $L$ (dashed curve) profiles for $(\alpha, i) = (30^\circ, 30^\circ)$ are shown in Figure \ref{m0w10lsintheta_dprofiles}. $I$ is normalised by its peak value. The magnetic colatitude $\theta_0(t)$ of the emission point (dotted curve) is also shown. The following trends are observed. (i) Along the $\alpha = i$ diagonal, where $\theta_0$ goes to zero at the centre of the pulse, the $\sin \theta_0$ dependence results in a double-peaked $L$ profile. Away from the diagonal, $L/I$ increases, reaching $0.8$ at $(\alpha, i) = (90^\circ, 10^\circ)$. (ii) Away from the diagonal, $L$ follows $I$, as in Section \ref{sec:dipolelcostheta}, resulting in similar phase portraits.  

\begin{figure*}
\includegraphics[scale=0.8]{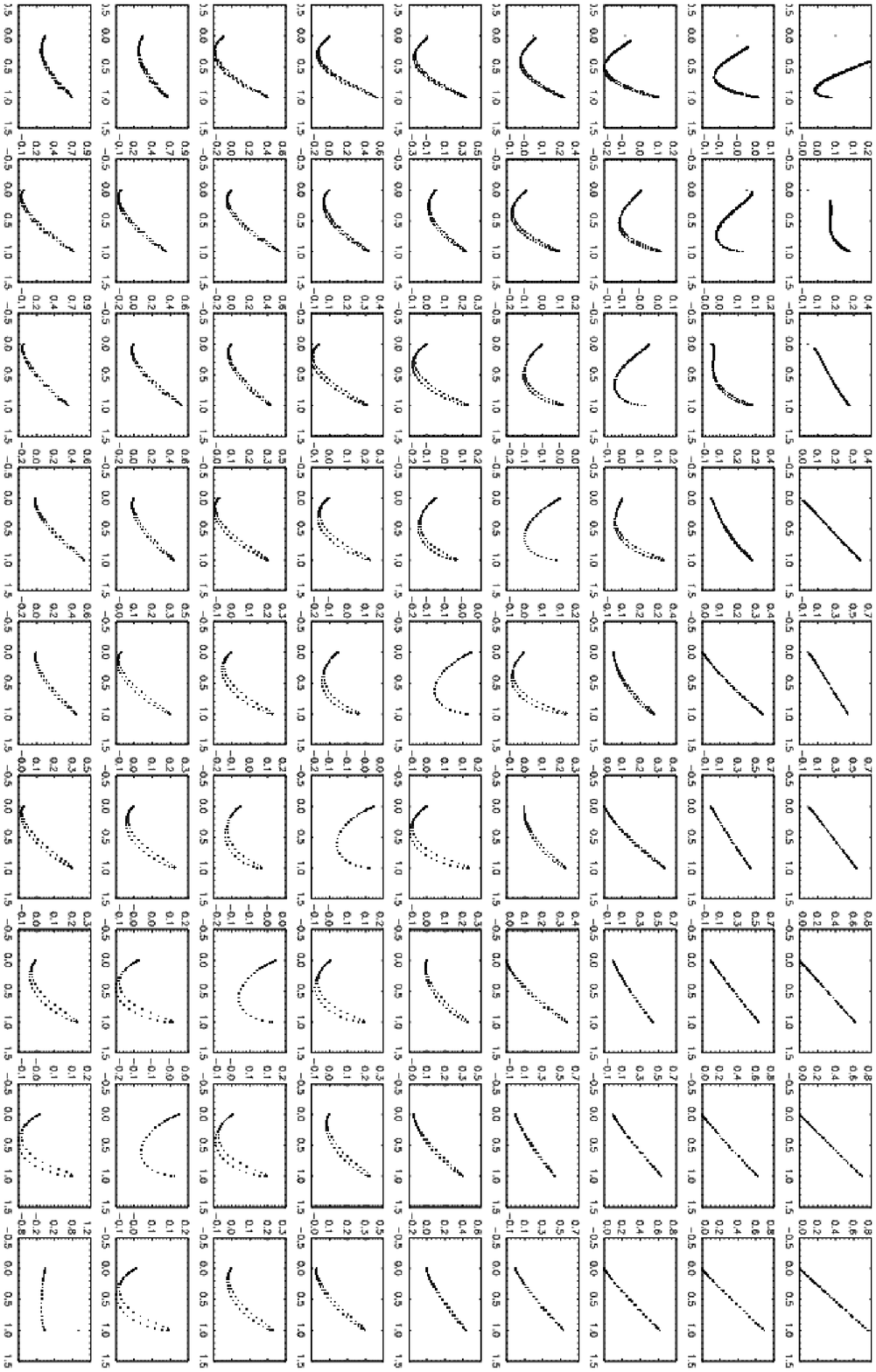}
\caption{Dipole field at $r \ll r_\text{LC}$. Look-up table of Stokes phase portraits in the $I$-$Q$ plane for a filled core beam with degree of linear polarization $L = I \sin \theta_0$, where $\theta_0$ is the emission point colatitude. The panels are organised in landscape mode, in order of increasing $10^\circ \leq i \leq 90^\circ$ (left--right) and $10^\circ \leq \alpha \leq 90^\circ$ (top--bottom) in intervals of $10^\circ$. $I$ is plotted on the horizontal axis and normalised by its peak value. $Q$ is plotted on the vertical axis.}
\label{m0w10lsintheta_dqvsi}
\end{figure*}

\begin{figure*}
\includegraphics[scale=0.8]{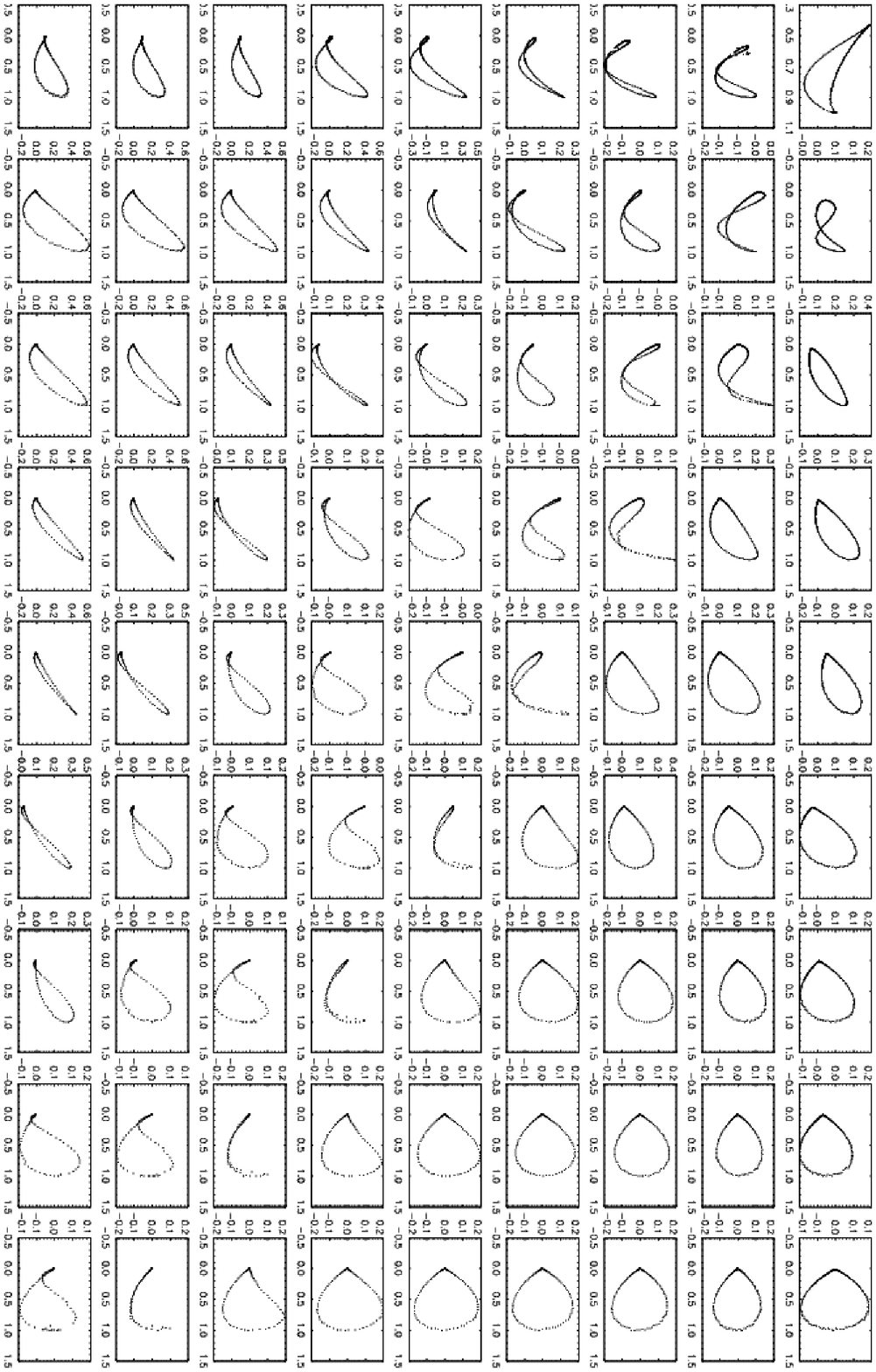}
\caption{Dipole field at $r \ll r_\text{LC}$. As for Figure \ref{m0w10lsintheta_dqvsi}, but for $I$-$U$ ($I$ on horizontal axis).}
\label{m0w10lsintheta_duvsi}
\end{figure*}

\begin{figure*}
\includegraphics[scale=0.8]{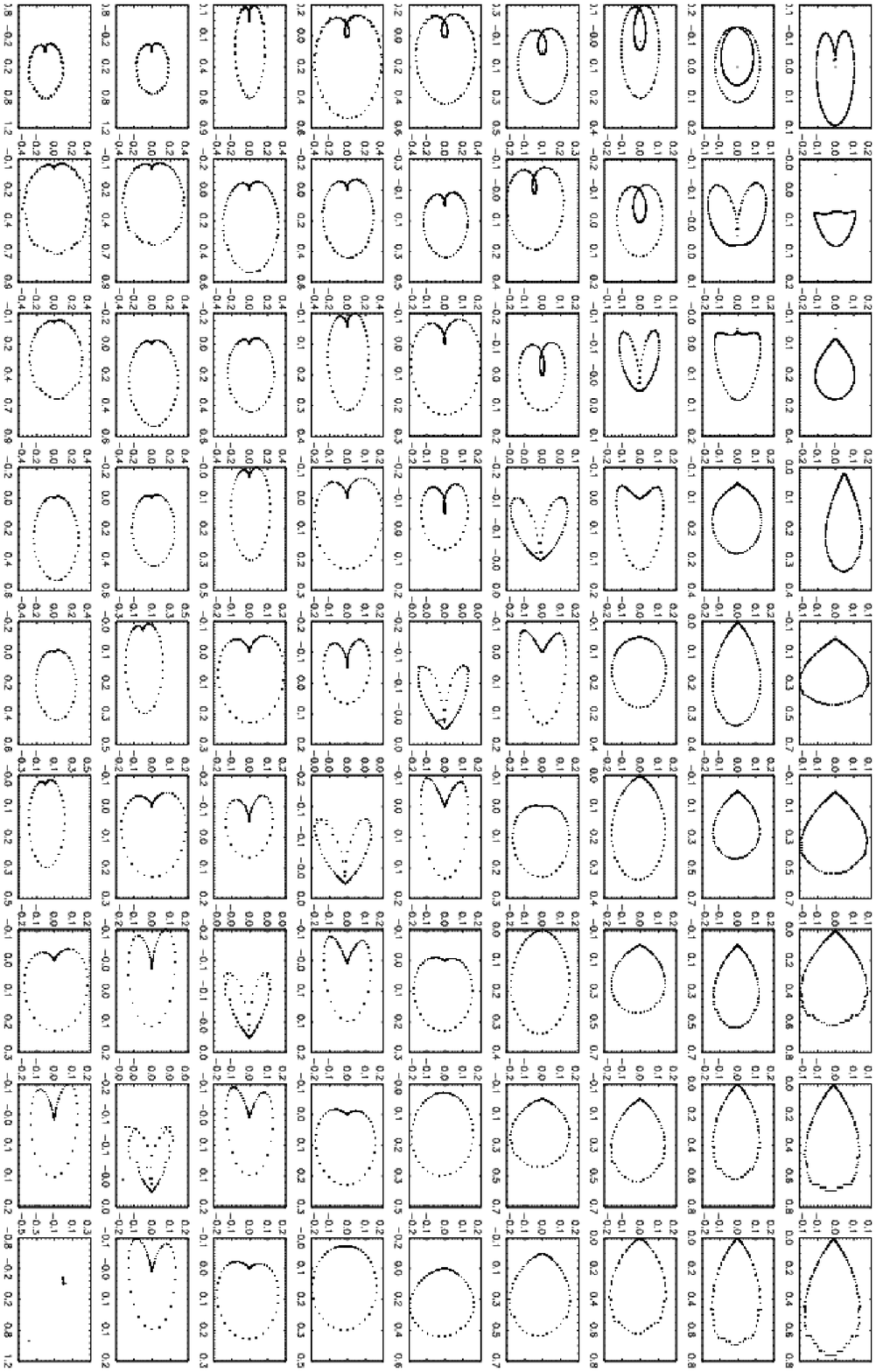}
\caption{Dipole field at $r \ll r_\text{LC}$. As for Figure \ref{m0w10lsintheta_dqvsi}, but for $Q$-$U$ ($Q$ on horizontal axis).}
\label{m0w10lsintheta_duvsq}
\end{figure*}

\begin{figure*}
\includegraphics[scale=0.8]{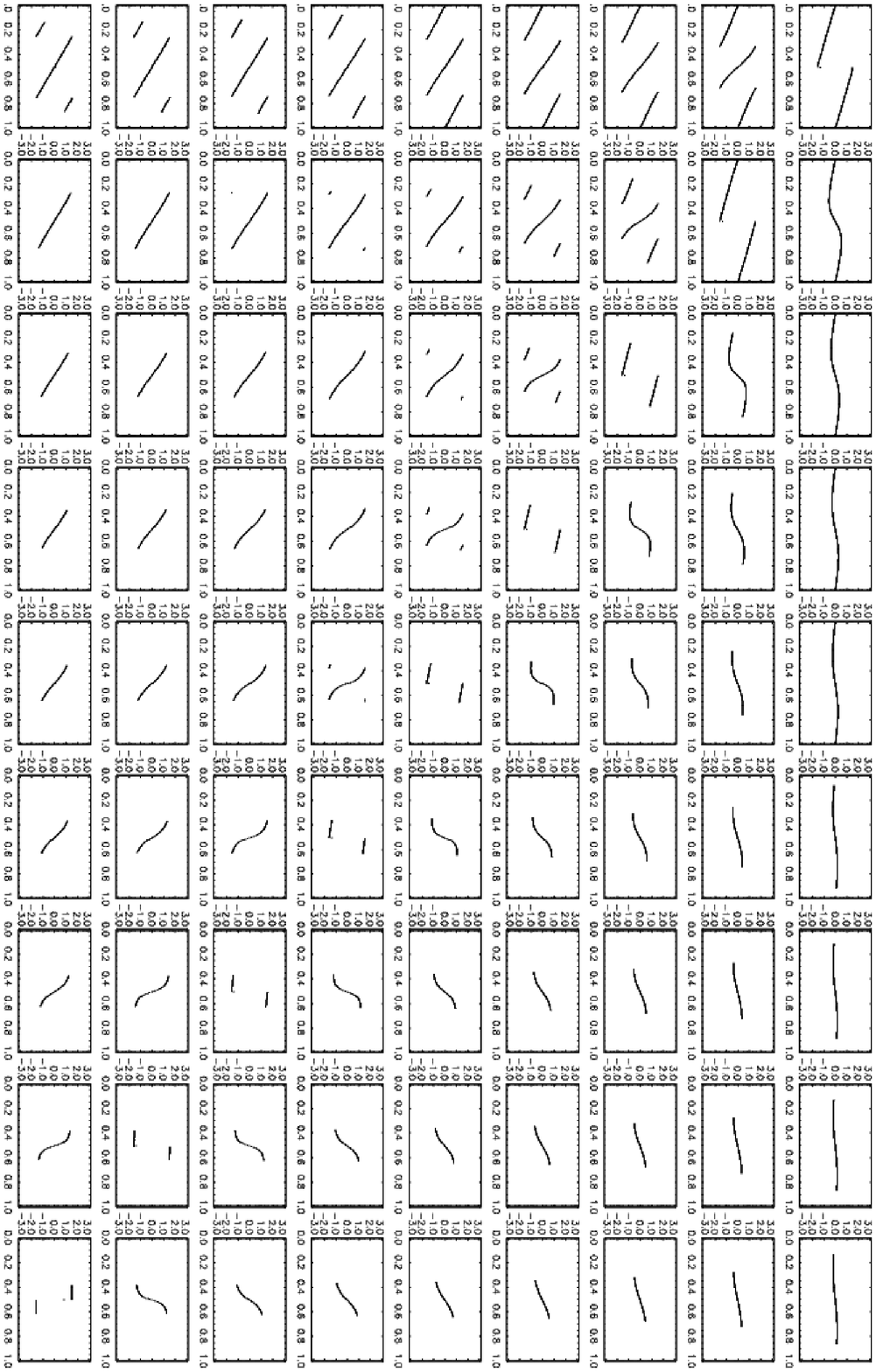}
\caption{Dipole field at $r \ll r_\text{LC}$. Layout as for Figure \ref{m25w10lcostheta_dqvsi}, but for position angle (on the vertical axis in landscape orientation, in units of radians) versus pulse longitude (on the horizontal axis, in units of $2 \pi$ radians).}
\label{m0w10lsintheta_dpa}
\end{figure*}

The Stokes phase portraits for a filled core with $L/I = \sin \theta_0$ are drawn in Figures \ref{m0w10lsintheta_dqvsi}--\ref{m0w10lsintheta_duvsq}. Comparing the Stokes phase portraits and PA swings to those of Section \ref{sec:dipolelcostheta}, we note the following differences. (i) In the $I$-$Q$ phase portrait (Figure \ref{m0w10lsintheta_dqvsi}), along the $\alpha = i$ diagonal, instead of a straight line, we see a `flipped' hockey stick which slopes in the opposite direction to the hockey sticks for $\alpha \neq i$. (ii) In the $I$-$U$ phase portrait (Figure \ref{m0w10lsintheta_duvsi}), along the $\alpha = i$ diagonal, there are broad balloon shapes instead of narrow ovals, and they extend leftwards from $I = 1$ as opposed to rightwards from $I = 0$. (iii) In the $Q$-$U$ plane (Figure \ref{m0w10lsintheta_duvsq}), instead of ovals, the diagonal now contains heart shapes with an extremely pronounced cusp, narrowing with increasing $\alpha = i$. 
There is no noticeable difference between the PA swings in this model (Figure \ref{m0w10lsintheta_dpa}) and those of Section \ref{sec:dipolelcostheta}. This is another example of how the Stokes phase portraits reveal differences in the beam geometry and polarization pattern which are not apparent from the PA swings.

In summary, a filled core beam produces single-peaked pulses, resulting in several unique shapes in the Stokes planes. In the $I$-$Q$ plane, a hockey stick shape is typically seen. In the $I$-$U$ plane, we see a balloon shape which twists into a figure-eight in the range $\alpha \leq 50^\circ, i \leq 20^\circ$. In the $Q$-$U$ plane, we see heart shapes for all $i < \alpha$, and balloons for $i - \alpha \geq 10^\circ$. 

\subsection{Hollow cone emission}
\label{sec:dipolecone}
\subsubsection{$L = I \cos \theta_0$}
\label{sec:dipoleconecostheta}
\begin{figure}
\includegraphics[scale=0.45]{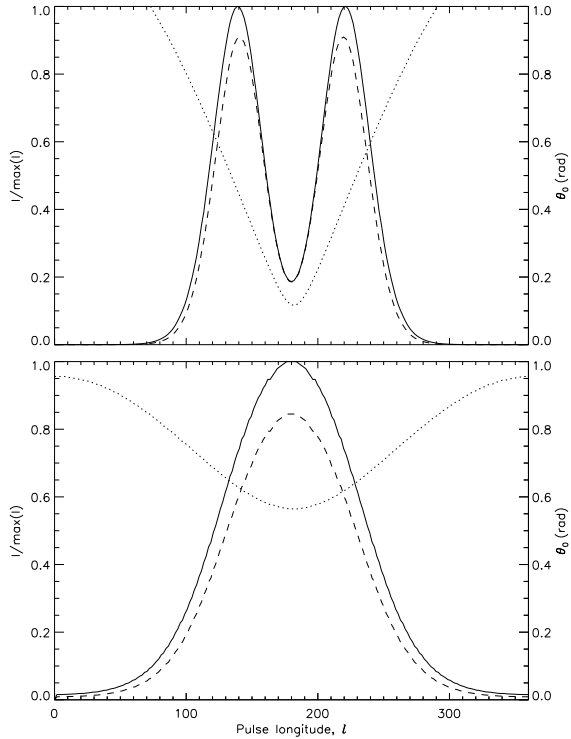}
\caption{Dipole field at $r \ll r_\text{LC}$. Examples of pulse profiles for a hollow cone with opening angle $25^\circ$ and linear polarization $L = I \cos \theta_0$ for two different orientations $(\alpha, i)$ (top and bottom panels). Solid, dashed and dotted curves represent the total polarized intensity $I$, degree of linear polarization $L$, and emission point colatitude $\theta_0$. Pulse longitude is measured in units of degrees. Top: double-peaked pulse with $(\alpha, i) = (70^\circ, 60^\circ)$; bottom: single-peaked pulse with $(\alpha, i) = (70^\circ, 20^\circ)$.}
\label{m25w10lcostheta_dprofiles}
\end{figure}

Examples of the $I$ (solid curve) and $L$ (dashed curve) profiles for $(\alpha, i) = (70^\circ, 20^\circ)$ and $(70^\circ, 60^\circ)$ are shown in Figure \ref{m25w10lcostheta_dprofiles}. $I$ is normalised by its peak value. The magnetic colatitude $\theta_0(t)$ of the emission point (dotted curve) is also shown. The following trends are observed. (i) Emission from a hollow cone results in double-peaked pulse profiles in the range $\lvert i - \alpha \rvert \leq 30^\circ$, where the path traced by $\hat{\mathbf{x}}_0(t)$ crosses the edge of the cone twice. (ii) For $\alpha < 45^\circ$, the central dip becomes shallower as $i$ increases. The reverse is true for $\alpha > 45^\circ$. (iii) We find $L/I \approx 1$ for $\alpha = i$, decreasing to $\approx$ 0.6 at $(\alpha, i) = (10^\circ, 90^\circ)$. The pulses are generally $\approx 50$\% broader than in Section \ref{sec:dipolecore}, even though $\sigma$ is the same in both instances.

\begin{figure*}
\includegraphics[scale=0.8]{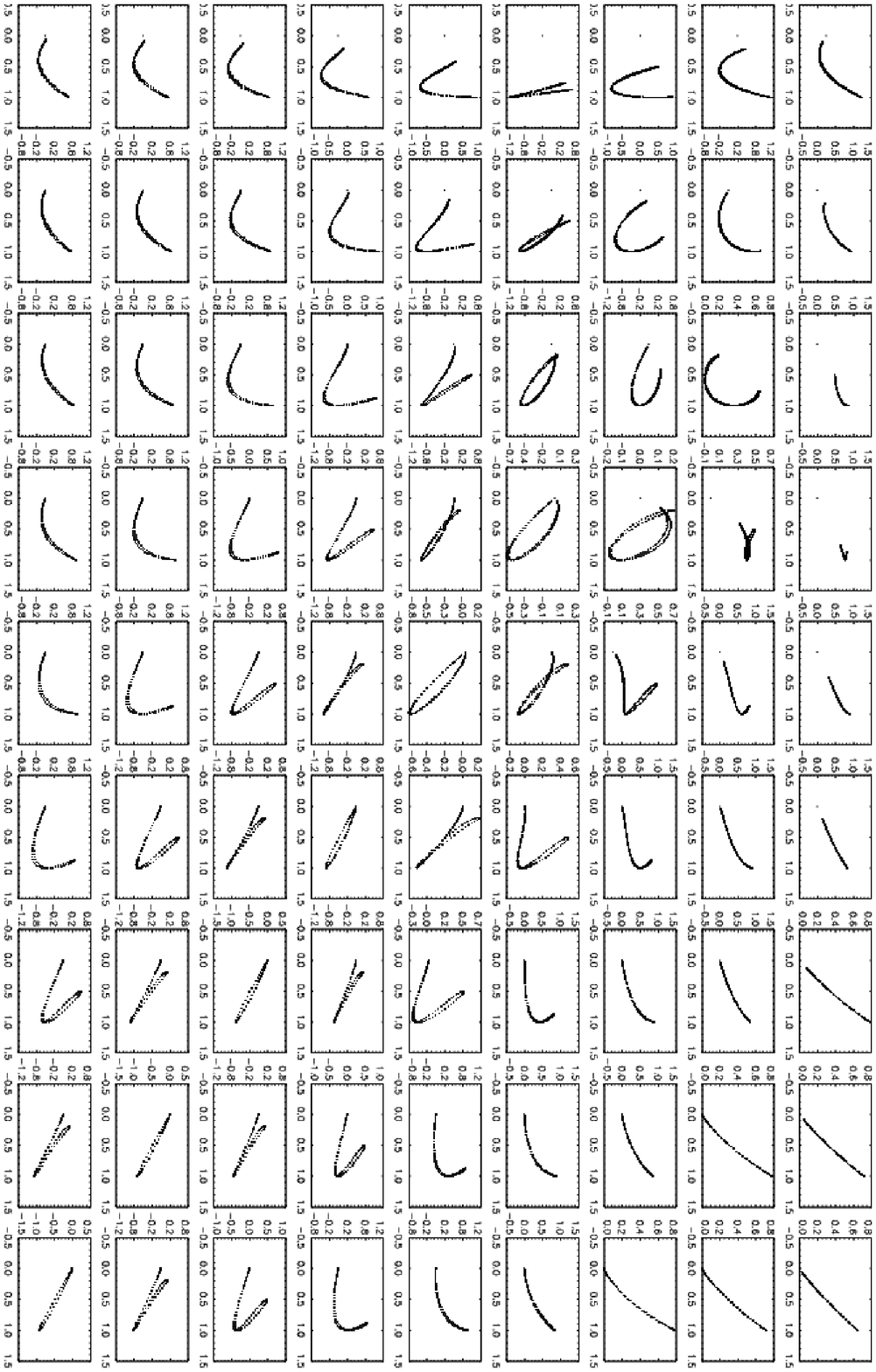}
\caption{Dipole field at $r \ll r_\text{LC}$. Look-up table of Stokes phase portraits in the $I$-$Q$ plane for a hollow cone with opening angle $25^\circ$ with degree of linear polarization $L = I \cos \theta_0$, where $\theta_0$ is the emission point colatitude. The panels are organised in landscape mode, in order of increasing $10^\circ \leq i \leq 90^\circ$ (left--right) and $10^\circ \leq \alpha \leq 90^\circ$ (top--bottom) in intervals of $10^\circ$. $I$ is plotted on the horizontal axis and normalised by its peak value. $Q$ is plotted on the vertical axis.}
\label{m25w10lcostheta_dqvsi}
\end{figure*}

\begin{figure*}
\includegraphics[scale=0.8]{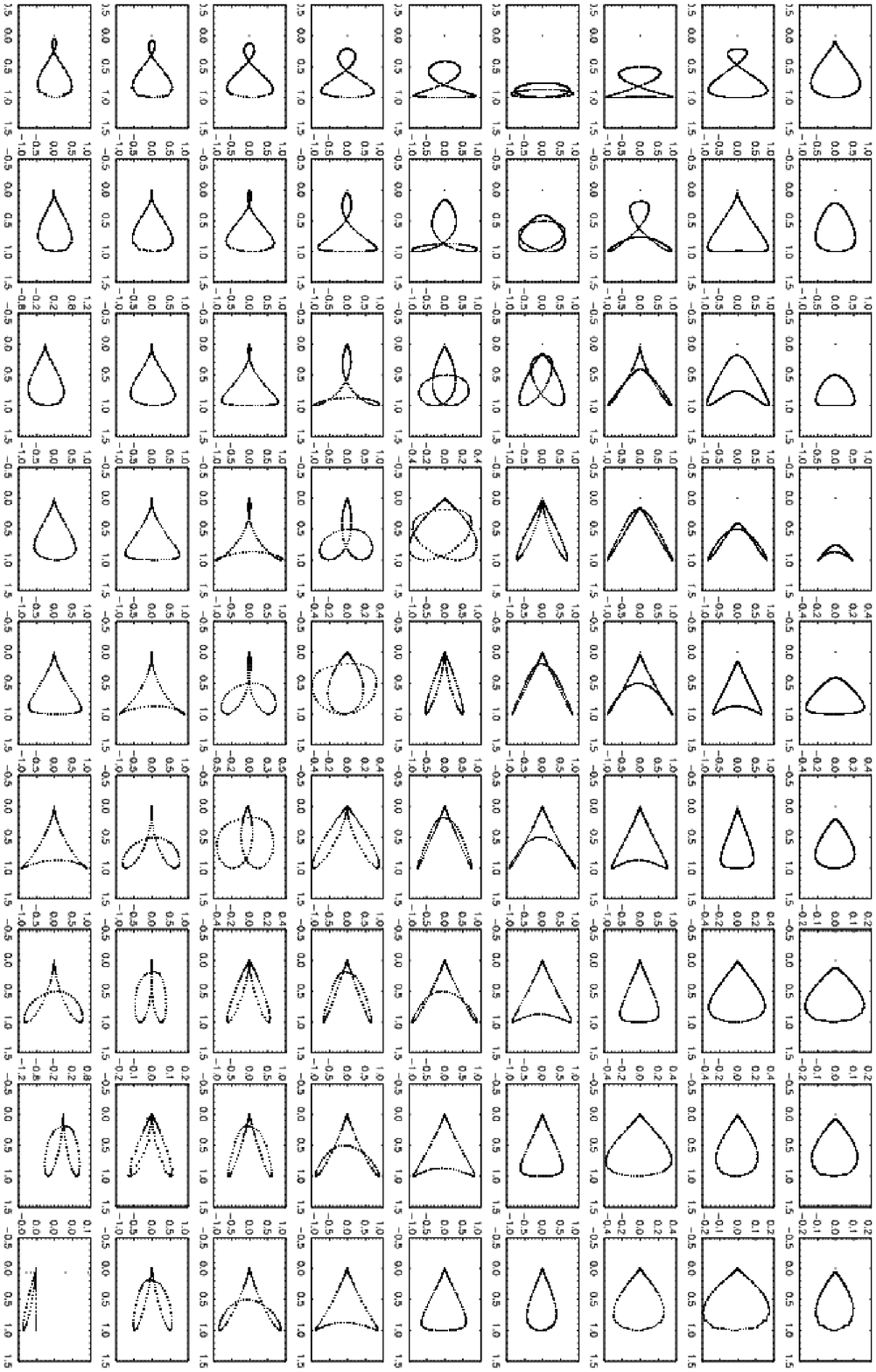}
\caption{Dipole field at $r \ll r_\text{LC}$. Layout as for Figure \ref{m25w10lcostheta_dqvsi}, but for $I$-$U$ ($I$ on horizontal axis). }
\label{m25w10lcostheta_duvsi}
\end{figure*}

\begin{figure*}
\includegraphics[scale=0.8]{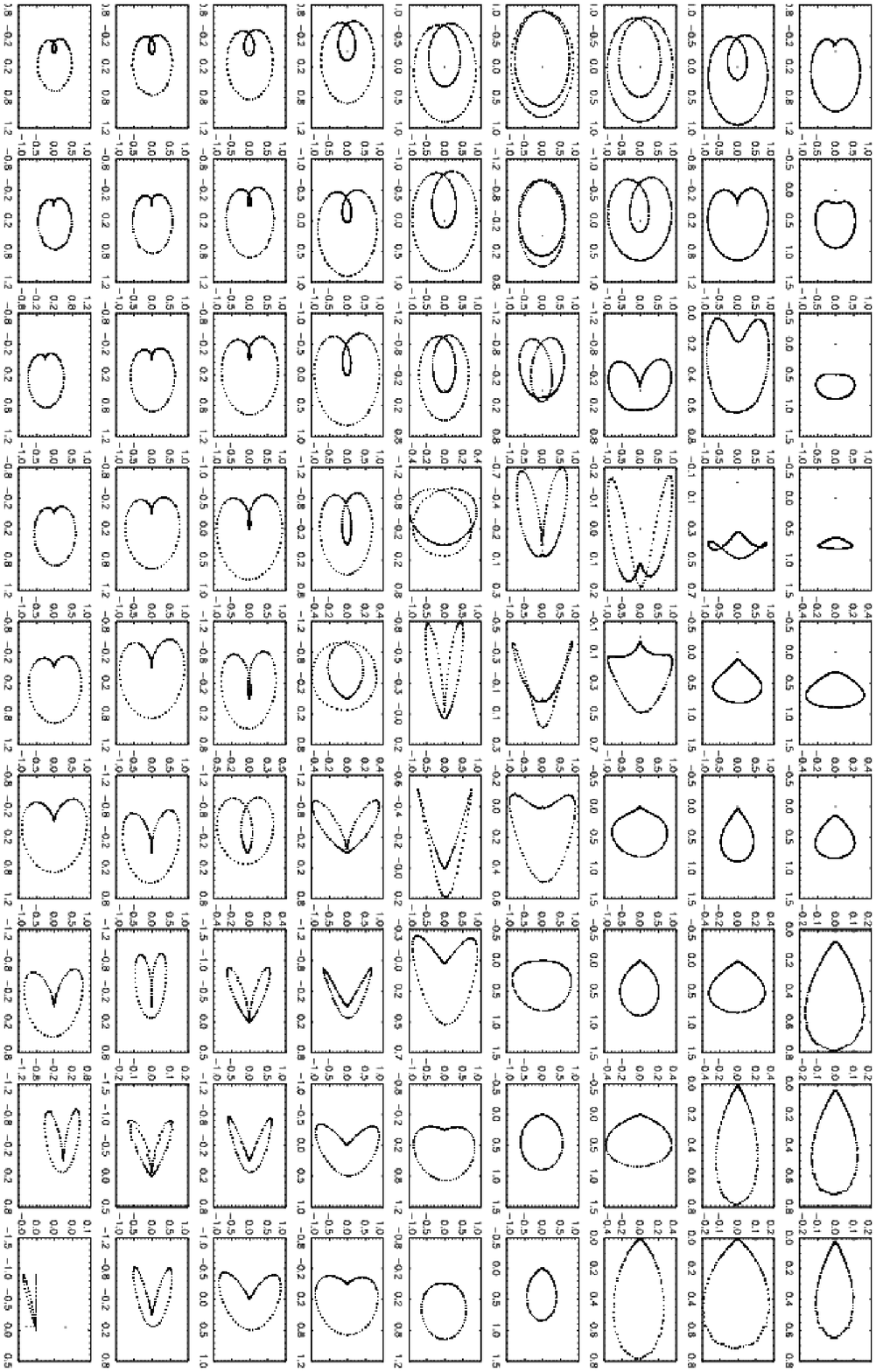}
\caption{Dipole field at $r \ll r_\text{LC}$. Layout as for Figure \ref{m25w10lcostheta_dqvsi}, but for $Q$-$U$ ($Q$ on horizontal axis).}
\label{m25w10lcostheta_duvsq}
\end{figure*}

The Stokes phase portraits for a hollow cone with $L/I = \cos \theta_0$ are drawn in Figures \ref{m25w10lcostheta_dqvsi}--\ref{m25w10lcostheta_duvsq}. The paths of $\hat{\mathbf{x}}_0(t)$ shown in Figure \ref{dipoleemission} still apply but they cross a different beam pattern with a bright band at $\theta_0 = \rho$. Hence the Stokes phase portraits look different to the left and right of the $\alpha = i$ diagonal. 

In the $I$-$Q$ phase portraits (Figure \ref{m25w10lcostheta_dqvsi}), we note the following trends. (i) Where the pulse is double-peaked, the hockey sticks twist into a tilted $\gamma$ shape. Where the pulse is single-peaked, the patterns are similar to Figure \ref{m0w10lcostheta_dqvsi}. (ii) For $\alpha = i$, the hockey stick twists into a $\gamma$-shape as $\alpha = i$ increases.

In the $I$-$U$ phase portraits (Figure \ref{m25w10lcostheta_duvsi}), we note the following trends. (i) Where the pulse profile is double-peaked, the $I$-$U$ portraits feature two interlocking loops forming a triangular shape with twisted corners.  Where the pulse is single-peaked, the patterns are similar to those of Figure \ref{m0w10lcostheta_duvsi}. (ii) For $\alpha = i$, the pattern evolves from a balloon to a triangular shape with an increasingly deep central cusp as $\alpha = i$ increases.

Additional twisting in the patterns is also seen in the $Q$-$U$ plane (Figure \ref{m25w10lcostheta_duvsq}), although the patterns are still symmetric about $U = 0$. (i) For $\alpha = i$, the patterns look qualitatively similar to those of Section \ref{sec:dipolelsintheta}, although they are $\sim 90$\% broader, covering $-1 \leq U \leq 1$. (ii) For $\alpha > i$, we see heart shapes and nested ovals, whereas for $\alpha < i$, we see simpler, convex balloons.

\begin{figure*}
\includegraphics[scale=0.8]{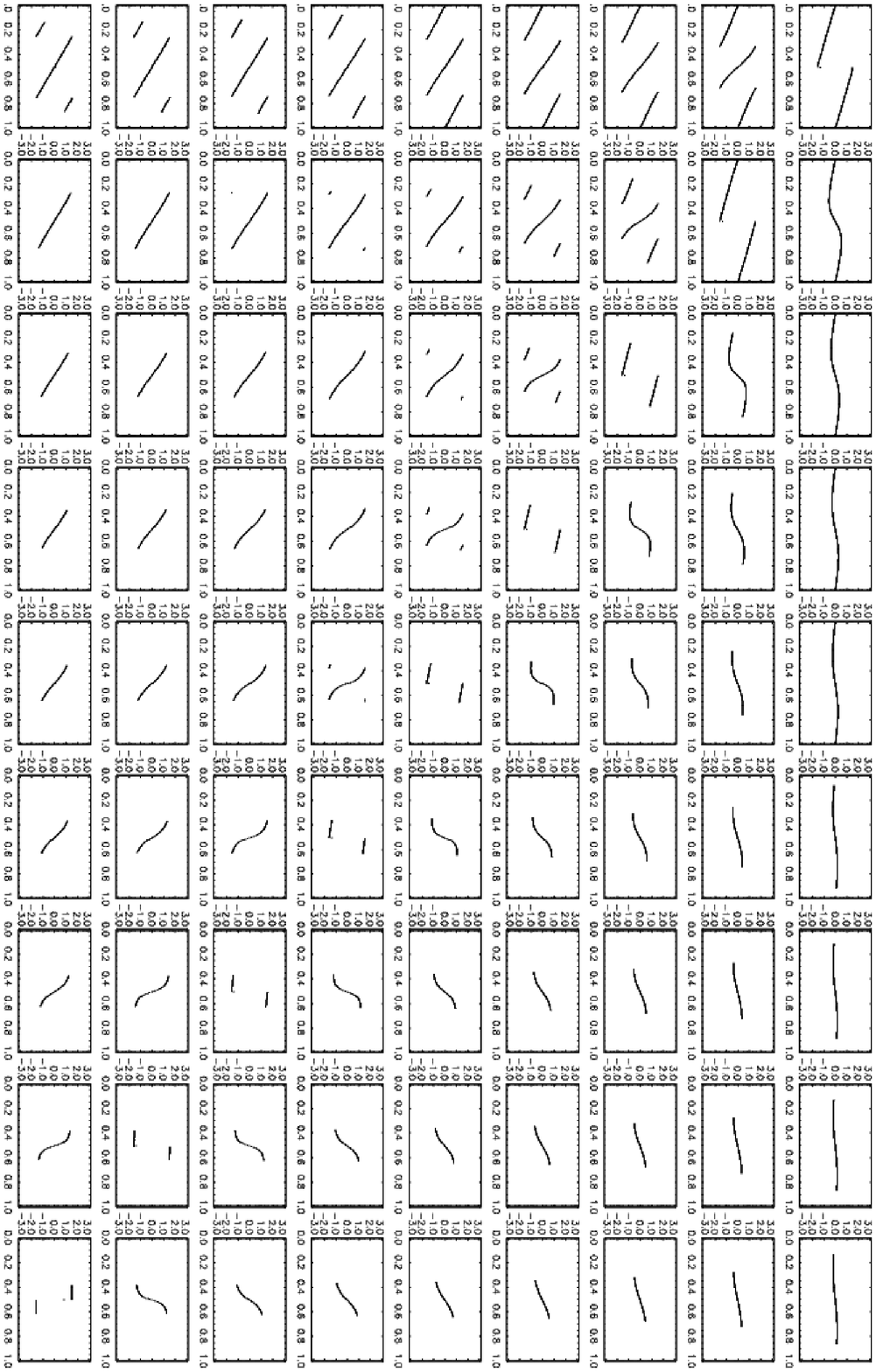}
\caption{Dipole field at $r \ll r_\text{LC}$. Layout as for Figure \ref{m25w10lcostheta_dqvsi}, but for position angle (on the vertical axis in landscape orientation, in units of radians) versus pulse longitude (on the horizontal axis, in units of $2 \pi$ radians).}
\label{m25w10lcostheta_dpa}
\end{figure*}

The PA swings (Figure \ref{m25w10lcostheta_dpa}) are identical to the previous cases, the only difference being that they are visible above a threshold intensity over a greater fraction of the pulse period due to the broader pulse profiles.

\subsubsection{$L = I \sin \theta_0$}
\label{sec:dipoleconesintheta}
\begin{figure}
\includegraphics[scale=0.45]{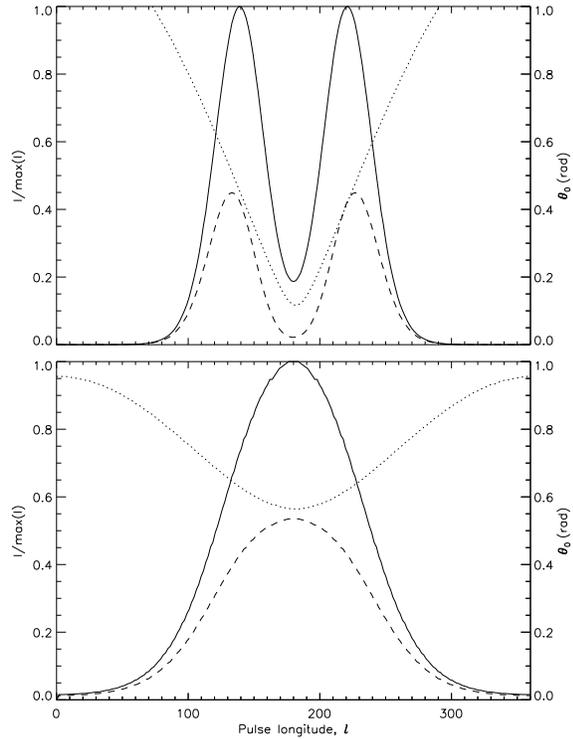}
\caption{As for Figure \ref{m25w10lcostheta_dprofiles}, but with linear polarization $L = I \sin \theta$.}
\label{m25w10lsintheta_dprofiles}
\end{figure}

Examples of the $I$ (solid curve) and $L$ (dashed curve) profiles for $(\alpha, i) = (70^\circ, 20^\circ)$ and $(70^\circ, 60^\circ)$ are shown in Figure \ref{m25w10lsintheta_dprofiles}. We note the following trends. (i) $L$ follows $I$ in the same way as Figure \ref{m25w10lcostheta_dprofiles}. However, $L/I$ peaks at $\approx$ 0.5 on the $\alpha = i$ diagonal. (ii) $L/I$ increases as $\lvert \alpha - i \rvert$ increases, reaching $\approx$ 0.8 for $\lvert \alpha - i \rvert = 80^\circ$.

\begin{figure*}
\includegraphics[scale=0.8]{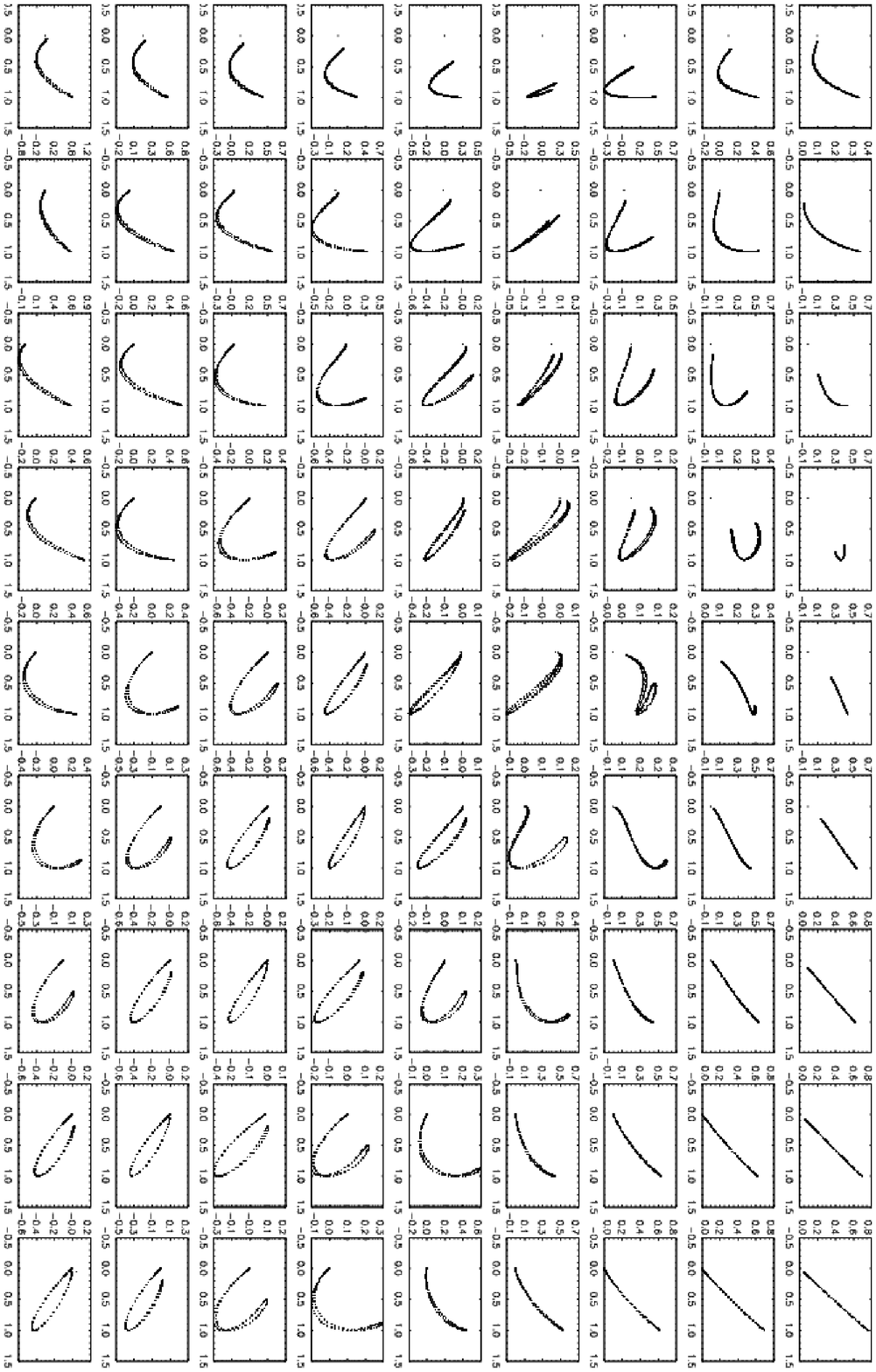}
\caption{Dipole field at $r \ll r_\text{LC}$. Look-up table of Stokes phase portraits in the $I$-$Q$ plane for a hollow cone with opening angle $25^\circ$ with degree of linear polarization $L = I \sin \theta_0$, where $\theta_0$ is the emission point colatitude. The panels are organised in landscape mode, in order of increasing $10^\circ \leq i \leq 90^\circ$ (left--right) and $10^\circ \leq \alpha \leq 90^\circ$ (top--bottom) in intervals of $10^\circ$. $I$ is plotted on the horizontal axis and normalised by its peak value. $Q$ is plotted on the vertical axis.}
\label{m25w10lsintheta_dqvsi}
\end{figure*}

\begin{figure*}
\includegraphics[scale=0.8]{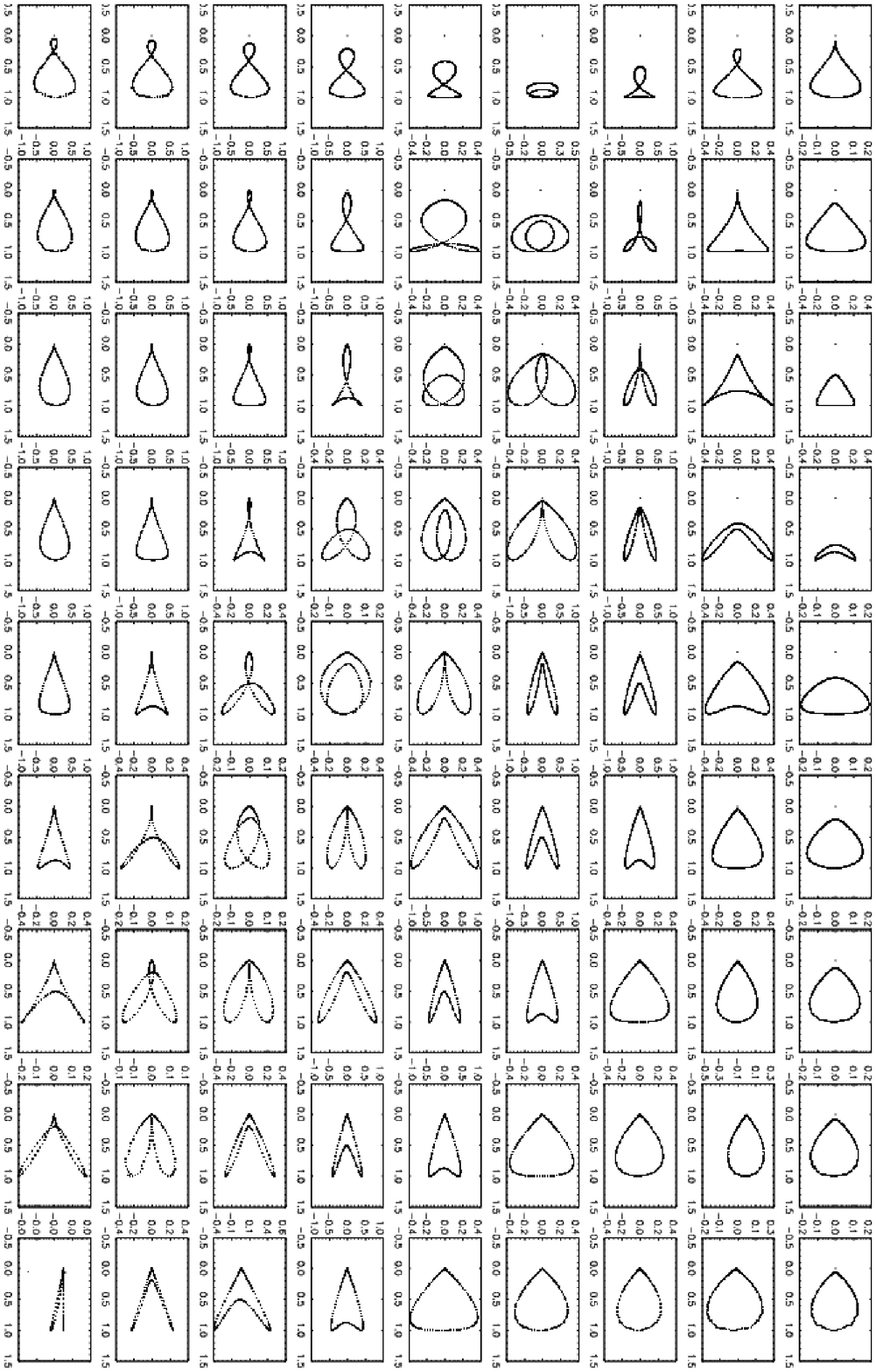}
\caption{Dipole field at $r \ll r_\text{LC}$. Layout as for Figure \ref{m25w10lsintheta_dqvsi}, but for $I$-$U$ ($I$ on horizontal axis).}
\label{m25w10lsintheta_duvsi}
\end{figure*}

\begin{figure*}
\includegraphics[scale=0.8]{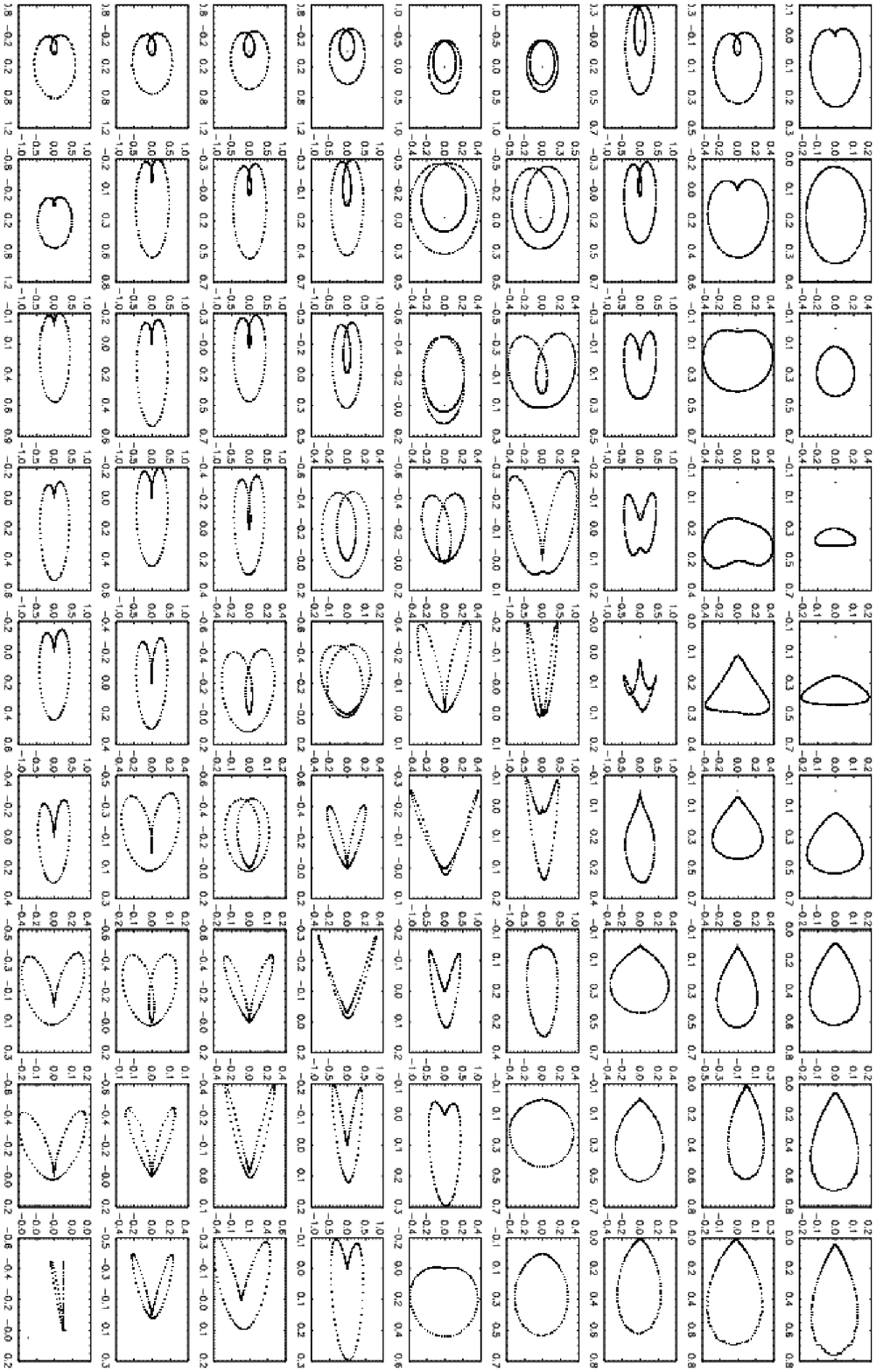}
\caption{Dipole field at $r \ll r_\text{LC}$. Layout as for Figure \ref{m25w10lsintheta_dqvsi}, but for $Q$-$U$ ($Q$ on horizontal axis).}
\label{m25w10lsintheta_duvsq}
\end{figure*}

\begin{figure*}
\includegraphics[scale=0.8]{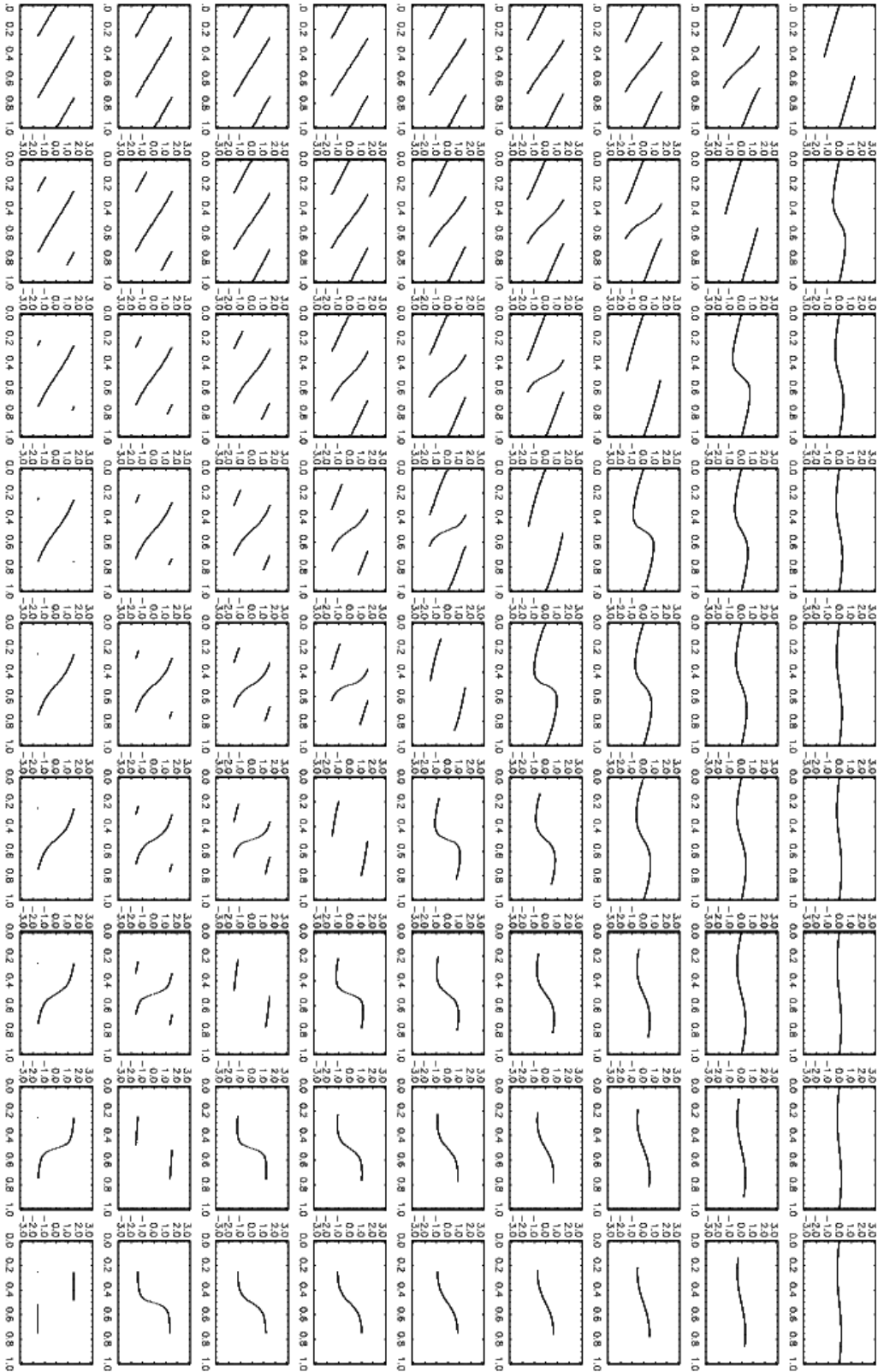}
\caption{Dipole field at $r \ll r_\text{LC}$. Layout as for Figure \ref{m25w10lsintheta_dqvsi}, but for position angle (on the vertical axis in landscape orientation, in units of radians) versus pulse longitude (on the horizontal axis, in units of $2 \pi$ radians).}
\label{m25w10lsintheta_dpa}
\end{figure*}

The $I$-$Q$, $I$-$U$ and $Q$-$U$ phase portraits are shown in Figures \ref{m25w10lsintheta_dqvsi}--\ref{m25w10lsintheta_duvsq}. The phase portraits are similar to Section \ref{sec:dipoleconecostheta}, except for orientations that yield double-peaked pulses. Here, the patterns are less twisted. Compared to Section \ref{sec:dipoleconecostheta}, the lower peak value of $L/I$ leads to a narrower range of $U$ and $Q$. For example, in the $Q$-$U$ plane, along the $\alpha = i$ diagonal, the heart shape for $(\alpha, i) = (10^\circ, 10^\circ)$ is approximately 80\% smaller than in Figure \ref{m25w10lcostheta_duvsq} along both the $U$ and $Q$ axes.
The PA swings (Figure \ref{m25w10lsintheta_dpa}) are exactly identical to Figure \ref{m25w10lcostheta_dpa}.

In summary, a hollow cone beam pattern produces Stokes phase portraits that differ significantly from those of the filled core beam, when the pulses are double-peaked (for $\lvert \alpha - i \rvert \leq 30^\circ$). In the $I$-$Q$ plane, double-peaked pulses trace out a tilted $\gamma$ shape, whereas single-peaked pulses trace out a hockey stick. In the $I$-$U$ plane, double-peaked pulses produce a triangle with twisted corners, whereas single-peaked pulses produce a balloon. In the $Q$-$U$ plane, the double-peaked pulses produce a mix of heart shapes and nested ovals, whereas the single-peaked pulses produce heart shapes (for $\alpha > i$) and balloons (for $\alpha < i$ diagonal).

\section{Dipole field at $\lowercase{r}$ = 0.1 $\lowercase{r}_\text{LC}$}
\label{sec:dipoleaberration}
Relativistic aberration affects the observed Stokes parameters in two ways. Firstly, the change in the emission point due to relativistic beaming [Eq. (\ref{eqn:tangent})] shifts the pulse centroid and hence $I(t)$. Beaming also shifts the PA swing in the same direction, resulting in a relative phase shift of $4 r/r_\text{LC}$ between the PA inflection point and the centroid of the pulse profile. Note that as we are only interested in the relative phase, we ignore time-delay effects, which phase-shift both the pulse profile and the PA swing by an additional $-r/r_\text{LC}$ \citep{dyks08}. 

In this section, we present look-up tables for a dipole field at an emission altitude $r = 0.1 r_\text{LC}$, where aberration distorts the Stokes phase portraits significantly while leaving the PA swings largely unchanged. 
We investigate the same beam patterns as in Section \ref{sec:dipole}, but for the sake of brevity, we present look-up tables for $L/I = \cos \theta_0$ only. For each beam pattern (core emission in Section \ref{sec:dipoleabcore}, conal emission in Section \ref{sec:dipoleabcone}) we construct three phase portraits ($I$-$Q$, $I$-$U$ and $Q$-$U$) for each pair of angles $(\alpha, i)$ at a fixed emission altitude ($r = 0.1\,r_\text{LC}$) and tabulate the results in Figures \ref{m0w10lcostheta_dabqvsi}--\ref{m25w10lcostheta_dabuvsq}.

\subsection{Filled core beam}
\label{sec:dipoleabcore}
\begin{figure*}
\includegraphics[scale=0.8]{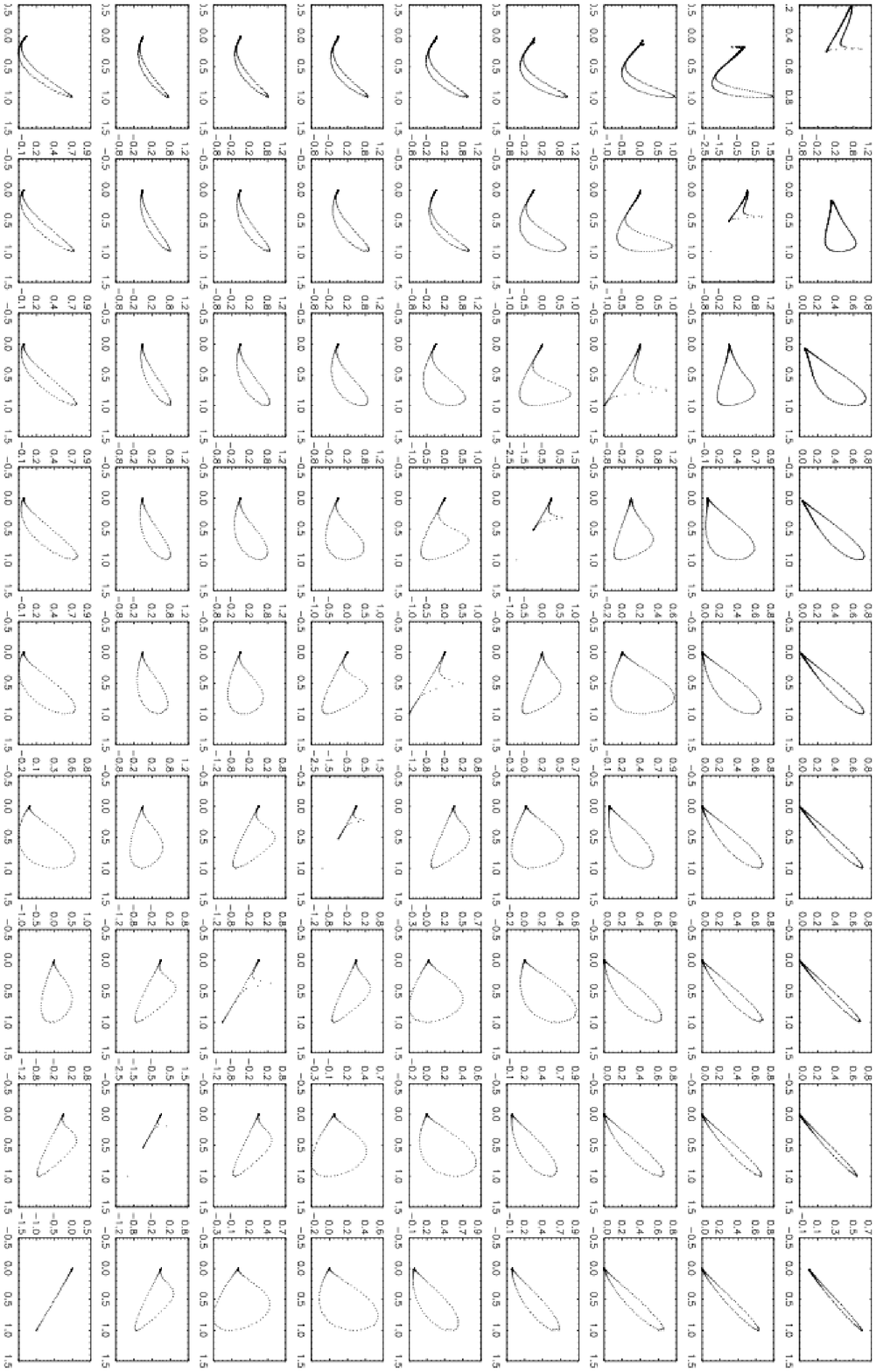}
\caption{Dipole field at $r = 0.1 r_\text{LC}$. Look-up table of Stokes phase portraits in the $I$-$Q$ plane for a filled core beam with degree of linear polarization $L = I \cos \theta_0$, where $\theta_0$ is the emission point colatitude. The panels are organised in landscape mode, in order of increasing $10^\circ \leq i \leq 90^\circ$ (left--right) and $10^\circ \leq \alpha \leq 90^\circ$ (top--bottom) in intervals of $10^\circ$. $I$ is plotted on the horizontal axis and normalised by its peak value. $Q$ is plotted on the vertical axis.}
\label{m0w10lcostheta_dabqvsi}
\end{figure*}

\begin{figure*}
\includegraphics[scale=0.8]{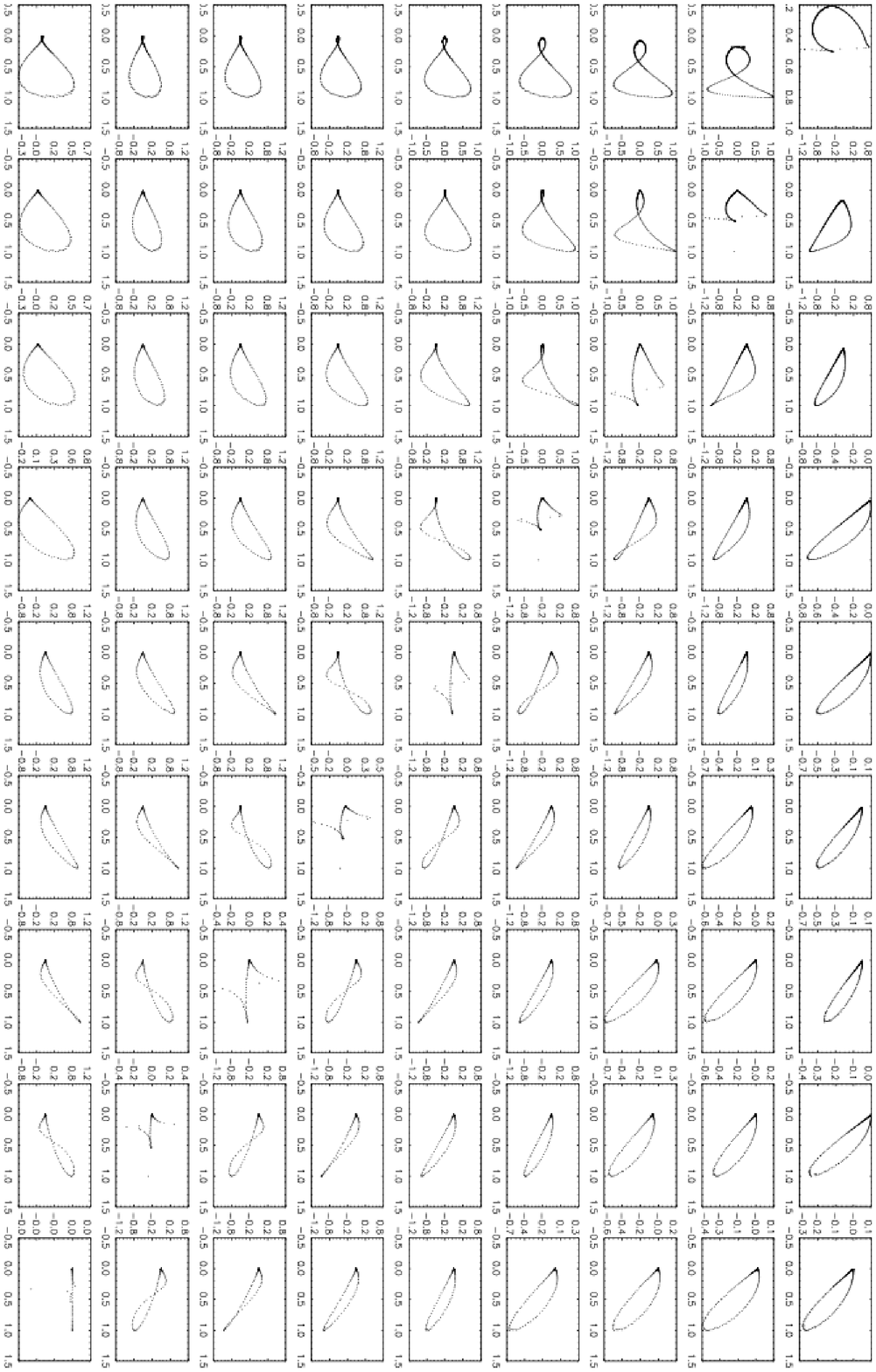}
\caption{Dipole field at $r = 0.1 r_\text{LC}$. Layout as for Figure \ref{m0w10lcostheta_dabqvsi}, but for $I$-$U$ ($I$ on the horizontal axis).}
\label{m0w10lcostheta_dabuvsi}
\end{figure*}

\begin{figure*}
\includegraphics[scale=0.8]{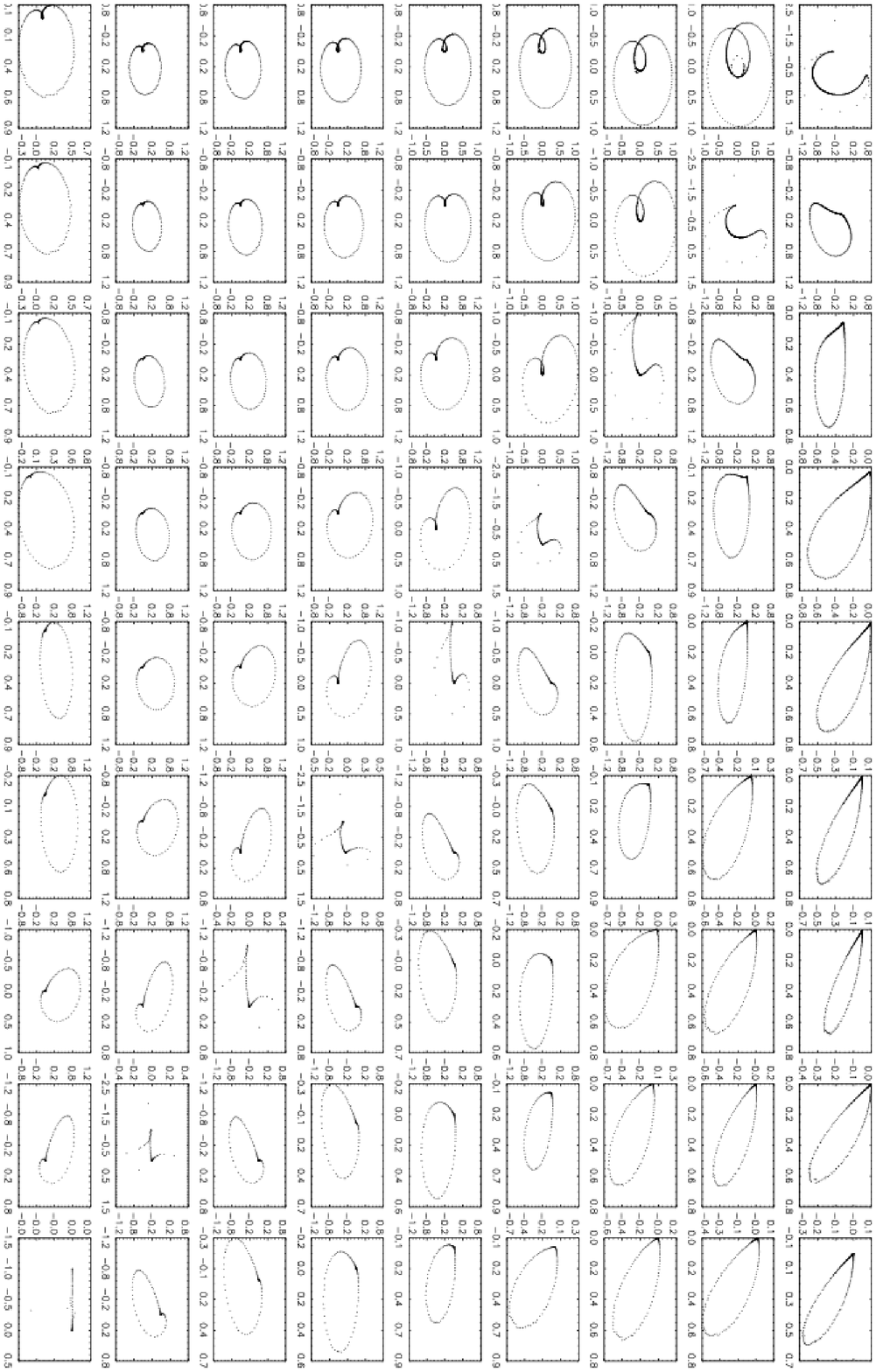}
\caption{Dipole field at $r = 0.1 r_\text{LC}$. Layout as for Figure \ref{m0w10lcostheta_dabqvsi}, but for $Q$-$U$ ($Q$ on the horizontal axis).}
\label{m0w10lcostheta_dabuvsq}
\end{figure*}

In this section, we point out the major differences between the Stokes phase portraits for a filled core beam and $L = I \cos \theta_0$ at $r = 0.1 r_\text{LC}$ (Figures \ref{m0w10lcostheta_dabqvsi}--\ref{m0w10lcostheta_dabuvsq}) and the corresponding non-aberrated situation at $r \ll r_\text{LC}$ (Figures \ref{m0w10lcostheta_dqvsi}--\ref{m0w10lcostheta_duvsq}).

In the $I$-$Q$ plane, the hockey sticks and straight lines in the non-aberrated case (Figure \ref{m0w10lcostheta_dqvsi}) broaden into balloons and ovals when aberration becomes important (Figure \ref{m0w10lcostheta_dabqvsi}). The broadening increases as the panels approach the $\alpha = i$ diagonal. 

In the $I$-$U$ plane (Figure \ref{m0w10lcostheta_dabuvsi}), the symmetry about $U = 0$ is broken, and the balloons and ovals are tilted. For $\alpha > i$ (below the diagonal), the balloons tilt upwards (major axis has $dU/dI > 0$), whereas for $\alpha < i$ (above the diagonal), the ovals tilt downwards (major axis has $dU/dI < 0$). Another difference is that for $\lvert \alpha - i \rvert = 10^\circ$, the phase portraits feature figure-eights instead of ovals. Along the $\alpha = i$ diagonal there is a distorted Z-shape, although we note that the sharp breaks in the pattern are numerical artifacts.

In the $Q$-$U$ plane (Figure \ref{m0w10lcostheta_dabuvsq}), the heart shapes in the $\alpha > i$ region (below the diagonal) are rotated counter-clockwise. As the panels approach the $\alpha = i$ diagonal, the top ventricle of the heart shape grows larger than the bottom ventricle. In the $\alpha < i$ region (above the diagonal), the ovals are rotated clockwise and are tilted downwards (major axis has $dU/dQ < 0$). Along the $\alpha = i$ diagonal, there are distorted heart shapes instead of ovals.

\subsection{Hollow cone}
\label{sec:dipoleabcone}
\begin{figure*}
\includegraphics[scale=0.8]{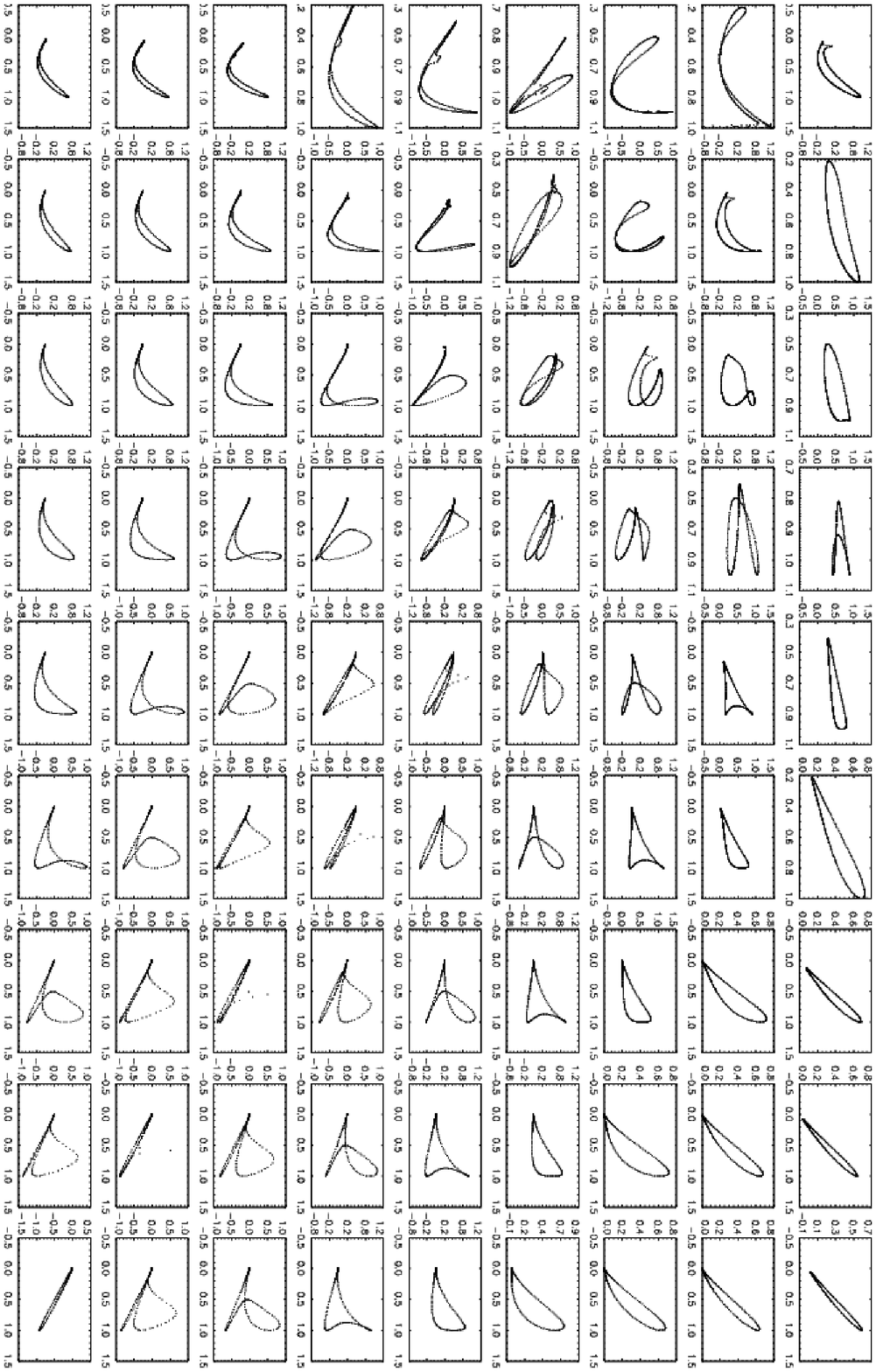}
\caption{Dipole field at $r = 0.1 r_\text{LC}$. Look-up table of Stokes phase portraits in the $I$-$Q$ plane for a hollow cone with opening angle $25^\circ$ with degree of linear polarization $L = I \cos \theta_0$, where $\theta_0$ is the emission point colatitude. The panels are organised in landscape mode, in order of increasing $10^\circ \leq i \leq 90^\circ$ (left--right) and $10^\circ \leq \alpha \leq 90^\circ$ (top--bottom) in intervals of $10^\circ$. $I$ is plotted on the horizontal axis and normalised by its peak value. $Q$ is plotted on the vertical axis.}
\label{m25w10lcostheta_dabqvsi}
\end{figure*}

\begin{figure*}
\includegraphics[scale=0.8]{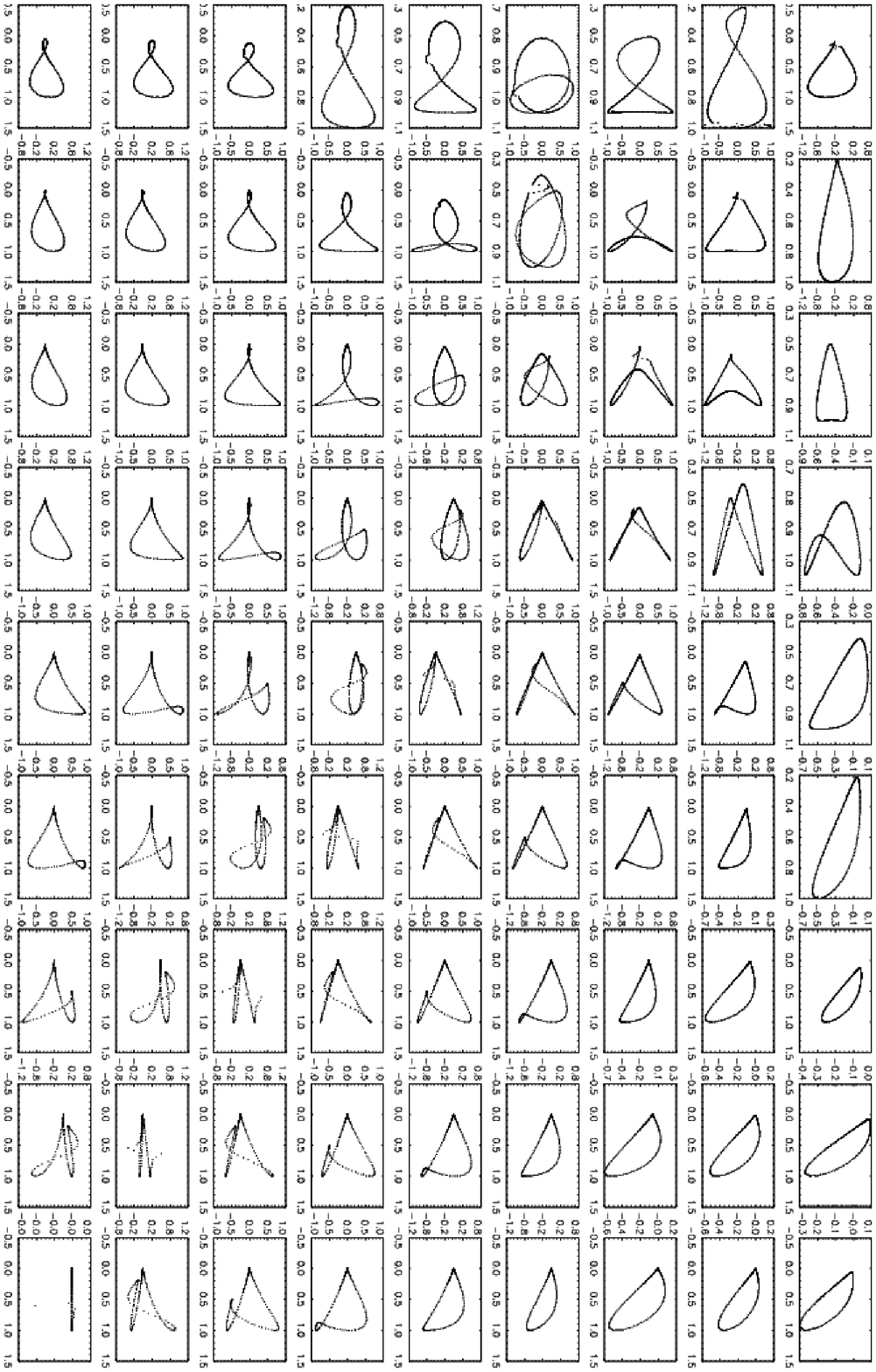}
\caption{Dipole field at $r = 0.1 r_\text{LC}$. Layout as for Figure \ref{m25w10lcostheta_dabqvsi}, but for $I$-$U$ ($I$ on the horizontal axis).}
\label{m25w10lcostheta_dabuvsi}
\end{figure*}

\begin{figure*}
\includegraphics[scale=0.8]{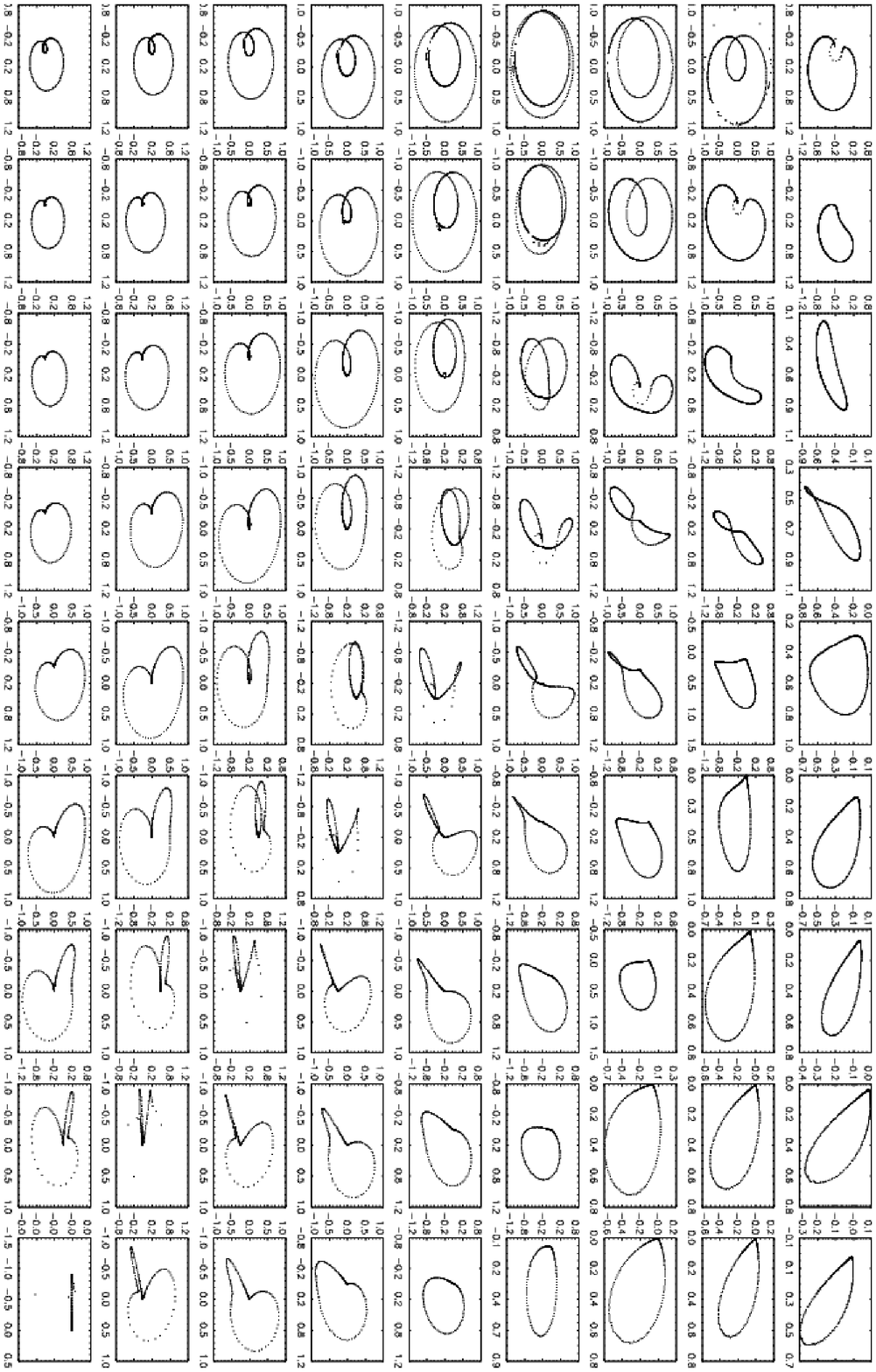}
\caption{Dipole field at $r = 0.1 r_\text{LC}$. Layout as for Figure \ref{m25w10lcostheta_dabqvsi}, but for $Q$-$U$ ($Q$ on the horizontal axis).}
\label{m25w10lcostheta_dabuvsq}
\end{figure*}

We now examine the major differences between the Stokes phase portraits for a hollow cone beam and $L = I \cos \theta_0$ at $r = 0.1 r_\text{LC}$ (Figures \ref{m25w10lcostheta_dabqvsi}--\ref{m25w10lcostheta_dabuvsq}) and the corresponding non-aberrated case (Figures \ref{m25w10lcostheta_dqvsi}--\ref{m25w10lcostheta_duvsq}).

Similar to the filled core beam, aberration broadens the shapes in the $I$-$Q$ plane (Figures \ref{m25w10lcostheta_dabqvsi}). At the orientations where the pulse profile is double-peaked, the $\gamma$-shapes at $\alpha \gtrsim 40^\circ$ and $i \gtrsim 40^\circ$ from Figure \ref{m25w10lcostheta_dqvsi} are replaced by mosquito shapes. For $\alpha \lesssim 30^\circ$ and $i \lesssim 30^\circ$, the $I$-$Q$ portraits maintain their general shape but are broader. An exception is $(\alpha, i) = (20^\circ, 30^\circ)$, which changes from a C-shape in the non-aberrated case to a figure-eight with aberration.

In the $I$-$U$ plane, at the orientations where the pulse profile is double-peaked, the trefoils, triangles, and mosquito shapes retain their general shape. However, they are distorted and asymmetric about $U = 0$, as the individual components of the patterns (e.g. the `wings' of the mosquitoes) tilt at different angles and have different widths.

The shapes in the $Q$-$U$ plane do change at those orientations where the pulse profile is double-peaked. In the $\alpha > i$ region (below the diagonal), the interlocking ovals become distorted, asymmetric heart shapes as $\alpha$ and $i$ increase.  In the $\alpha > i$ region (above the diagonal), there are figure-eights [e.g. ($\alpha, i$) = $(20^\circ, 40^\circ)$] which become distorted heart shapes as $\alpha$ and $i$ increase. Along the $\alpha = i$ diagonal, distorted heart shapes evolve into distorted mosquito shapes as $\alpha$ and $i$ increase.

\subsection{How widespread are low emission altitudes?}
\label{sec:compdipole}
As our first application of the above theory to observational data, we ask: what proportion of pulsars with superficially dipole-like characteristics (e.g. S-shaped PA swings) are truly consistent with a pure dipole field and low, non-aberrated emission altitude (Section \ref{sec:dipole}), when we examine their Stokes phase portraits? 

To this end, we select two sets of superficially dipole-like pulsars from the 366 objects presented in \citet{gould98} and \citet{manchester98}. Set I contains pulsars that follow closely the $W \propto P^{-1/2}$ relation \citep{rankin93} and hence are likely prima facie to have an approximately dipolar field in the emission region, with or without aberration. They are drawn from near the diagonal line in Figure 9 of \citet{gould98}; as a rule, their pulse width at 0.1$I_\text{max}$ (where $I_\text{max}$ is the peak intensity) is less than $20^\circ$. Set II contains pulsars that have a clean, S-shaped PA swing, after allowing for phase wrapping. From the suitable pulsars, we choose 12 at random in each set.

\begin{figure*}
\includegraphics[scale=0.8]{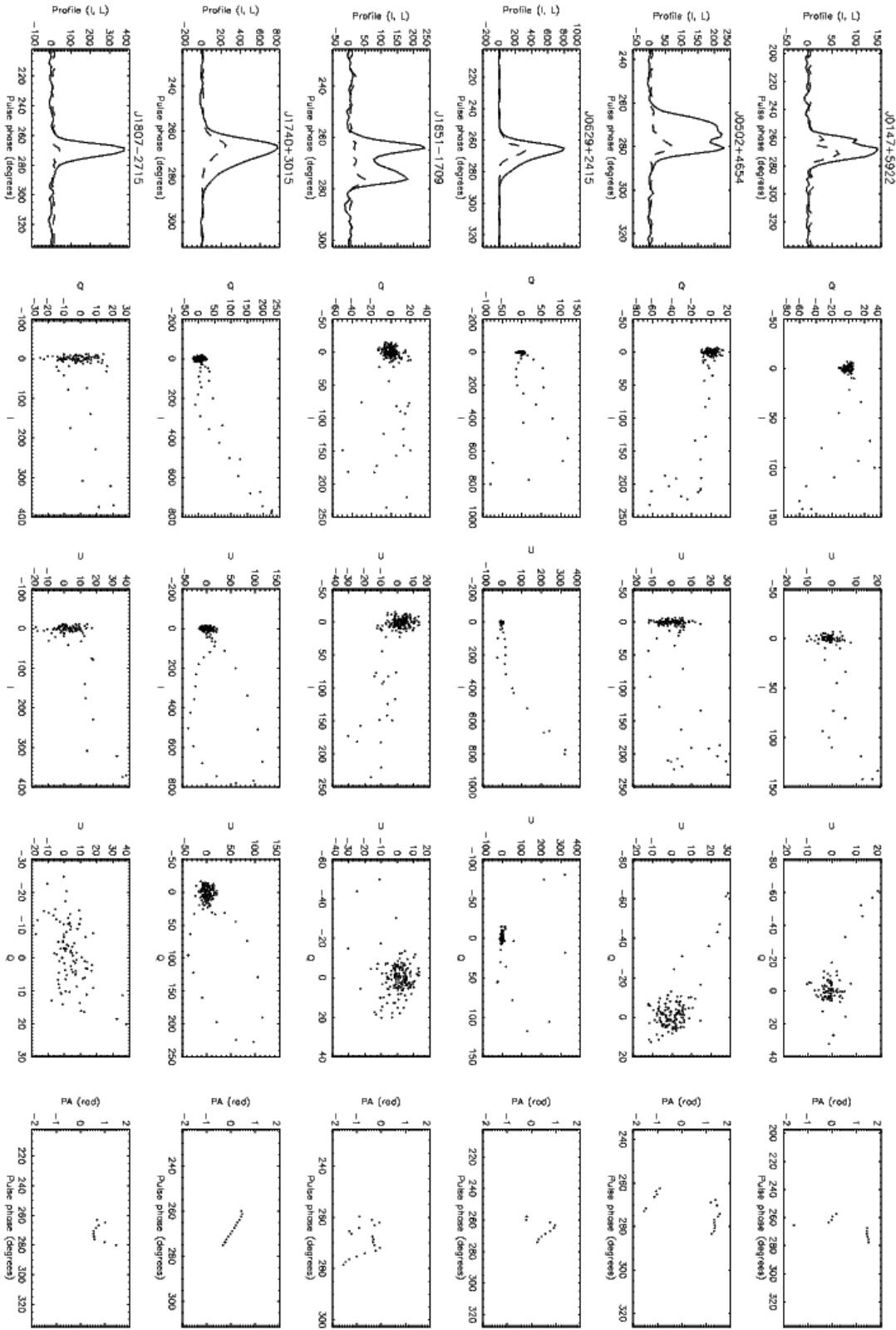}
\caption{Pulse profiles, Stokes phase portraits and PA swings for a selection of six superficially dipolar pulsars that closely follow the pulse-width-period relation at 610\,MHz \citep{gould98}. The data for each pulsar occupy a row each in landscape mode. From left to right, the columns show: (1) $I$ (mJy, solid curve) and $L$ (mJy, dashed curve), (2) the $I$-$Q$ phase portrait, (3) the $I$-$U$ phase portrait, (4) the $Q$-$U$ phase portrait, and (5) the PA swing (data points with $L \geq 0.1 L_\text{max}$ and $I \geq 0.1 I_\text{max}$ plotted only). Data are presented courtesy of the EPN online archive.}
\label{compdipole1}
\end{figure*}

\begin{figure*}
\includegraphics[scale=0.8]{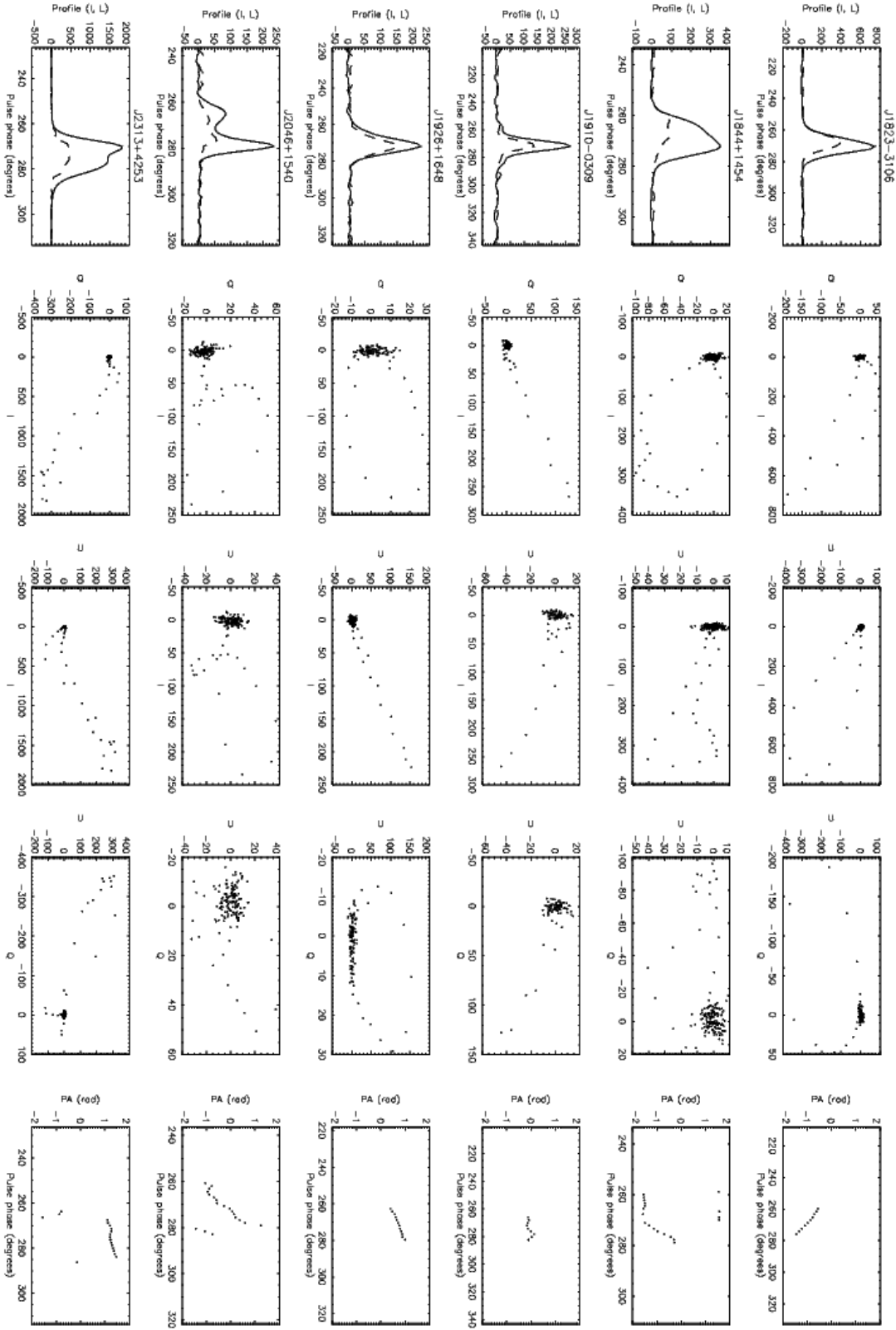}
\caption{As for Figure \ref{compdipole1}, for six other objects selected according to the same criterion.}
\label{compdipole2}
\end{figure*}

\begin{figure*}
\includegraphics[scale=0.8]{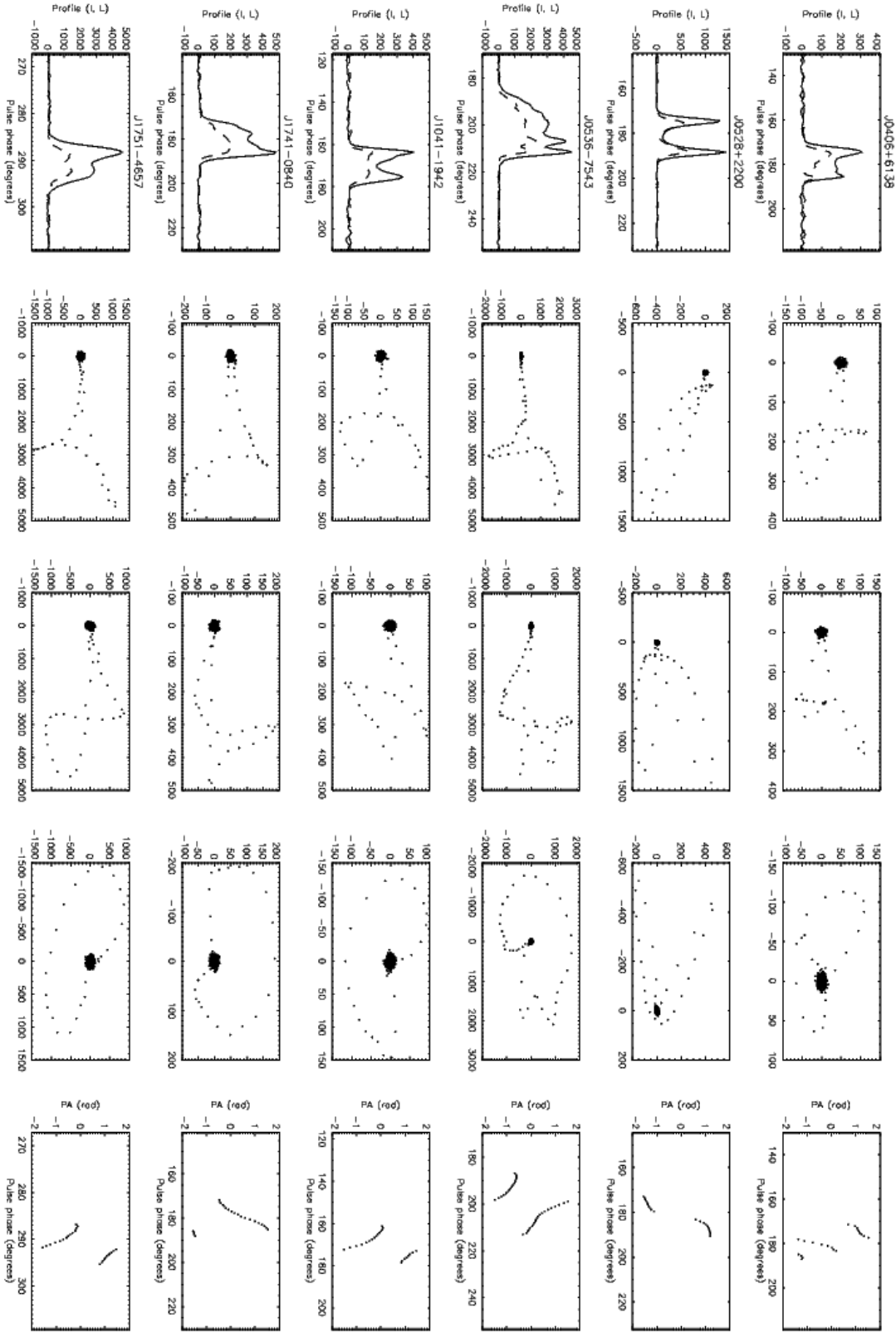}
\caption{Pulse profiles, Stokes phase portraits and PA swings for a selection of six superficially dipolar pulsars that exhibit a clean, S-shaped PA swing. All pulsars are observed at 610\,MHz \citep{gould98}, except for PSR J0536$-$7543 (663\,MHz) and PSR J1751$-$4657 (433\,MHz) \citep{manchester98}. The data for each pulsar occupy a row each in landscape mode. From left to right, the columns show: (1) $I$ (mJy, solid curve) and $L$ (mJy, dashed curve), (2) the $I$-$Q$ phase portrait, (3) the $I$-$U$ phase portrait, (4) the $Q$-$U$ phase portrait, and (5) the PA swing (data points with $L \geq 0.1 L_\text{max}$ and $I \geq 0.1 I_\text{max}$ plotted only). Data are presented courtesy of the EPN online archive.}
\label{compdipole3}
\end{figure*}

\begin{figure*}
\includegraphics[scale=0.8]{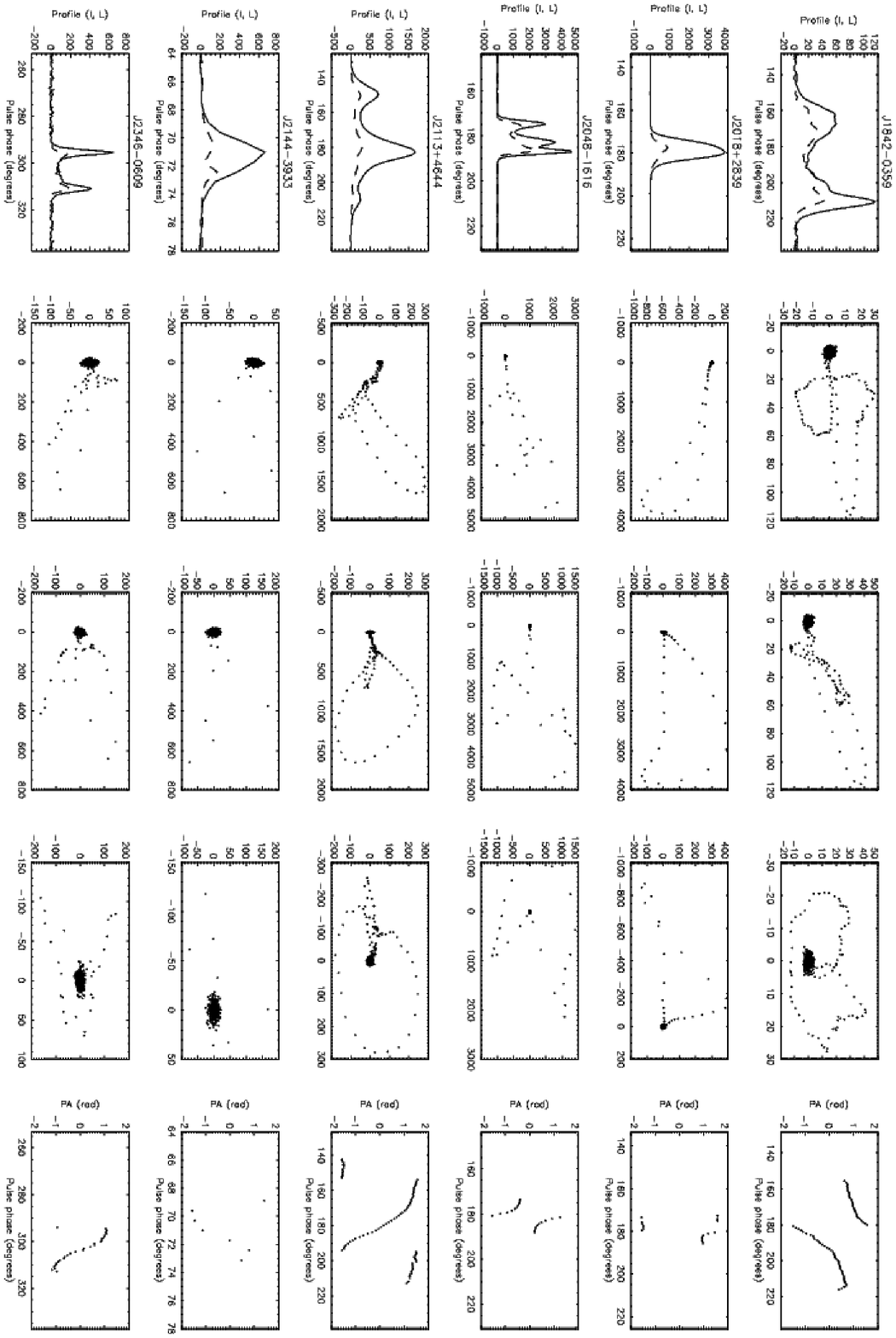}
\caption{As for Figure \ref{compdipole1} for six other objects selected according to the same criterion. All pulsars are observed at 610\,MHz \citep{gould98}, except for PSR J1842$-$0359 (1.408\,GHz) \citep{gould98}, PSR J2144$-$3933 (659\,MHz) and PSR J2346$-$0609 (435\,MHz) \citep{manchester98}.}
\label{compdipole4}
\end{figure*}

We present the pulse profiles, Stokes phase portraits, and PA swings for Set I at 610\,MHz in Figures \ref{compdipole1} and \ref{compdipole2}. PA data points are only plotted for $L \geq 0.1 L_\text{max}$ and $I \geq 0.1 I_\text{max}$. The data are sourced from the EPN online archive \citep{lorimer98}. The 12 objects are calibrated separately, so the polarization basis varies from pulsar to pulsar. For the purposes of this section, we do not attempt to determine $\beta$ for each object; we focus on the \textit{shape} of the $Q$-$U$ patterns, not their orientations. From Sections \ref{sec:dipole} and \ref{sec:dipoleaberration}, we note that one clear fingerprint of low-altitude emission from a pure dipole is the symmetry of the $Q$-$U$ pattern about $U = 0$ (rotated in general by an angle $2 \beta$). Hence, in Figures \ref{compdipole1}--\ref{compdipole4}, any pattern which is not symmetric about any line either (i) cannot originate from a low altitude where aberration is negligible, or (ii) is the result of a largely tilted axis of symmetry, as discussed in Section \ref{sec:displaced}. It is not possible to disentangle these effects solely from linear polarization data. Circular polarization phase portraits may assist, but their analysis lies outside the scope of this paper. 

Of the 12 objects in Figures \ref{compdipole1} and \ref{compdipole2}, four have Stokes phase portraits which roughly resemble a dipole at low altitudes. PSR J0629+2415 and PSR J1740$-$3015 (Figure \ref{compdipole1}, third and fifth rows) match a core beam for large $i$ and small $\alpha$. Both $L = I \cos \theta_0$ and $L = I \sin \theta_0$ give similar phase portraits for these angles. PSR J1823$-$3106 and J1926+1648 (Figure \ref{compdipole2}, first and fourth rows) match a core beam with $\alpha \leq i$ and $L = I \cos \theta_0$. 
The other eight pulsars have $Q$-$U$ portraits which are significantly asymmetric. This is reflected in their PA swings, which are not as clean as the other four pulsars. For example, PSR J2046+1540 (Figure \ref{compdipole2}, fifth row) features an asymmetric mosquito in the $Q$-$U$ plane. Assuming that the asymmetry is the result of relativistic aberration, such a pattern cannot be reconciled with emission from low altitudes. 
It matches a dipole field at $r = 0.1 r_\text{LC}$ along $\alpha = i$ for a hollow cone with $L = I \cos \theta_0$ (see the look-up table in Figure \ref{m25w10lcostheta_dabuvsq}), where aberration is important. Alternatively, if the asymmetry is the result of a tilted axis of beam symmetry, it is possible that the emission originates instead from a low altitude, and that aberration does not play a part in distorting the phase portraits.

Figures \ref{compdipole3} and \ref{compdipole4} display the pulse profiles, Stokes phase portraits, and PA swings for Set II. This set comprises data at multiple frequencies \citep{gould98, manchester98}, sourced from the EPN online archive. Four of the 12 pulsars in this set feature approximately symmetric mosquito and heart shapes matching a dipole at low emission altitudes. PSR J0528+2200 (Figure \ref{compdipole3}, second row), PSR J2048$-$1616, and J2346$-$0609 (Figure \ref{compdipole4}, third and sixth rows) match the $Q$-$U$ portraits for a hollow cone with $\alpha = i \geq 30^\circ$ ($L = I \cos \theta_0$ and $L = I \sin \theta_0$ give similar answers for these angles). PSR J2113+4644 (Figure \ref{compdipole4}, fourth row) matches all three phase portraits for a filled core beam for $0^\circ \leq \alpha - i \leq 10^\circ$ (again, $L = I \cos \theta_0$ and $L = I \sin \theta_0$ give similar answers for these angles).
In Figure \ref{compdipole3} and \ref{compdipole4}, the mosquito and heart shapes of the other eight pulsars are noticeably asymmetric. Again, we note that the asymmetry may be the result of a tilted axis of beam symmetry, or it may indicate that the emission originates from higher altitudes, where aberration takes effect.   

For a dipole field (or indeed any axisymmetric field), aberration changes the relative phase between the PA swing and pulse profile, but it does not cause the shape of the PA swing to change with emission frequency (unlike the pulse width). In reality, multi-frequency observations show that PA swings do change with frequency \citep{johnston08}. We return to this issue in Section \ref{sec:rmapping}, where we find that the Stokes phase portraits for magnetic geometries with an appreciable toroidal component at $r \gtrsim 0.1 r_\text{LC}$ evolve with frequency dramatically and in an informative way. 

A key tenet of pulsar radio emission theory holds that core and cone emission mostly comes from low altitudes, where departures from a dipole field are small. The results in Figures \ref{compdipole1}-\ref{compdipole4} imply that in a clear majority ($>$ 60\%) of objects, the emission may well originate from high altitudes, where aberration and/or the toroidal field component are significant. 
We now turn to explore this important issue in Section \ref{sec:toroidal}.

\section{Current-modified dipole field}
\label{sec:toroidal}

In this section, we examine the pulse profiles, Stokes phase portraits, and PA swings associated with a current-modified magnetic field composed of a dipole with symmetry axis $\mathbf{e}_3$ plus a toroidal component (cylindrically symmetric about $\mathbf{e}_3$) with magnitude
\begin{equation}
\label{eq:bphi}
B_\phi = - B_p \cos \alpha \sin \theta r/r_\text{LC},
\end{equation}
where $B_p = (B_r^2 + B_\theta^2)^{1/2}$ is the poloidal field strength.  Equation (\ref{eq:bphi}) is the scaling expected for $r \lesssim r_\text{LC}$ if $B_\phi$ is generated by a field-aligned Goldreich-Julian current in a plasma-filled magnetosphere \citep{hibschman01}. If $B_\phi$ is generated by the displacement current, as in a vacuum rotator, it scales as $r^2$; we save this case for future analysis \citep{deutsch55, melatos97}. 

We investigate the same beam and polarization patterns as in Section \ref{sec:dipole}. The emission point $\mathbf{x}_0(t)$ traces out a different locus in the current-modified dipole as compared to the pure dipole, and the locus varies differently with altitude. In Sections \ref{sec:toroidcore} and \ref{sec:toroidcone}, we present two-dimensional look-up tables of Stokes phase portraits for orientations $10^\circ \leq i \leq 90^\circ$ and $10^\circ \leq \alpha \leq 90^\circ$. For each beam pattern (core emission in Section \ref{sec:toroidcore}, conal emission in Section \ref{sec:toroidcone}) and polarization pattern ($L/I = \cos \theta_0,\, \sin \theta_0$), we construct three phase portraits ($I$-$Q$, $I$-$U$ and $Q$-$U$) for each pair of angles $(\alpha, i)$ at a fixed emission altitude ($r = 0.1\, r_\text{LC}$). In Section \ref{sec:rmapping}, we show some examples of how the phase portraits change with altitude.

\subsection{Filled core emission}
\label{sec:toroidcore}
\subsubsection{$L = I \cos \theta_0$}
\label{sec:toroidlcostheta}
\begin{figure}
\includegraphics[scale=0.5]{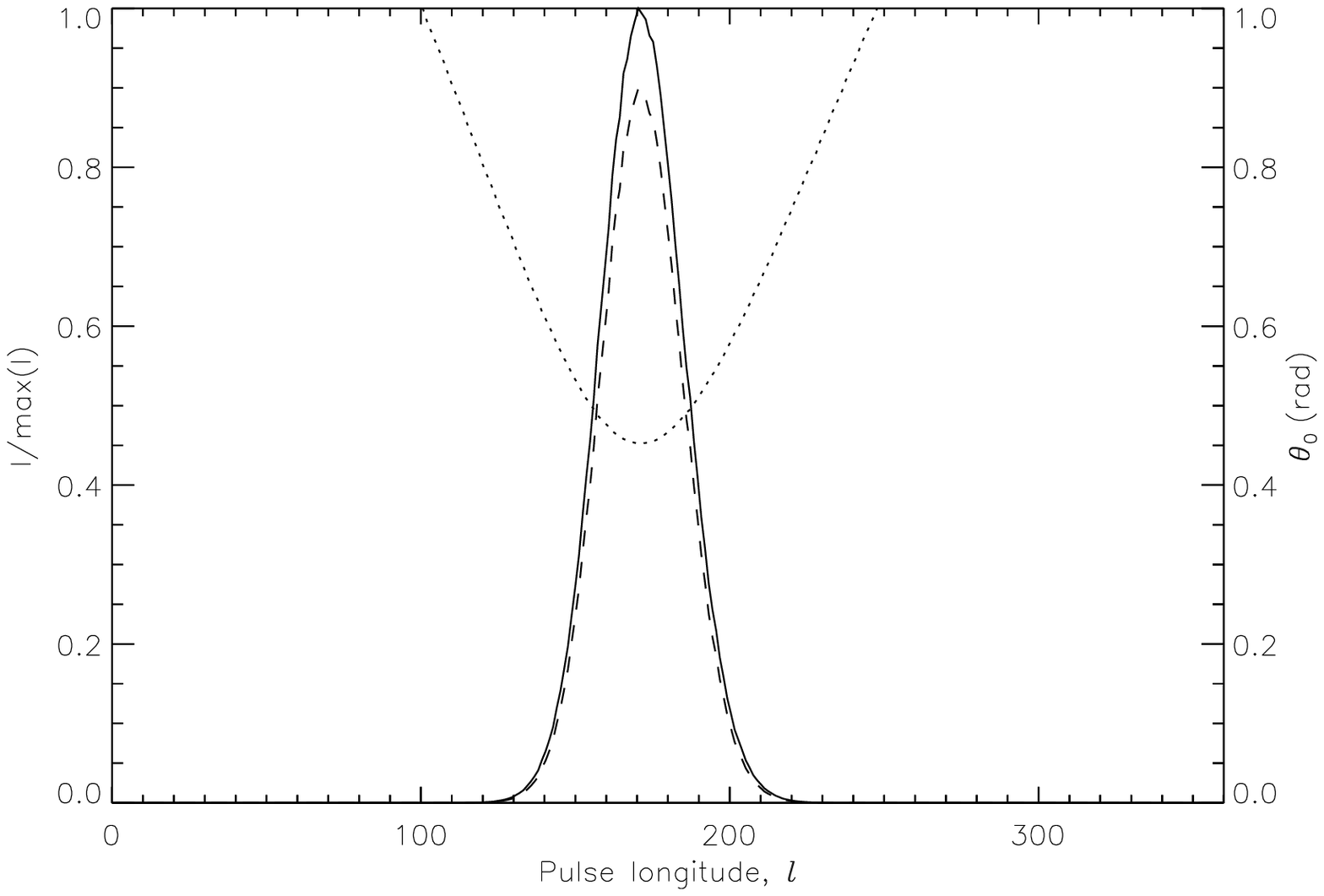}
\caption{Current-modified dipole field at $r = 0.1 r_\text{LC}$. Example of a pulse profile for a filled core beam with linear polarization $L = I \cos \theta_0$ and $(\alpha, i) = (70^\circ, 30^\circ)$. Solid, dashed and dotted curves represent the total polarized intensity $I$, degree of linear polarization $L$, and emission point colatitude $\theta_0$. Pulse longitude $l$ is measured in units of degrees.}
\label{m0w10lcostheta_tprofilesr25}
\end{figure}

An example of the $I$ (solid curve) and $L$ (dashed curve) profiles for a single-peaked pulse with ($\alpha, i$) = ($70^\circ, 30^\circ$) are shown in Figure \ref{m0w10lcostheta_tprofilesr25}. $I$ is normalised by its peak value. The dotted curve shows how the magnetic colatitude $\theta_0(t)$ of the emission point varies across one pulse period. We note the following trends. (i) The emission points and pulse profiles are phase-shifted by approximately $-r/r_\text{LC}$ radians due to aberration. (ii) The profiles behave similarly to the pure dipole (Section \ref{sec:dipolelcostheta}), narrowing with increasing $\alpha$ and $i$. (iii) For $\lvert \alpha - i \rvert \leq 10^\circ$, $\theta_0$ remains near zero for some time [$\approx 0.5$ phase units at $(\alpha, i) = (10^\circ, 10^\circ)$ and $\approx 0.13$ phase units at $(\alpha, i) = (90^\circ, 90^\circ)$]. This is a numerical artifact caused by the finite grid resolution near $\theta_0 = 0$.

\begin{figure}
\centering
\includegraphics[scale=0.5]{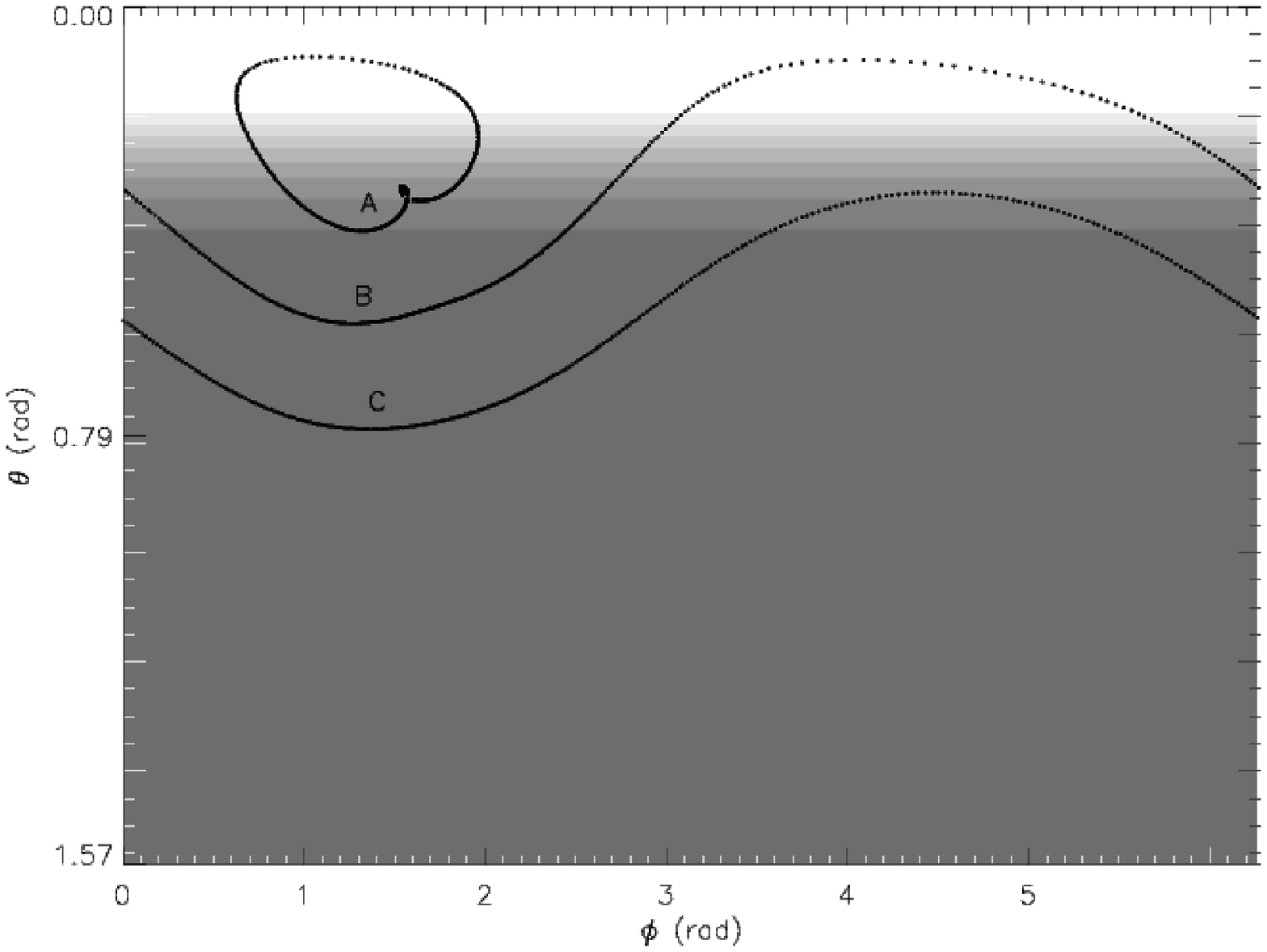}
\caption{Current-modified dipole field at $r = 0.1 r_\text{LC}$. Examples of the path traced by the emission point $\hat{\mathbf{x}}_0(t)$ in the body frame (curves), overplotted on the intensity map for a filled core beam (greyscale). The beam is brightest at $\theta = 0$ (corresponding to the $\mathbf{e}_3$ axis). Curves A: $(\alpha, i) = (20^\circ, 10^\circ)$, B: $(\alpha, i) = (20^\circ, 30^\circ)$, and C: $(\alpha, i) = (20^\circ, 50^\circ)$.}
\label{toroidemission}
\end{figure}

Figure \ref{toroidemission} shows the path traced by $\hat{\mathbf{x}}_0(t)$ in the body frame across one pulse period for three cases: (A) $(\alpha, i) = (20^\circ, 10^\circ)$, (B) $(20^\circ, 30^\circ)$, and (C) $(20^\circ, 50^\circ)$. As in Figure \ref{dipoleemission}, the path changes from an undulation for $\alpha < i$ (curve C) to an ellipse for $\alpha > i$ (curve A). Note that curve A features a cusp near $(\theta, \phi) \approx (0.4\,\text{rad}, 1.6\,\text{rad})$. 
The toroidal magnetic field twists the paths, so that they are not reflection symmetric about some longitude, unlike in Figure \ref{dipoleemission}. The asymmetry distorts the Stokes phase portraits, but the clear distinction between $\alpha < i$ and $\alpha > i$ persists. 

\begin{figure*}
\includegraphics[scale=0.8]{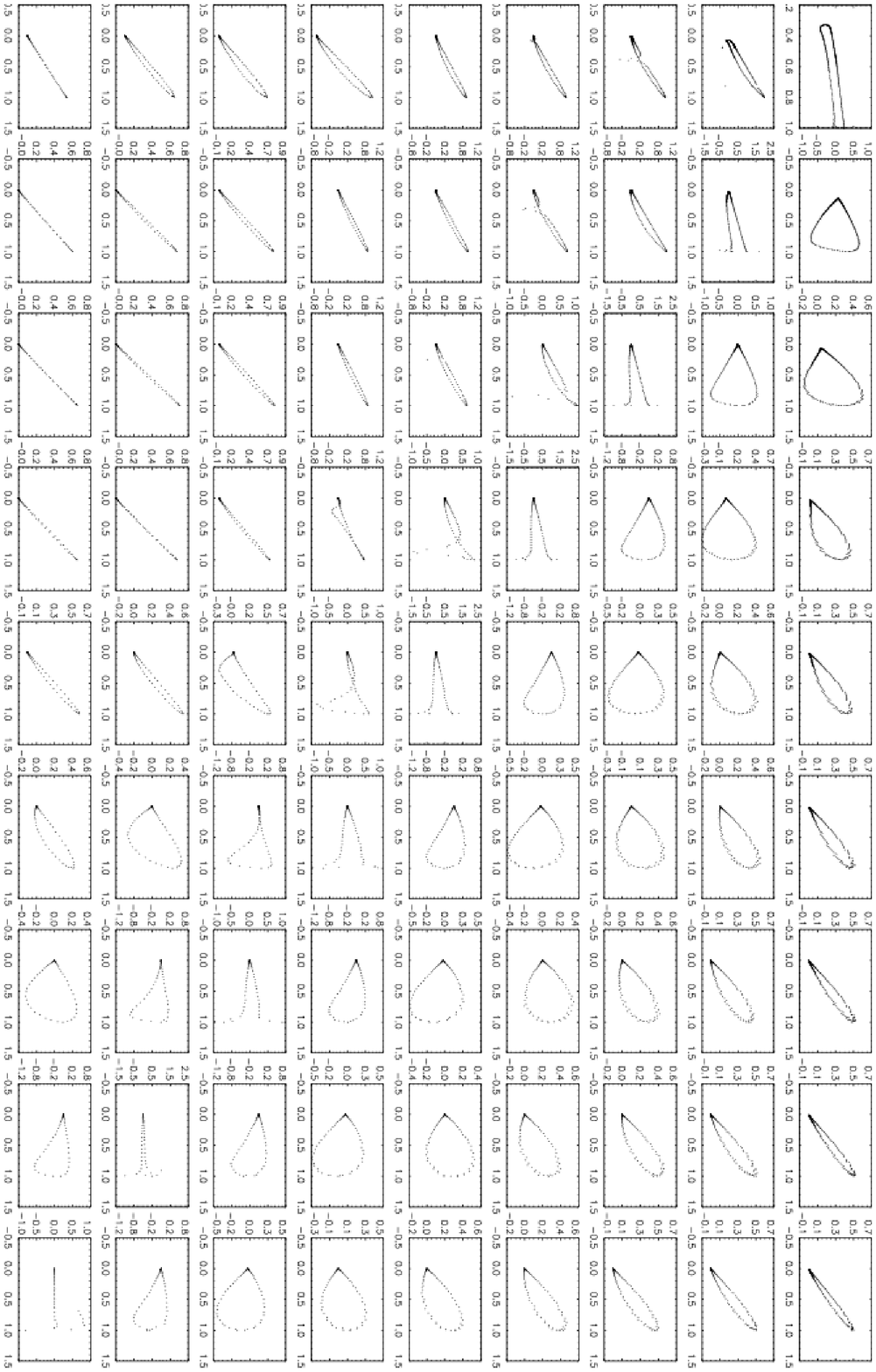}
\caption{Current-modified dipole field at $r = 0.1 r_\text{LC}$. Look-up table of Stokes phase portraits in the $I$-$Q$ plane for a filled core beam with degree of linear polarization $L = I \cos \theta_0$, where $\theta_0$ is the emission point colatitude. The panels are organised in landscape mode, in order of increasing $10^\circ \leq i \leq 90^\circ$ (left-right) and $10^\circ \leq \alpha \leq 90^\circ$ (top-bottom) in intervals of $10^\circ$. $I$ is plotted on the horizontal axis and normalised by its peak value. $Q$ is plotted on the vertical axis.}
\label{m0w10lcostheta_tqvsir25}
\end{figure*}

\begin{figure*}
\includegraphics[scale=0.8]{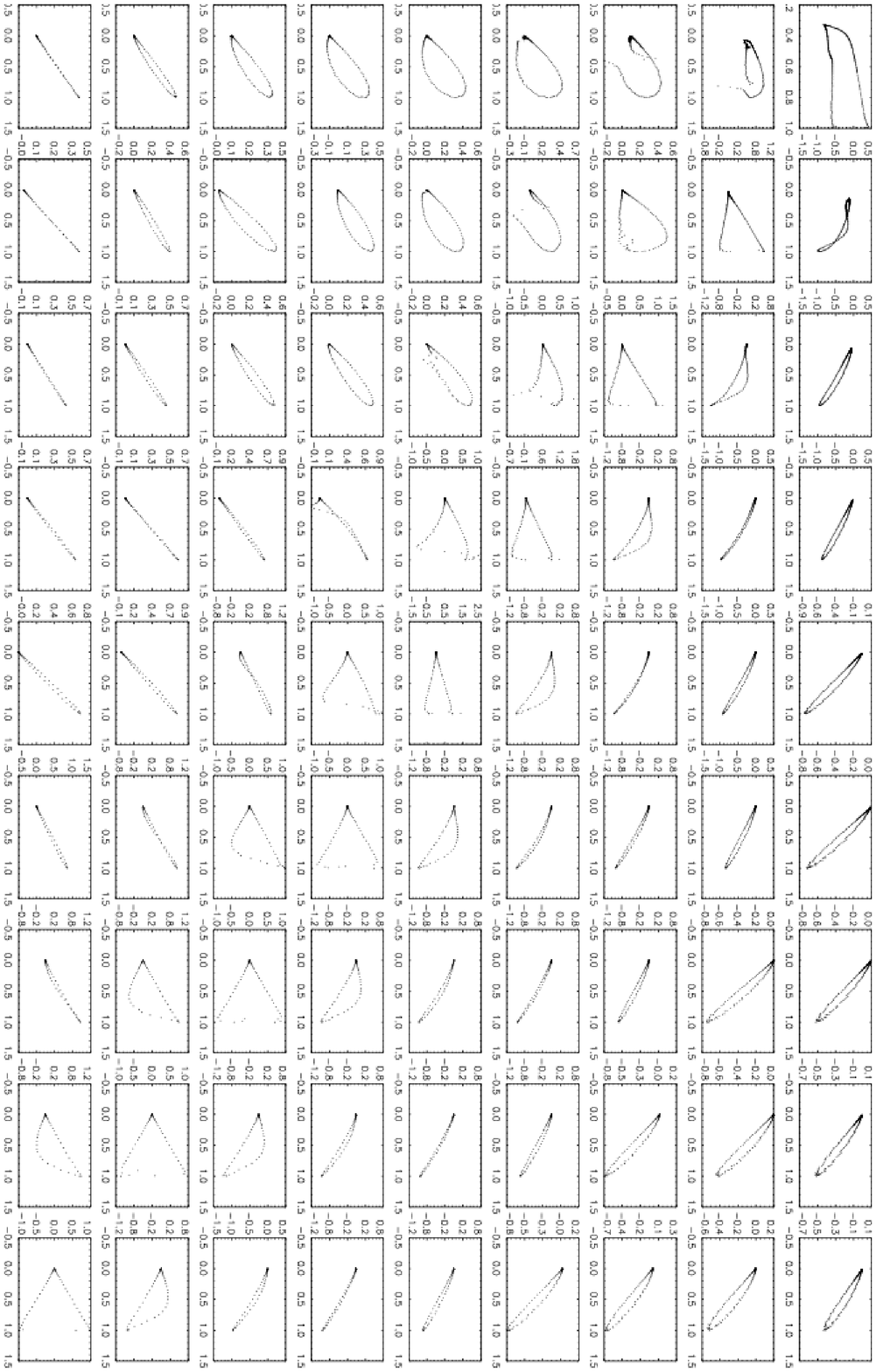}
\caption{As for Figure \ref{m0w10lcostheta_tqvsir25}, but for $I$-$U$ ($I$ on the horizontal axis).}
\label{m0w10lcostheta_tuvsir25}
\end{figure*}

\begin{figure*}
\includegraphics[scale=0.8]{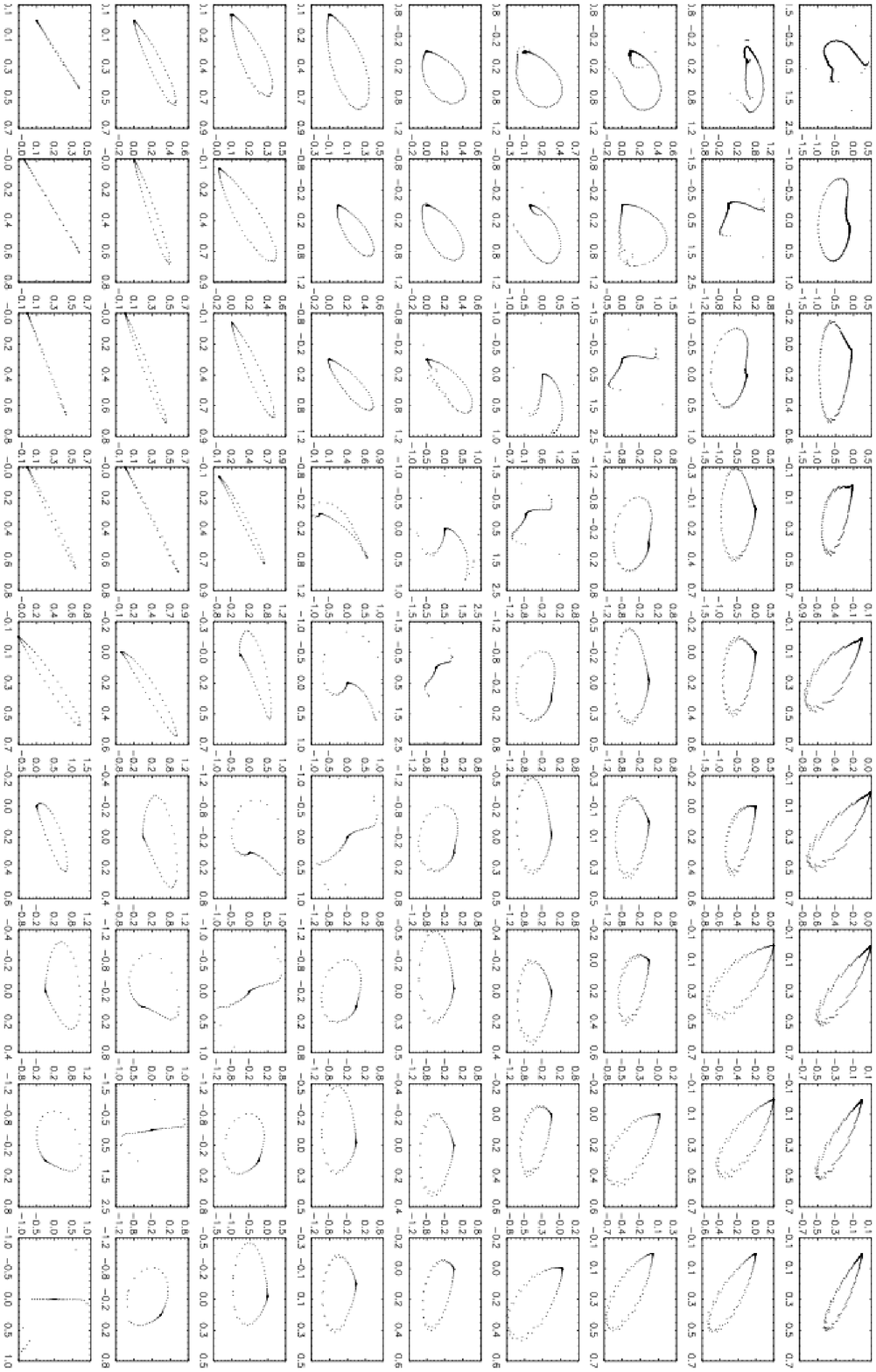}
\caption{As for Figure \ref{m0w10lcostheta_tqvsir25}, but for $Q$-$U$ ($Q$ on the horizontal axis).}
\label{m0w10lcostheta_tuvsqr25}
\end{figure*}

The Stokes phase portraits are drawn in Figures \ref{m0w10lcostheta_tqvsir25}--\ref{m0w10lcostheta_tuvsqr25}. 
The phase portraits differ from those of the pure dipole in both Sections \ref{sec:dipole} and \ref{sec:dipoleaberration}.
In the $I$-$Q$ plane (Figure \ref{m0w10lcostheta_tqvsir25}), we note the following behaviour. (i) For $\alpha = i$, the pulse traces out a triangular shape. The cutoff of the shapes at $I = 1$ is a numerical artifact caused by the finite grid resolution near $\theta_0 = 0$. (ii) For $\alpha > i$, we see tilted, narrow ovals. The ovals tilt upwards (major axis has slope $dQ/dI > 0$). For $i \geq 20^\circ$ and $\alpha \gtrsim 45^\circ$, the balloon twists into a figure-eight. (iii) For $\alpha < i$, we see tilted balloons for $\lvert \alpha - i \rvert \leq 45^\circ$, which narrow into tilted ovals as $i$ increases. The balloons tilt downwards (major axis has $dQ/dI < 0$) for $i - \alpha \lesssim 25^\circ$ and upwards (major axis has $dQ/dI > 0$) for $i - \alpha \gtrsim 25^\circ$.

In the $I$-$U$ plane (Figure \ref{m0w10lcostheta_tuvsir25}), we note the following behaviour. (i) For $\alpha = i$, we obtain a triangular shape. (ii) For $\alpha > i$ (below the diagonal in Figure \ref{m0w10lcostheta_tuvsir25}), the pulse traces out a balloon which narrows into a straight line as $\alpha$ increases.  (iii) For $\alpha < i$ (above the diagonal), we obtain balloons for $i - \alpha \lesssim 10^\circ$, and narrow ovals for $i - \alpha \gtrsim 20^\circ$. 

In the $Q$-$U$ plane (Figure \ref{m0w10lcostheta_tuvsqr25}), we see a mix of balloons, heart shapes and ovals with the following properties. (i) Along the $\alpha = i$ diagonal, we observe a `C' shape for $(\alpha, i) = (10^\circ, 10^\circ)$, which evolves into a distorted heart shape as $\alpha = i$ increases. (ii) For $\alpha > i$ (below the diagonal in Figure  \ref{m0w10lcostheta_tuvsqr25}) and $\lvert\alpha - i \rvert < 25^\circ$, we see distorted heart shapes with a cusp at $(Q, U) = (0, 0)$. For $\lvert \alpha - i \rvert \gtrsim 25^\circ$, we see tilted balloons which narrow into a straight line as $\alpha$ increases. (iii) For $\alpha < i$ (above the diagonal), we see heart shapes for $\lvert \alpha - i \rvert \leq 10^\circ$. These shapes evolve into tilted balloons as $\lvert \alpha - i \rvert$ increases. (iv) As in Figure \ref{m0w10lcostheta_tuvsir25}, the patterns are broadest for $\lvert \alpha - i \rvert \leq 10^\circ$ and narrow as $\lvert \alpha - i \rvert$ increases.

Generally, the shapes in Figure \ref{m0w10lcostheta_tuvsqr25} are not symmetric about $U = 0$ or $Q = 0$ and have a higher degree of rotation than those in Section \ref{sec:dipoleaberration}. The amount by which the shapes rotate depend on $\lvert \alpha - i \rvert$. Geometrically, this happens because $\theta_0(t)$ increases with $\lvert \alpha - i \rvert$. The degree of rotation decreases as $\lvert \alpha - i \rvert$ increases.  

\begin{figure*}
\includegraphics[scale=0.8]{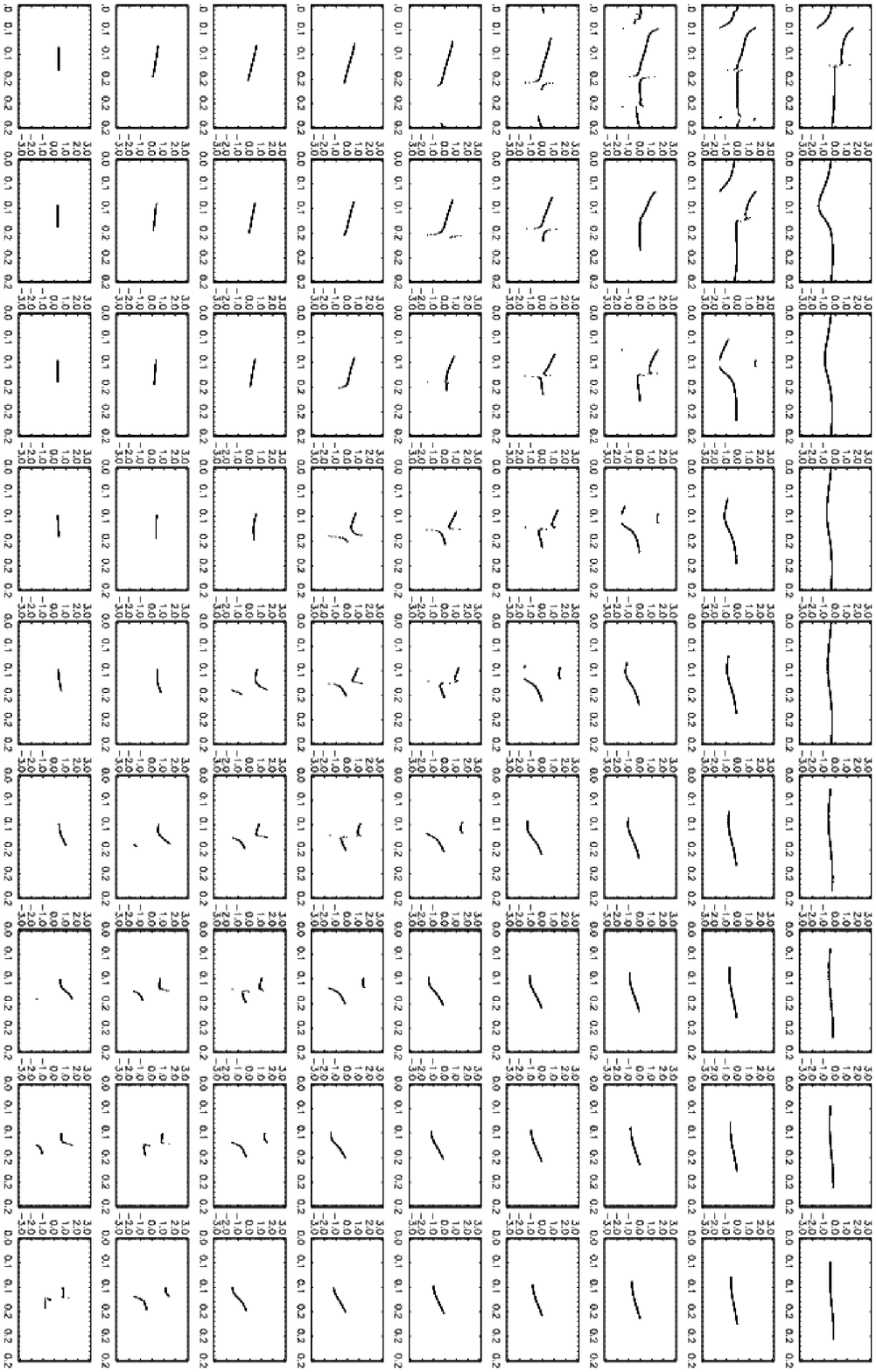}
\caption{Current-modified dipole field at $r = 0.1 r_\text{LC}$. Layout as for Figure \ref{m0w10lcostheta_tqvsir25}, but for position angle (on the vertical axis in landscape orientation, in units of radians) versus pulse longitude (on the horizontal axis, in units of $2\pi$ radians).}
\label{m0w10lcostheta_tpar25}
\end{figure*}

In Figure \ref{m0w10lcostheta_tpar25}, we plot the PA swings corresponding to each panel in Figures \ref{m0w10lcostheta_tqvsir25}--\ref{m0w10lcostheta_tuvsqr25}. We plot only the parts of the swing that are illuminated by the pulse, i.e. when $L \geq 10^{-2}$. We note the following behaviour. (i) There are clear distortions in the S-shape for $\lvert \alpha - i \rvert \leq 10^\circ$. Along $\alpha = i$, where the $I \approx 1$ saturation occurs, $d(PA)/dl$ changes discontinuously. A good example is $(\alpha, i) = (20^\circ, 20^\circ)$. The kinks around the discontinuities are numerical artifacts, but the smoothly varying curves leading up to the kinks are real. (ii) For $\lvert \alpha - i \rvert \geq 20^\circ$, the shape of the PA swings appear similar to those of Section \ref{sec:dipole}. The toroidal field enhances the phase shift of the PA swings in Figure \ref{m0w10lcostheta_tpar25}. For example, at $(\alpha, i) = (40^\circ, 70^\circ)$, the relative phase shift between the PA swing and the pulse profile is $\approx 5 r/r_\text{LC}$. The phase shift is altitude dependent; we discuss altitudinal variations in Section \ref{sec:rmapping}. The similarity of the PA swings for the pure and current-modified dipoles reinforces the necessity to supplement PA swings with Stokes phase portraits when diagnosing the pulsar magnetosphere.

\subsubsection{$L = I \sin \theta_0$}

\begin{figure}
\includegraphics[scale=0.5]{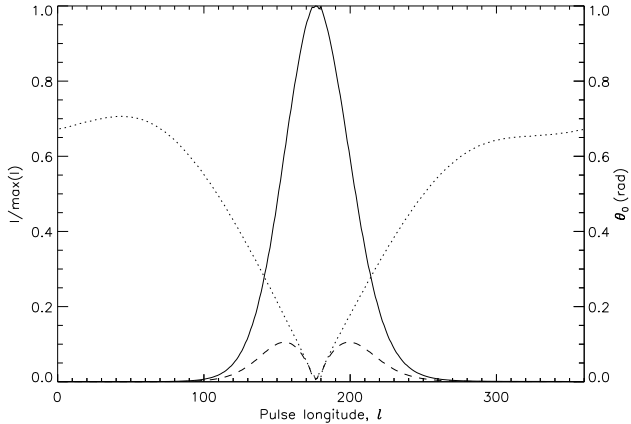}
\caption{Current-modified dipole field at $r = 0.1 r_\text{LC}$. Example of a pulse profile for a filled core beam with linear polarization $L = I \sin \theta_0$ and $(\alpha, i) = (30^\circ, 30^\circ)$. Solid, dashed and dotted curves represent the total polarized intensity $I$, degree of linear polarization $L$, and emission point colatitude $\theta_0$. Pulse longitude $l$ is measured in units of degrees.}
\label{m0w10lsintheta_tprofilesr25}
\end{figure}

An example of the $I$ (solid curve) and $L$ (dashed curve) profiles for $(\alpha, i) = (30^\circ, 30^\circ)$ is shown in Figure \ref{m0w10lsintheta_tprofilesr25}. The profiles are similar to those of the pure dipole in Section \ref{sec:dipolelsintheta}, displaying the same double-peaked $L$ profile along the $\alpha = i$ diagonal. The only difference is that the path traced by the emission point is not symmetric about the pulse centroid. 

\begin{figure*}
\includegraphics[scale=0.8]{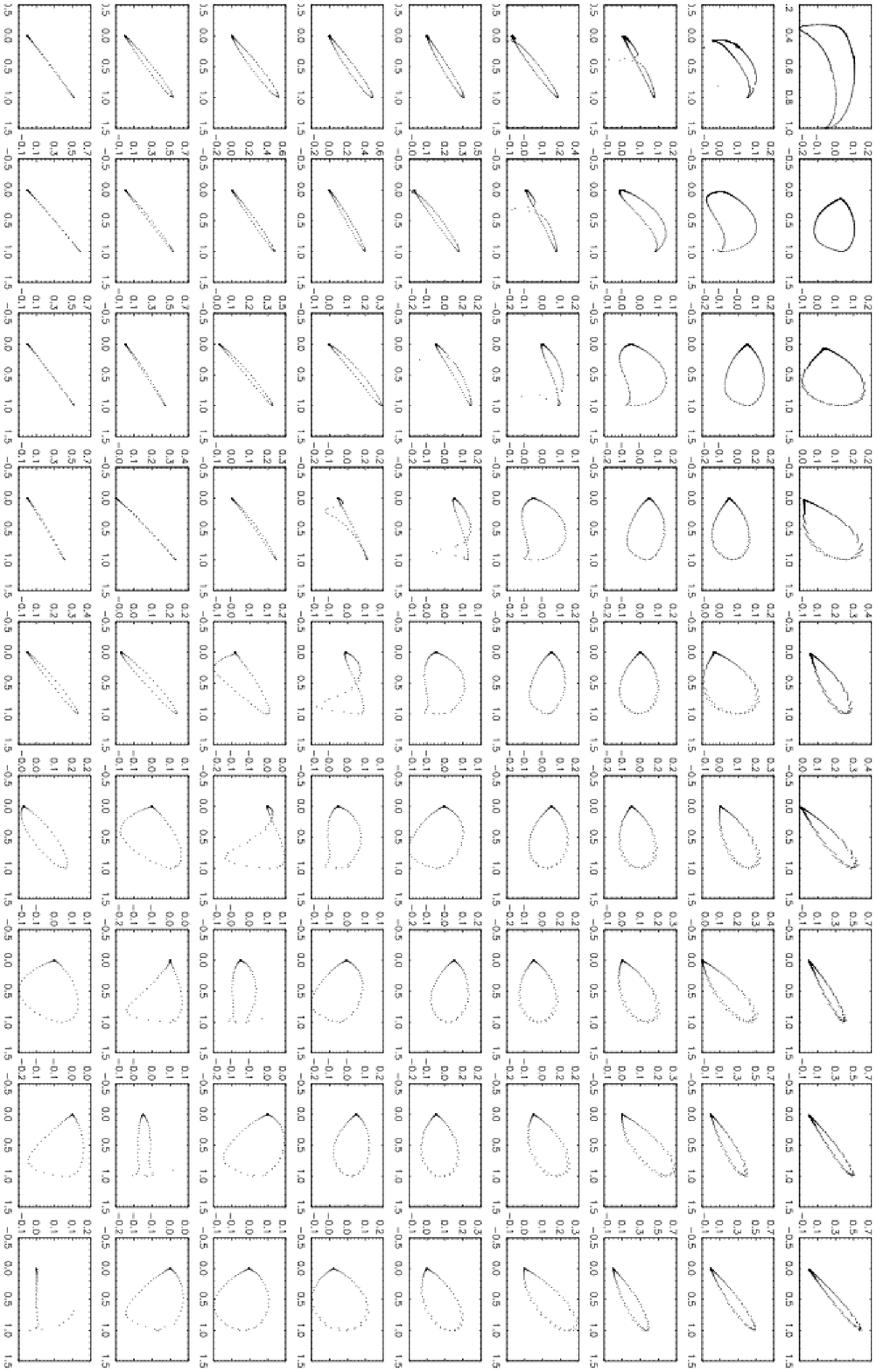}
\caption{Current-modified dipole field at $r = 0.1 r_\text{LC}$. Look-up table of Stokes phase portraits in the $I$-$Q$ plane for a filled core beam and degree of linear polarization $L = I \sin \theta_0$, where $\theta_0$ is the emission point colatitude. The panels are organised in landscape mode, in order of increasing $10^\circ \leq i \leq 90^\circ$ (left-right) and $10^\circ \leq \alpha \leq 90^\circ$ (top-bottom) in intervals of $10^\circ$. $I$ is plotted on the horizontal axis and normalised by its peak value. $Q$ is plotted on the vertical axis.}
\label{m0w10lsintheta_tqvsir25}
\end{figure*}

\begin{figure*}
\includegraphics[scale=0.8]{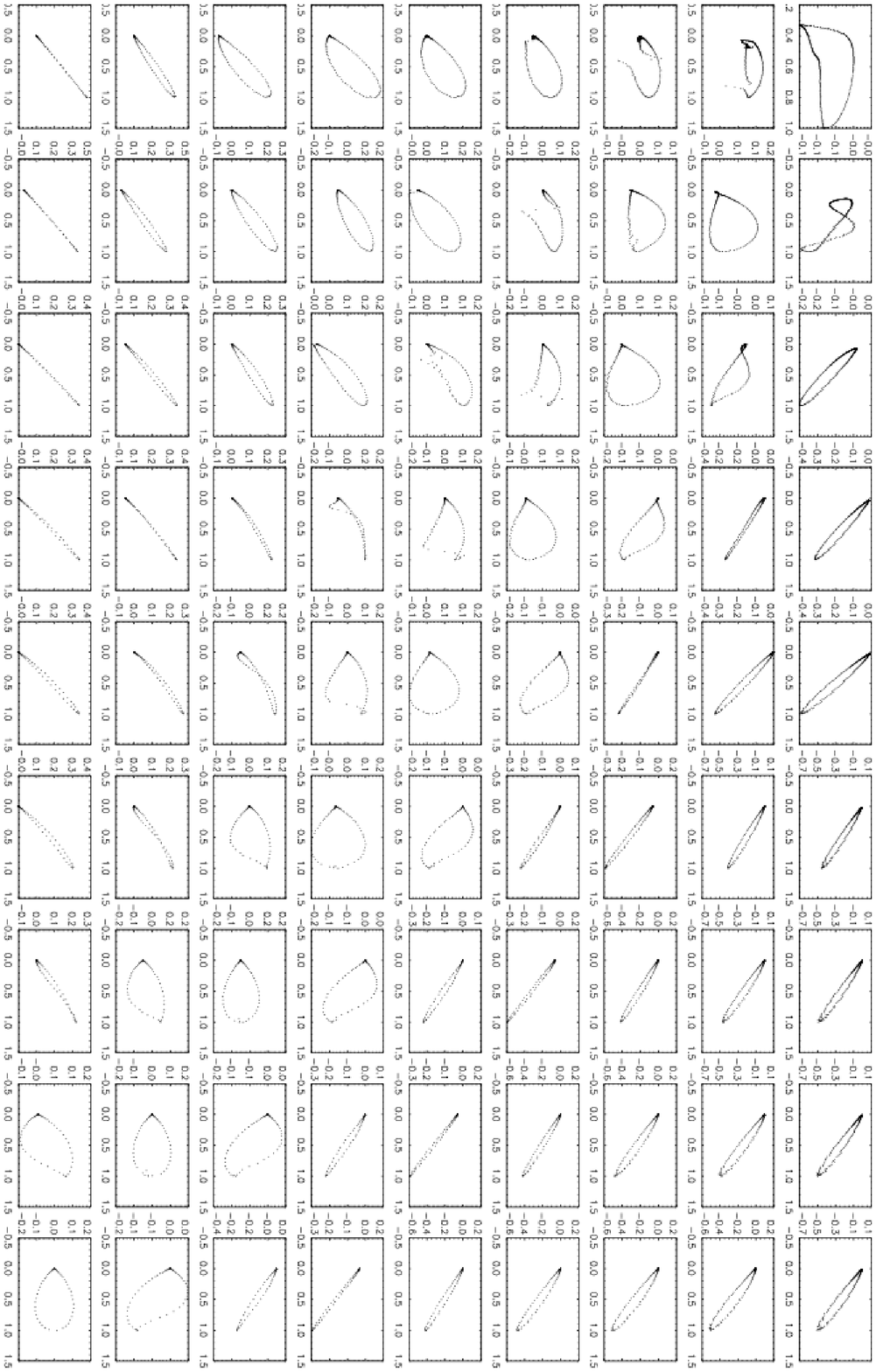}
\caption{As for Figure \ref{m0w10lsintheta_tqvsir25}, but for $I$-$U$ ($I$ on the horizontal axis).}
\label{m0w10lsintheta_tuvsir25}
\end{figure*}

\begin{figure*}
\includegraphics[scale=0.8]{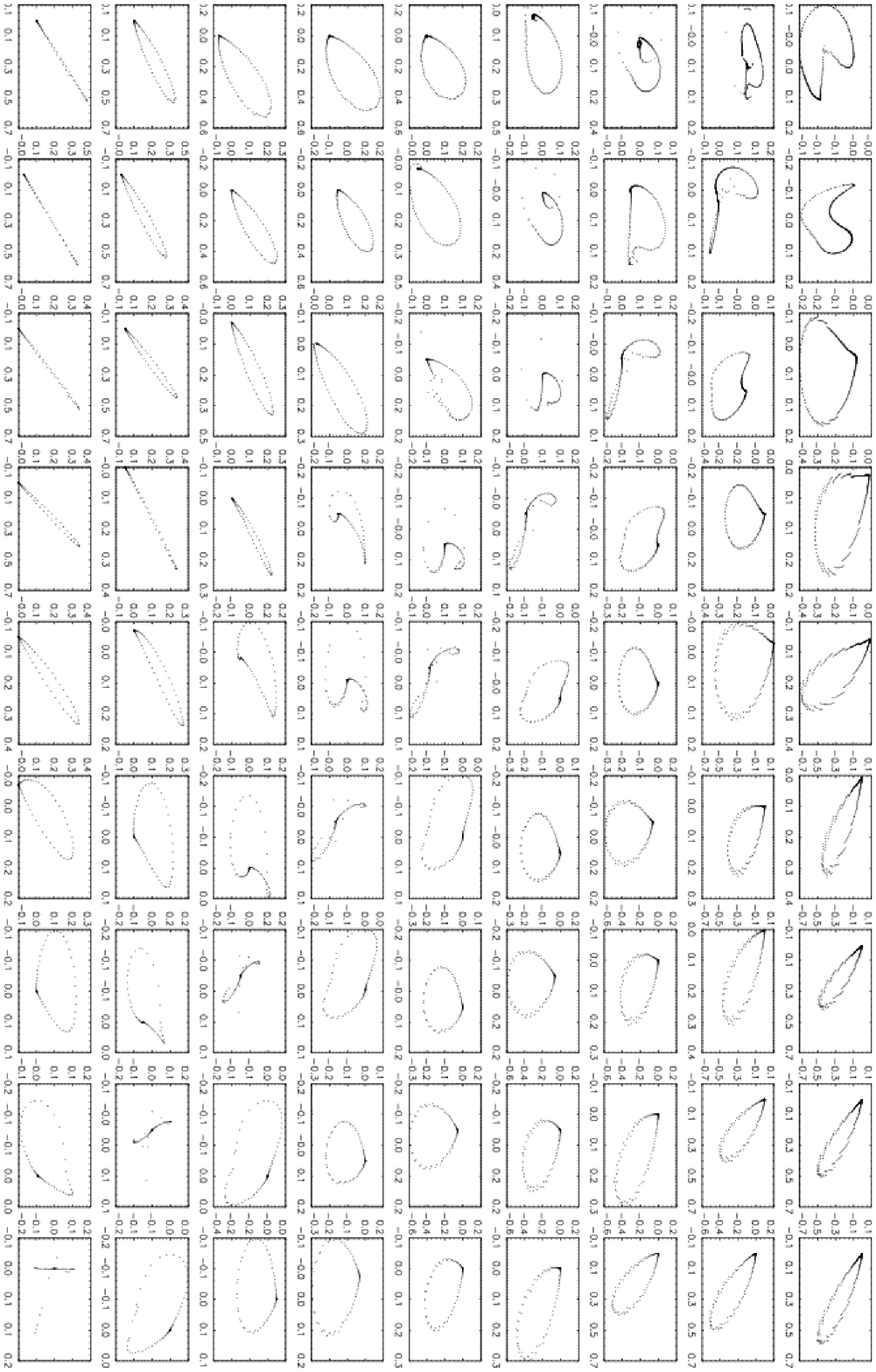}
\caption{As for Figure \ref{m0w10lsintheta_tqvsir25}, but for $Q$-$U$ ($Q$ on the horizontal axis).}
\label{m0w10lsintheta_tuvsqr25}
\end{figure*}

\begin{figure*}
\includegraphics[scale=0.8]{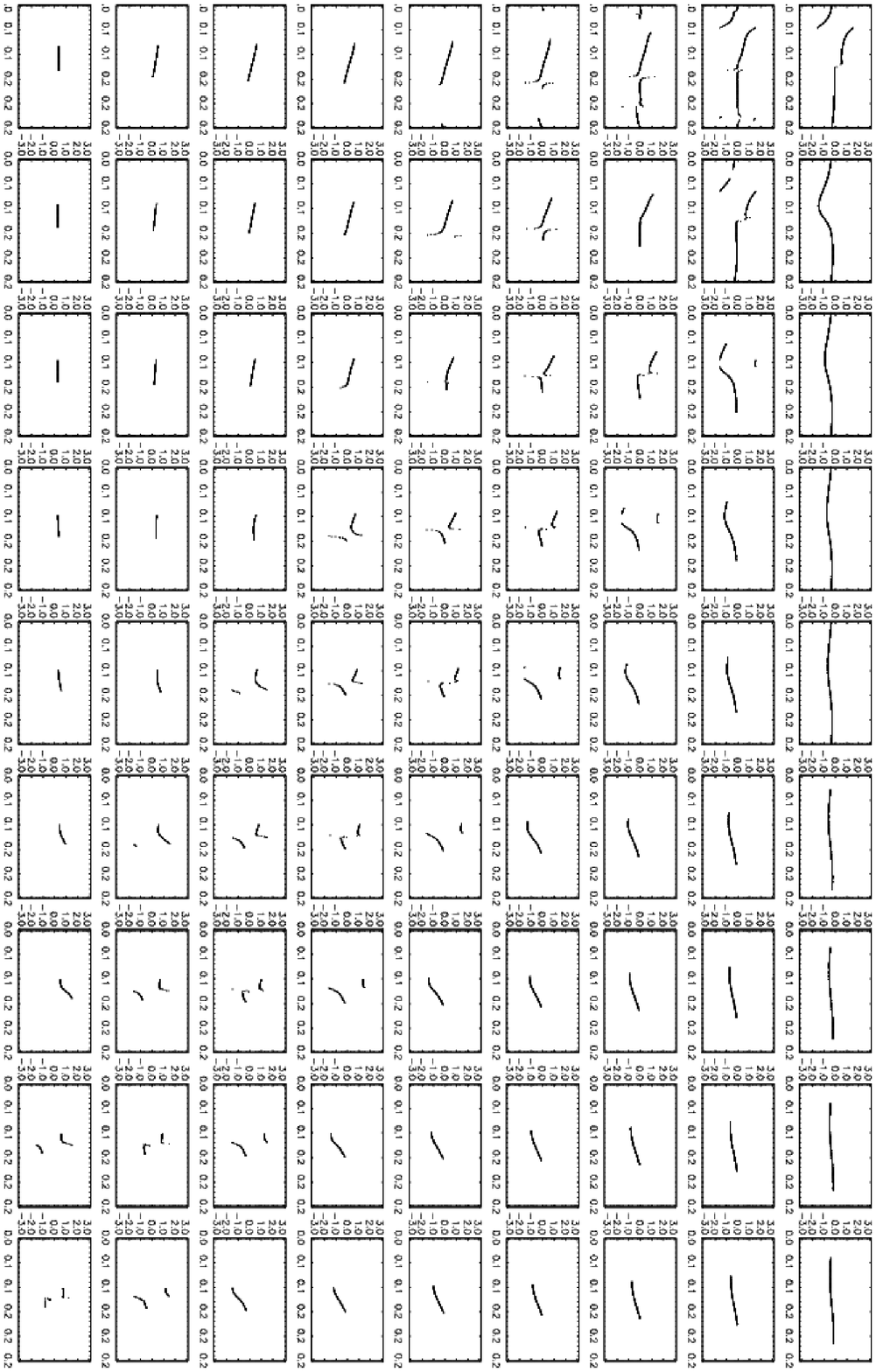}
\caption{Current-modified dipole field at $r = 0.1 r_\text{LC}$. Layout as for Figure \ref{m0w10lsintheta_tqvsir25}, but for position angle (on the vertical axis in landscape orientation, in units of radians) versus pulse longitude (on the horizontal axis, in units of $2\pi$ radians).}
\label{m0w10lsintheta_tpar25}
\end{figure*}

The Stokes phase portraits for a filled core with $L/I = \sin \theta_0$ are drawn in Figures \ref{m0w10lsintheta_tqvsir25}--\ref{m0w10lsintheta_tuvsqr25}. 
Comparing the Stokes phase portraits and PA swings to those in Section \ref{sec:toroidlcostheta}, we note the following differences. (i) In the $I$-$Q$ phase portrait (Figure \ref{m0w10lsintheta_tqvsir25}), along the $\alpha = i$ diagonal, there is a balloon instead of a triangle. The panels where $L/I$ is smaller than in Section \ref{sec:toroidlcostheta} exhibit narrower shapes than their counterparts in Figure \ref{m0w10lcostheta_tqvsir25}; for example, along the diagonal, the shapes are $\approx 10$ times narrower than in Figure \ref{m0w10lcostheta_tqvsir25}. (ii) In the $I$-$U$ phase portrait, (Figure \ref{m0w10lsintheta_tuvsir25}), the $I = 1$ wall seen in Figure \ref{m0w10lcostheta_tuvsir25} for $\lvert \alpha - i \rvert \leq 10^\circ$ is no longer present. For $\alpha = i$, instead of triangles, we see balloons. In the adjacent panels, we see tilted balloons which twist into a figure-eight for $\alpha \leq 20^\circ$ and $\alpha \geq 60^\circ$. Again, the shapes along the $\alpha = i$ diagonal are $\approx 10$ times narrower than their counterparts in Figure \ref{m0w10lcostheta_tuvsir25}. (iii) In the $Q$-$U$ plane (Figure \ref{m0w10lsintheta_tuvsqr25}), the main difference between Figures \ref{m0w10lcostheta_tuvsqr25} and \ref{m0w10lsintheta_tuvsqr25} occurs for $\lvert \alpha - i \rvert = 10^\circ$, where the heart shapes are distorted. Along the $\alpha = i$ diagonal, we see a distorted figure-eight. 
There is no noticeable difference between the PA swings in this model (Figure \ref{m0w10lsintheta_tpar25}) and the model in Section \ref{sec:toroidlcostheta}.

In summary, the Stokes phase portraits for a filled core beam are significantly different for pure and current-modified dipoles. Depending on its orientation, the toroidal field can either enhance or diminish the effects of aberration. The oval, balloon, and heart shapes are not symmetric about any axis and are rotated about $(U, Q) = (0, 0)$. The PA swings are distorted for $\lvert \alpha - i \rvert \leq 10^\circ$ but are similar to the pure dipole for other orientations apart from the phase shift. 

\subsection{Hollow cone emission}
\label{sec:toroidcone}
\subsubsection{$L = I \cos \theta_0$}
\label{sec:toroidconecostheta}

\begin{figure}
\includegraphics[scale=0.5]{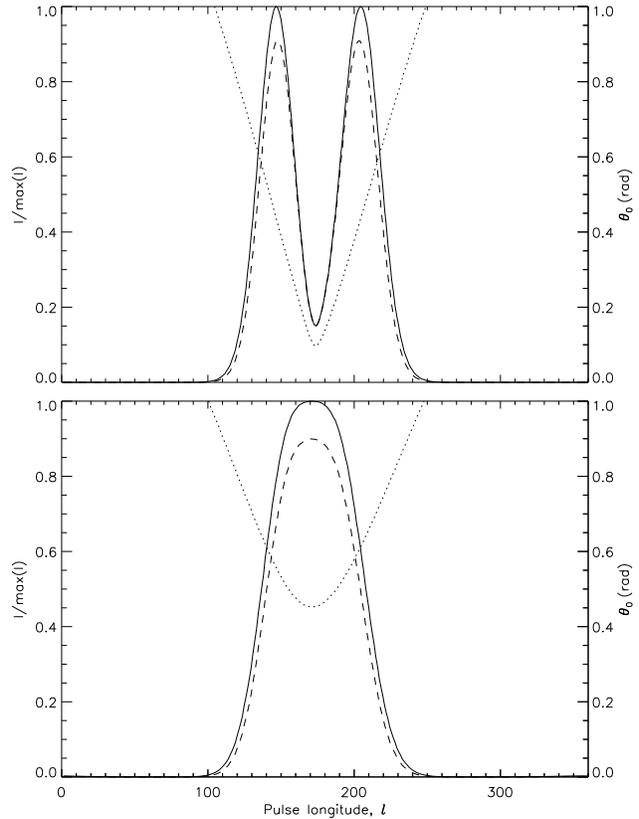}
\caption{Current-modified dipole field at $r = 0.1\, r_\text{LC}$. Examples of pulse profiles for a hollow cone with opening angle $25^\circ$ and degree of linear polarization $L = I \cos \theta_0$. Solid, dashed and dotted curves represent the total polarized intensity $I$, degree of linear polarization $L$, and emission point colatitude $\theta_0$. Pulse longitude $l$ is measured in units of degrees. Top: double-peaked pulse with $(\alpha, i) = (70^\circ, 60^\circ)$; bottom: single-peaked pulse with $(\alpha, i) = (70^\circ, 30^\circ)$.}
\label{m25w10lcostheta_tprofilesr25}
\end{figure}

Examples of the $I$ (solid curve) and $L$ (dashed curve) profiles for $(\alpha, i) =  (70^\circ, 30^\circ)$ and $(70^\circ, 60^\circ)$ are shown in Figure \ref{m25w10lcostheta_tprofilesr25}. $I$ is normalised by its peak value. The magnetic colatitude $\theta_0(t)$ of the emission point (dotted curve) is also shown. The profiles are double-peaked for $\lvert \alpha - i \rvert \leq 30^\circ$ and behave similarly to the pure dipole at low altitudes (Section \ref{sec:dipoleconecostheta}), with the exception of the $-r/r_\text{LC}$ phase shift. 

\begin{figure*}
\includegraphics[scale=0.8]{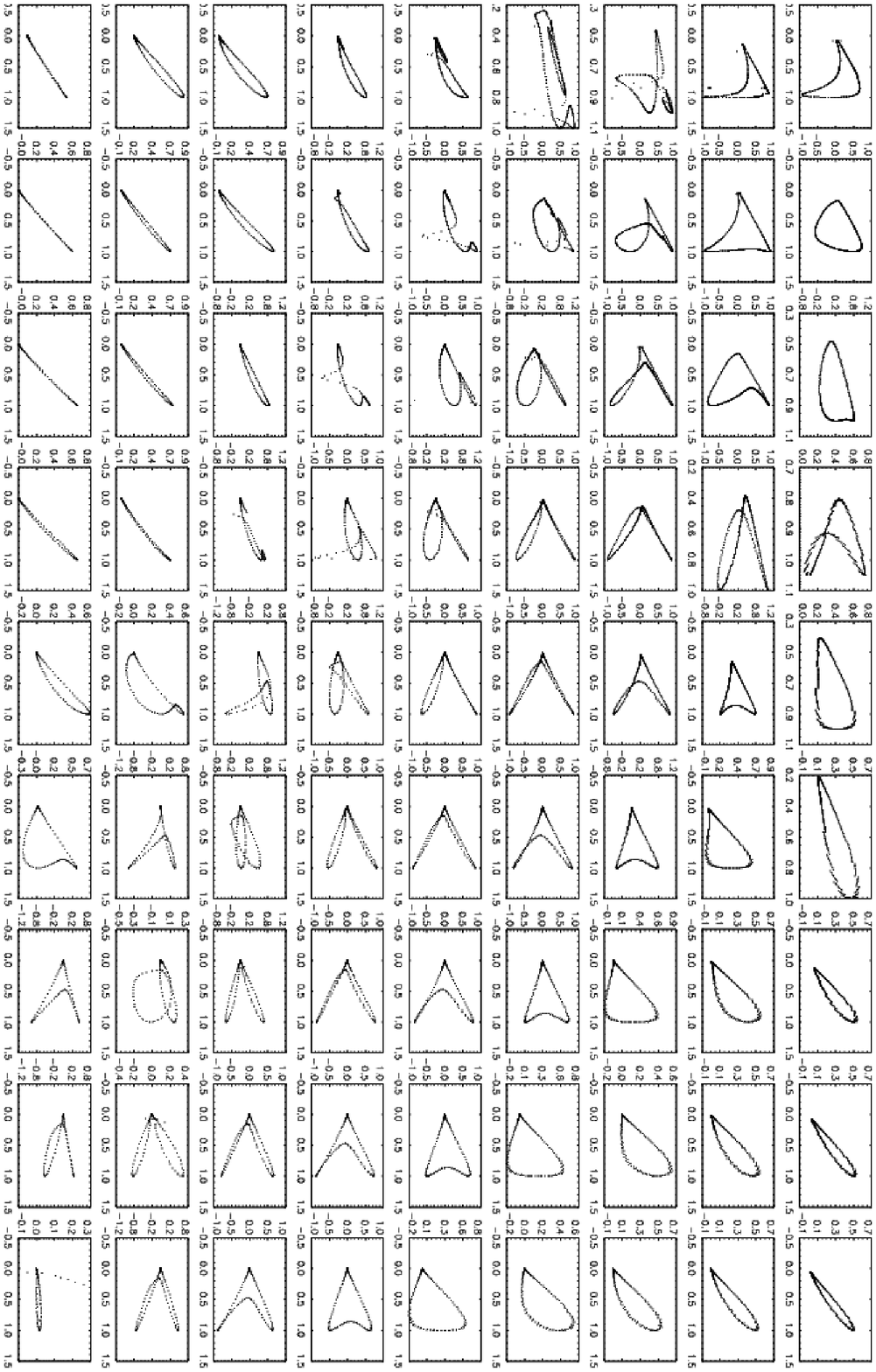}
\caption{Current-modified dipole field at $r = 0.1\, r_\text{LC}$. Look-up table of Stokes phase portraits in the $I$-$Q$ plane for a hollow cone with opening angle $25^\circ$ and degree of linear polarization $L = I \cos \theta_0$, where $\theta_0$ is the emission point colatitude. The panels are organised in landscape mode, in order of increasing $10^\circ \leq i \leq 90^\circ$ (left-right) and $10^\circ \leq \alpha \leq 90^\circ$ (top-bottom) in intervals of $10^\circ$. $I$ is plotted on the horizontal axis and normalised by its peak value. $Q$ is plotted on the vertical axis.}
\label{m25w10lcostheta_tqvsir25}
\end{figure*}

\begin{figure*}
\includegraphics[scale=0.8]{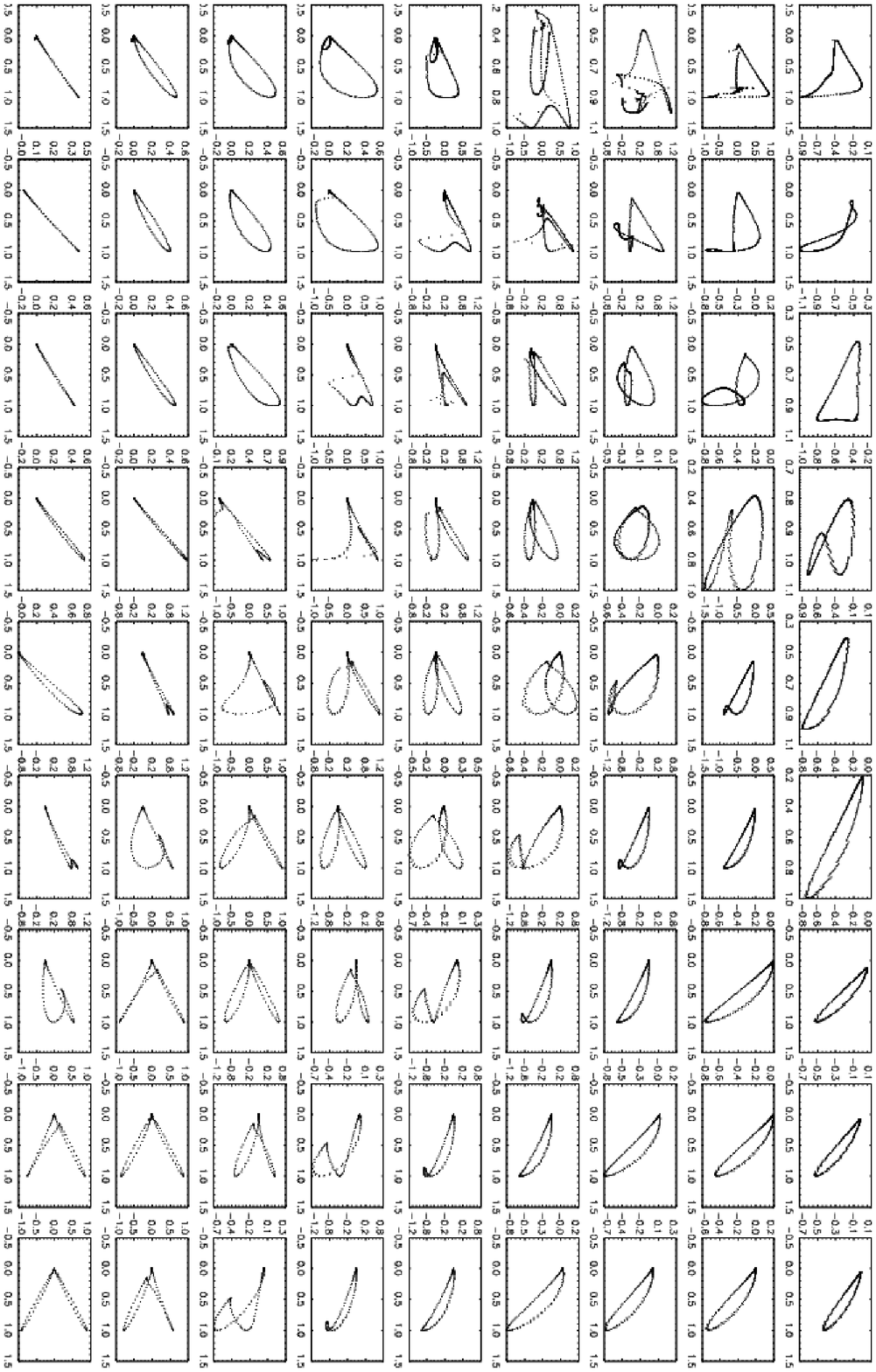}
\caption{As for Figure \ref{m25w10lcostheta_tqvsir25}, but for $I$-$U$ ($I$ on the horizontal axis).}
\label{m25w10lcostheta_tuvsir25}
\end{figure*}

\begin{figure*}
\includegraphics[scale=0.8]{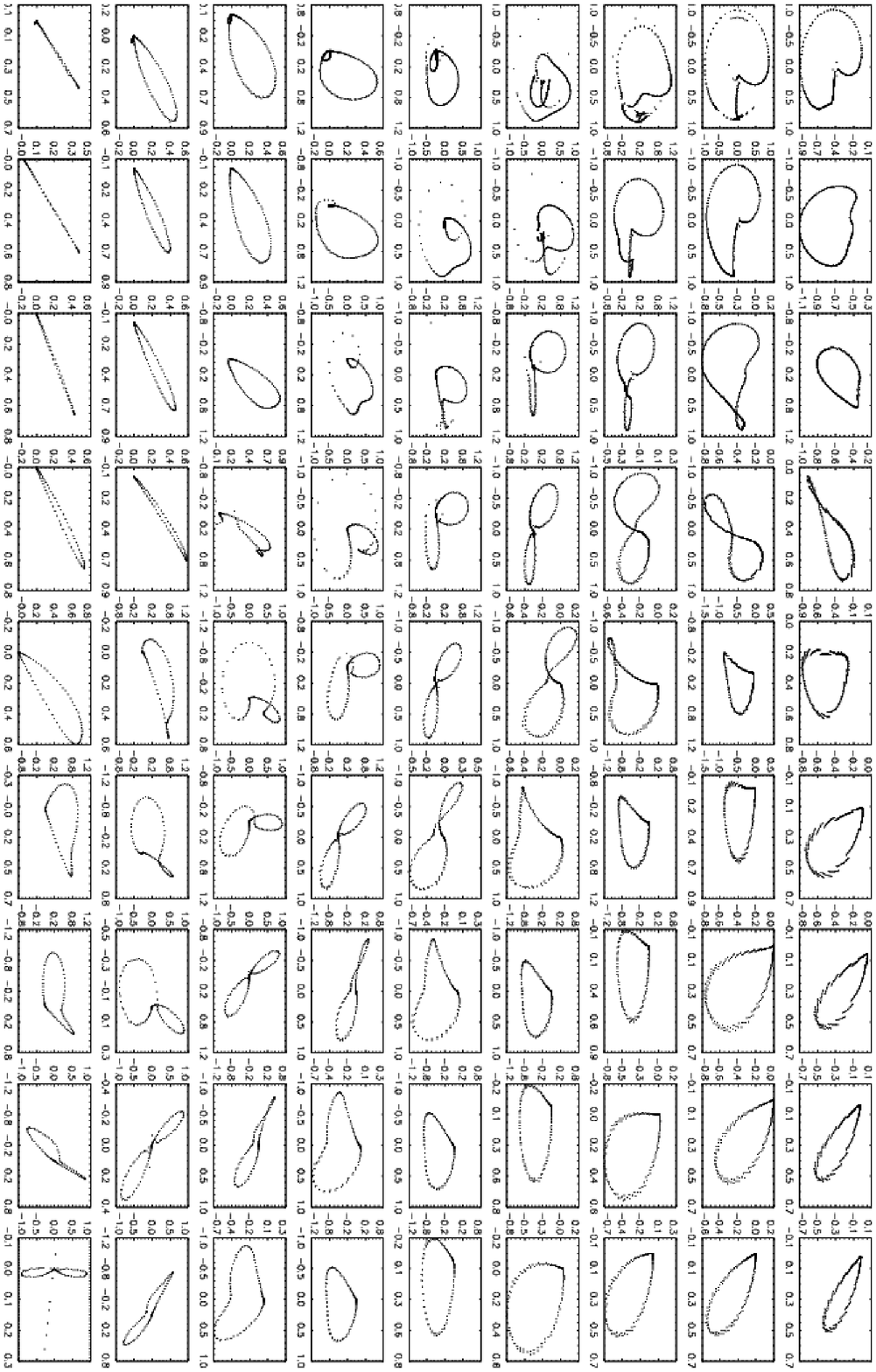}
\caption{As for Figure \ref{m25w10lcostheta_tqvsir25}, but for $Q$-$U$ ($Q$ on the horizontal axis).}
\label{m25w10lcostheta_tuvsqr25}
\end{figure*}

The Stokes phase portraits for a hollow cone with $L/I = \cos \theta_0$ are drawn in Figures \ref{m25w10lcostheta_tqvsir25}--\ref{m25w10lcostheta_tuvsqr25}. 
In the $I$-$Q$ phase portraits (Figure \ref{m25w10lcostheta_tqvsir25}), we note the following trends. (i) For $\alpha = i$, we obtain a tilted crescent at $(\alpha, i) = (10^\circ, 10^\circ)$, which evolves into an asymmetric mosquito shape as $\alpha = i$ increases. (ii) For $\alpha > i$ (below the diagonal in Figure \ref{m25w10lcostheta_tqvsir25}), where the pulse is double-peaked, we obtain a mix of figure-eights and distorted mosquito shapes. (iii) For $\alpha < i$ (above the diagonal), where the pulse is double-peaked, the pulse traces mosquito shapes (for $\alpha \gtrsim 25^\circ$) and a mix of balloons, figure-eights and triangular shapes (for $\alpha \lesssim 25^\circ$).

In the $I$-$U$ phase portraits (Figure \ref{m25w10lcostheta_tuvsir25}), we note the following trends. (i) Along the $\alpha = i$ diagonal, we obtain a tilted crescent at $(\alpha, i) = (10^\circ, 10^\circ)$ which twists into a mosquito shape as $\alpha = i$ increases. (ii) For $\alpha > i$ (below the diagonal), where the pulse is double-peaked, we obtain heavily distorted, complex shapes at $\alpha \leq 30^\circ$, which evolve into a distorted mosquito shape as $\alpha$ increases. (iii) For $\alpha < i$ (above the diagonal), where the pulse is double-peaked, we obtain a tilted figure-eight at $\alpha = 10$, which evolves into twisted, interlocking ovals as $\alpha$ increases.

In the $Q$-$U$ phase portraits (Figure \ref{m25w10lcostheta_tuvsqr25}), we note the following trends. (i) Along the $\alpha = i$ diagonal, the pattern evolves from a tilted heart shape to a figure-eight as $\alpha = i$ increases. (ii) Interestingly, in the panels to the left of the $\alpha = i$ diagonal, one of the ventricles of the heart shape twists into a secondary loop which rotates counter-clockwise as $\alpha$ increases [compare $(\alpha, i) = (40^\circ, 30^\circ), (60^\circ, 50^\circ)$, and $(90^\circ, 80^\circ)$]. (iii) For $\alpha < i$, where the pulses are double-peaked, a tilted figure-eight pattern emerges. (iv) The phase portraits for orientations which produce single-peaked pulses behave like in Section \ref{sec:toroidlcostheta}; i.e. they rotate about $(U, Q) = (0, 0)$ as $\lvert \alpha - i \rvert$ increases. 

\begin{figure*}
\includegraphics[scale=0.8]{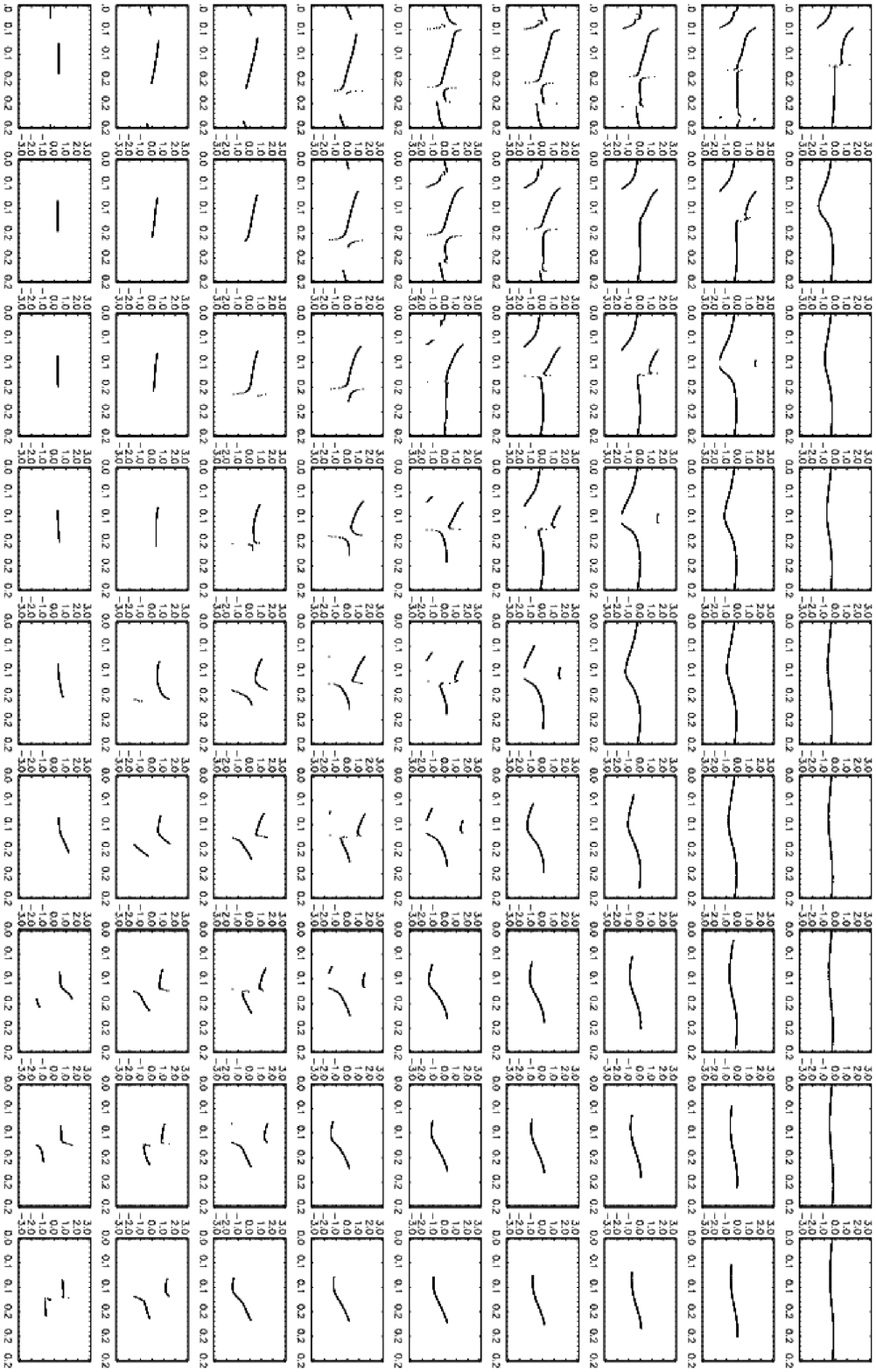}
\caption{Current-modified dipole field at $r = 0.1\, r_\text{LC}$. Layout as for Figure \ref{m25w10lcostheta_tqvsir25}, but for position angle (on the vertical axis in landscape orientation, in units of radians) versus pulse longitude (on the horizontal axis, in units of $2\pi$ radians).}
\label{m25w10lcostheta_tpar25}
\end{figure*}

The PA swings (Figure \ref{m25w10lcostheta_tpar25}) are identical to the previous cases for all orientations, the only difference being that they are visible above a threshold intensity over a greater fraction of the pulse period due to the broader pulse profiles.

\subsubsection{$L = I \sin \theta_0$}

\begin{figure}
\includegraphics[scale=0.5]{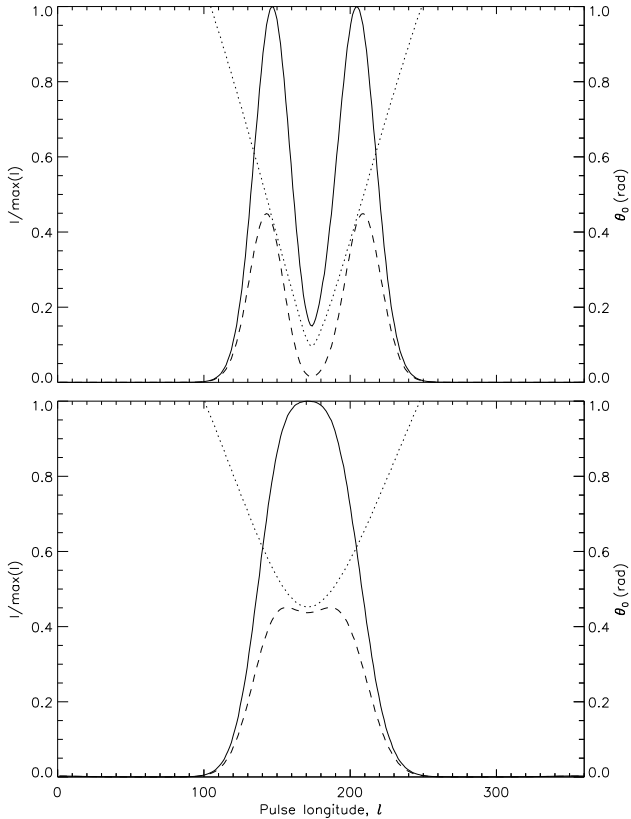}
\caption{Current-modified dipole field at $r = 0.1\, r_\text{LC}$. Examples of pulse profiles for a hollow cone with opening angle $25^\circ$ and degree of linear polarization $L = I \sin \theta_0$. Solid, dashed and dotted curves represent the total polarized intensity $I$, degree of linear polarization $L$, and emission point colatitude $\theta_0$. Pulse longitude $l$ is measured in units of degrees. Top: double-peaked pulse with $(\alpha, i) = (70^\circ, 60^\circ)$; bottom: single-peaked pulse with $(\alpha, i) = (70^\circ, 30^\circ)$.}
\label{m25w10lsintheta_tprofilesr25}
\end{figure}

Examples of the $I$ (solid curve) and $L$ (dashed curve) profiles for $(\alpha, i) =  (70^\circ, 30^\circ)$ and $(70^\circ, 60^\circ)$ are shown in Figure \ref{m25w10lsintheta_tprofilesr25}. The profiles are similar to those of Section \ref{sec:dipoleconesintheta}.

\begin{figure*}
\includegraphics[scale=0.8]{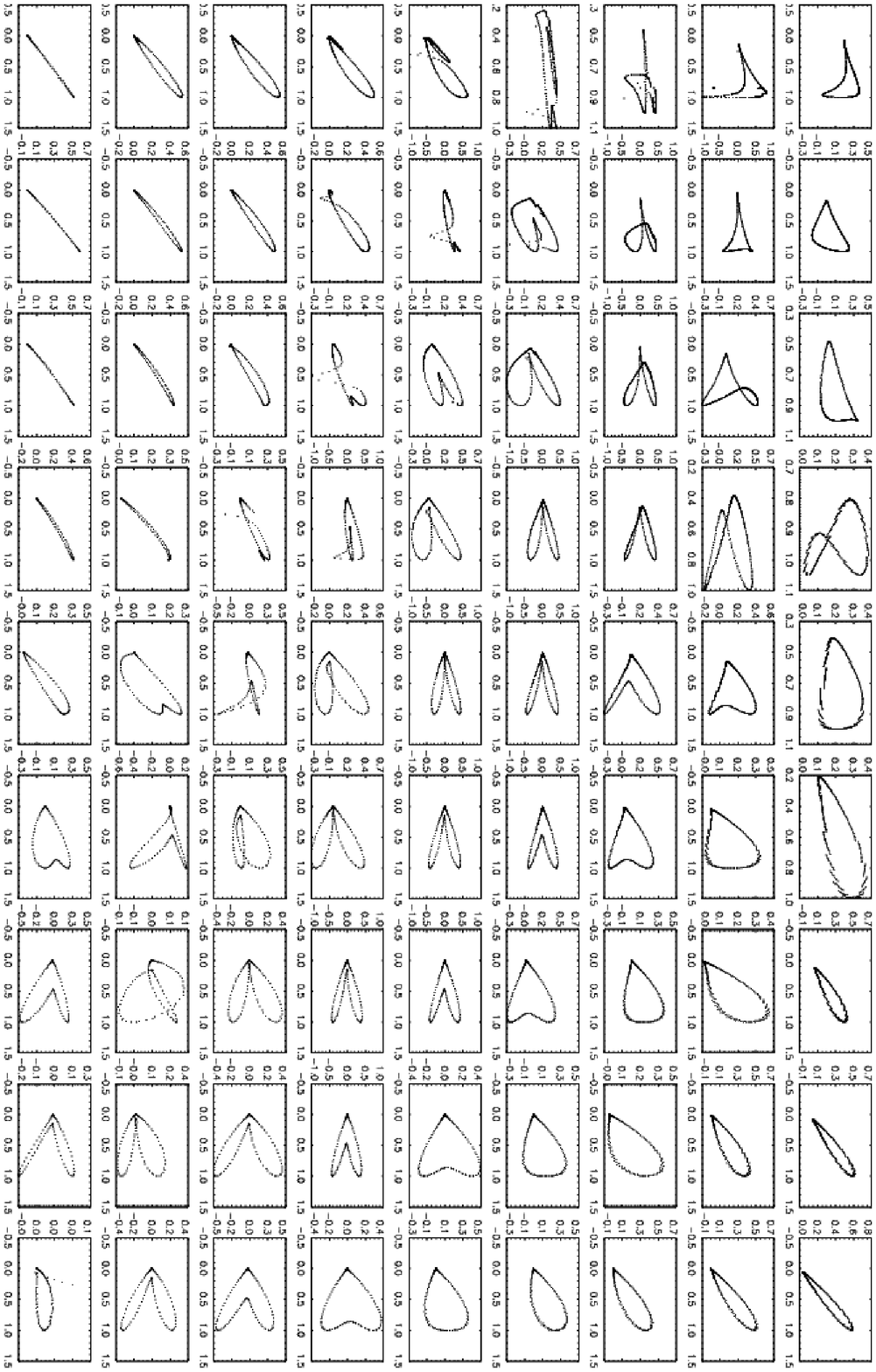}
\caption{Current-modified dipole field at $r = 0.1\, r_\text{LC}$. Look-up table of Stokes phase portraits in the $I$-$Q$ plane for a hollow cone with opening angle $25^\circ$ and degree of linear polarization $L = I \sin \theta_0$, where $\theta_0$ is the emission point colatitude. The panels are organised in landscape mode, in order of increasing $10^\circ \leq i \leq 90^\circ$ (left-right) and $10^\circ \leq \alpha \leq 90^\circ$ (top-bottom) in intervals of $10^\circ$. $I$ is plotted on the horizontal axis and normalised by its peak value. $Q$ is plotted on the vertical axis.}
\label{m25w10lsintheta_tqvsir25}
\end{figure*}

\begin{figure*}
\includegraphics[scale=0.8]{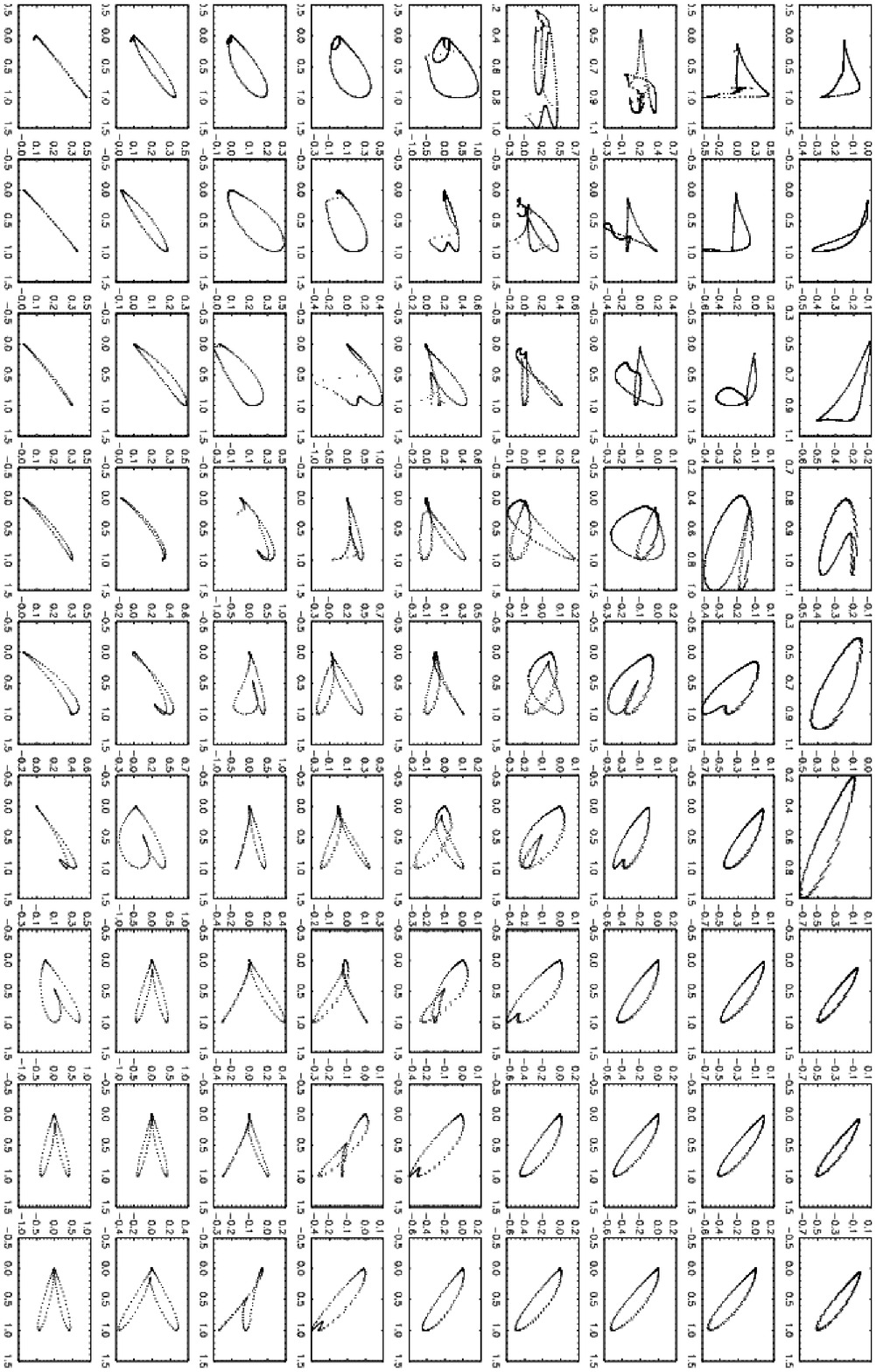}
\caption{As for Figure \ref{m25w10lsintheta_tqvsir25}, but for $I$-$U$ ($I$ on the horizontal axis).}
\label{m25w10lsintheta_tuvsir25}
\end{figure*}

\begin{figure*}
\includegraphics[scale=0.8]{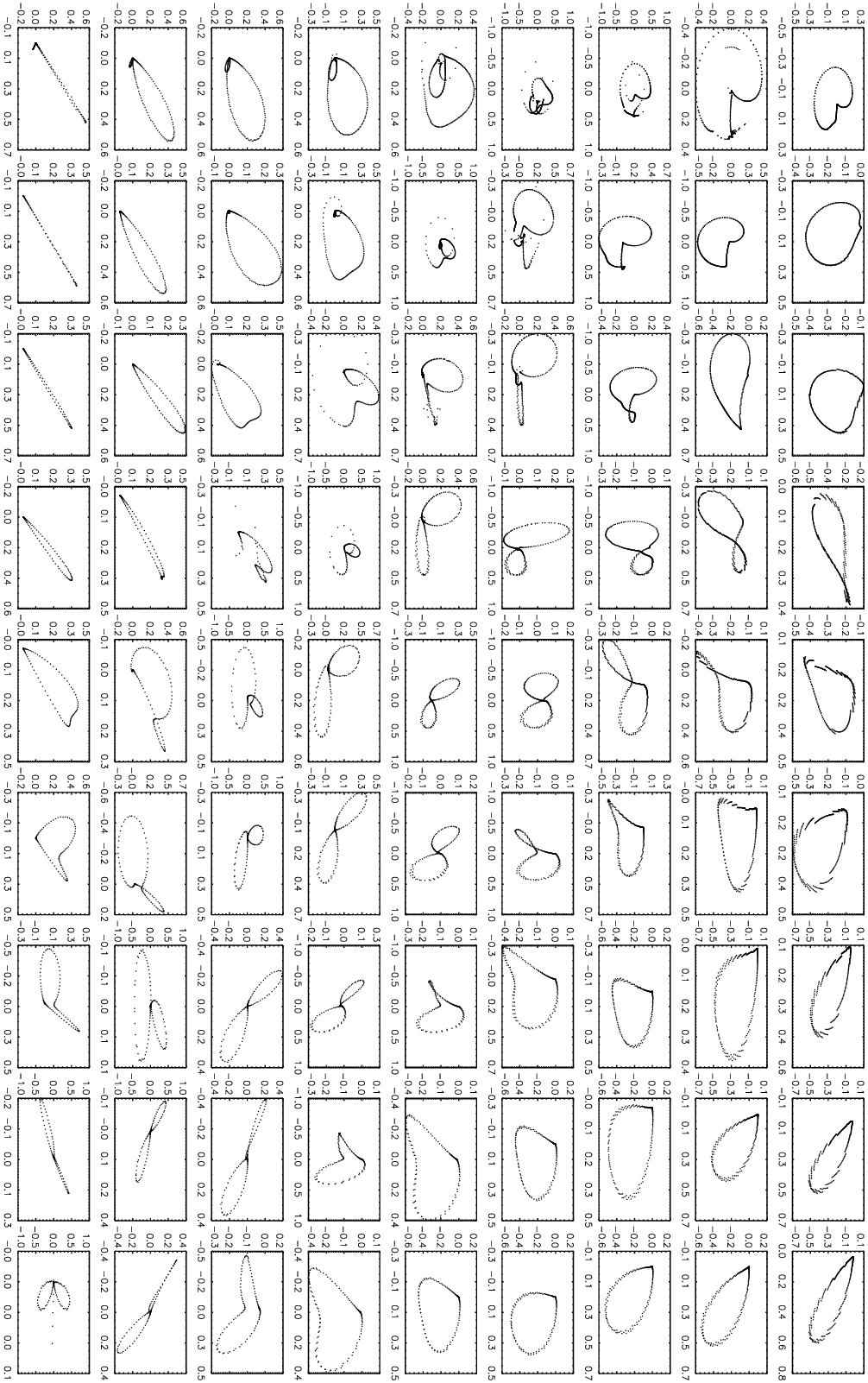}
\caption{As for Figure \ref{m25w10lsintheta_tqvsir25}, but for $Q$-$U$ ($Q$ on the horizontal axis).}
\label{m25w10lsintheta_tuvsqr25}
\end{figure*}

The $I$-$Q$, $I$-$U$, and $Q$-$U$ phase portraits are shown in Figures \ref{m25w10lsintheta_tqvsir25}--\ref{m25w10lsintheta_tuvsqr25}. 
All three phase portraits are similar to Section \ref{sec:toroidconecostheta}, except for orientations that produce double-peaked pulses. Here, the patterns are less twisted. The lower peak value of $L/I$ as compared to Section \ref{sec:toroidconecostheta} is accompanied by a narrower range of $U$ and $Q$. For example, in the $Q$-$U$ plane (Figure \ref{m25w10lsintheta_tuvsqr25}) for $(\alpha, i) = (10^\circ, 10^\circ)$, the oval has approximately 20\% the size of its counterpart in Figure \ref{m25w10lcostheta_tuvsqr25} along both the $U$ and $Q$ axes. 

\begin{figure*}
\includegraphics[scale=0.8]{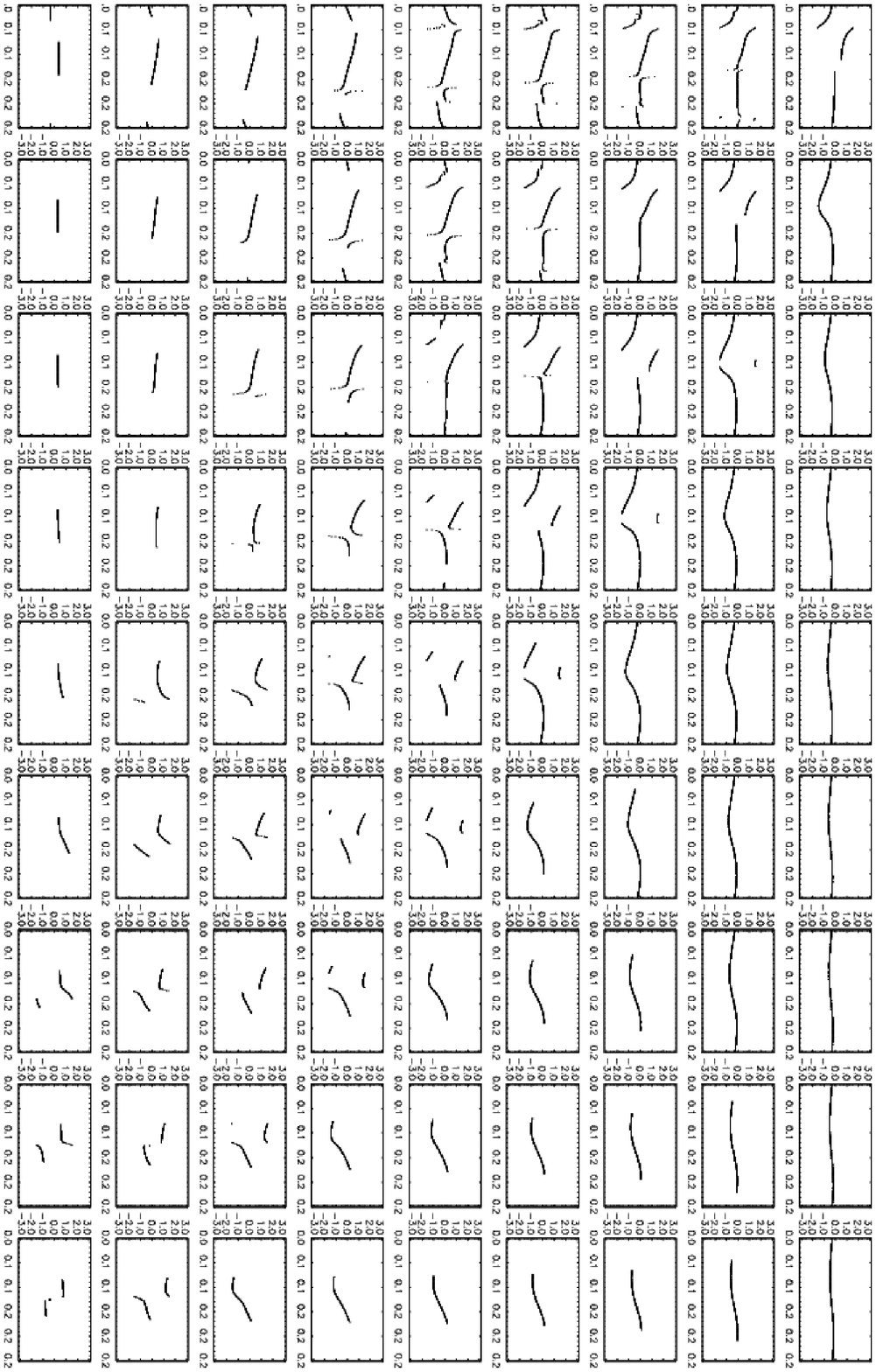}
\caption{Current-modified dipole field at $r = 0.1\, r_\text{LC}$. Layout as for Figure \ref{m25w10lsintheta_tqvsir25}, but for position angle (on the vertical axis in landscape orientation, in units of radians) versus pulse longitude (on the horizontal axis, in units of $2\pi$ radians).}
\label{m25w10lsintheta_tpar25}
\end{figure*}

The PA swings in Figure \ref{m25w10lsintheta_tpar25} are identical to Figure \ref{m25w10lcostheta_tpar25}, except in the squared-off valleys where $\theta_0$ is near zero and we do not plot the curve ($L < 10^{-2}$).

\subsection{Emission altitude}
\label{sec:rmapping}
\begin{figure*}
\includegraphics[scale=0.8]{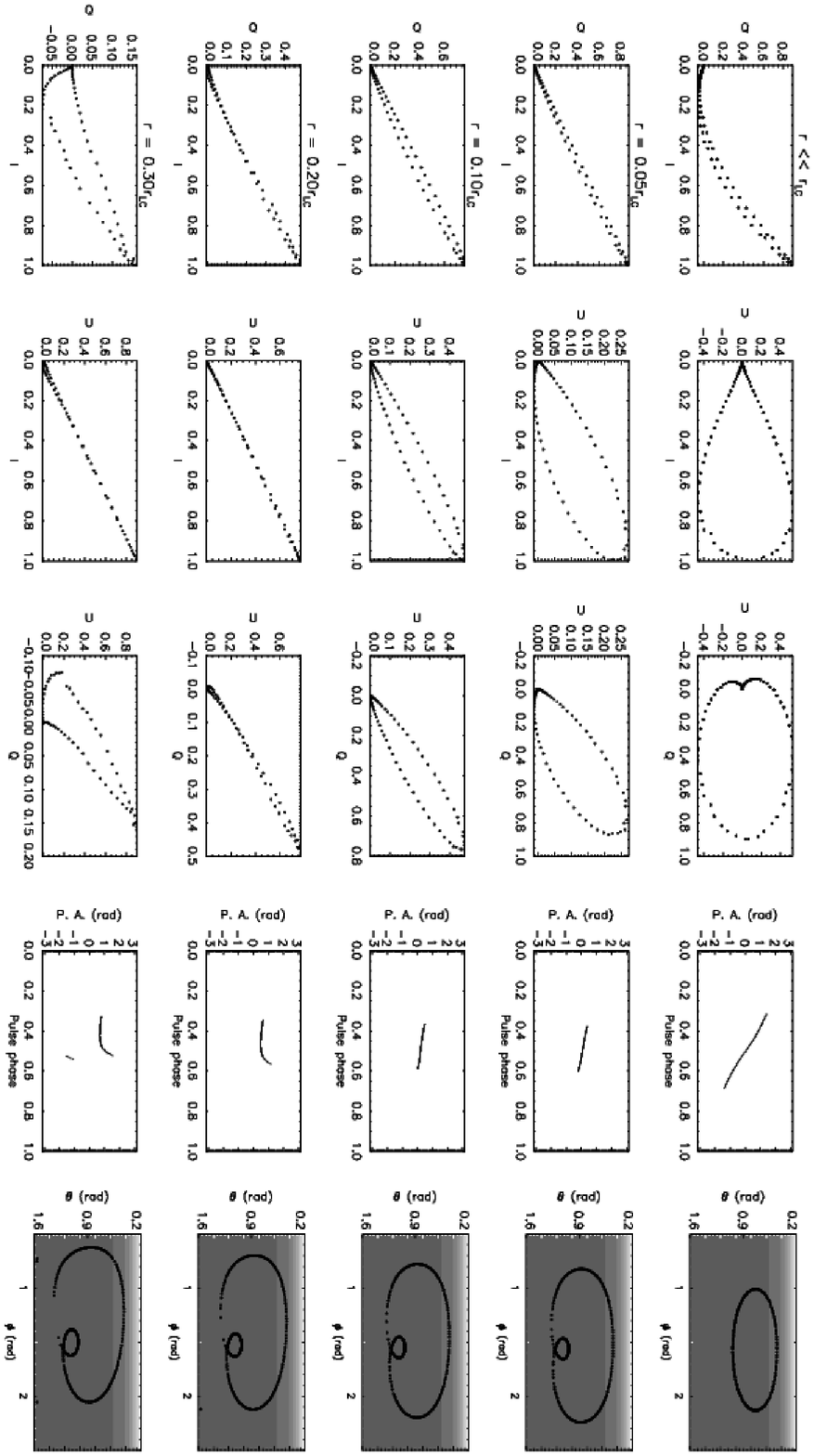}
\caption{Stokes phase portraits, PA swings and emission point locus $\hat{\mathbf{x}}_0(t)$ in the body frame for a filled core beam with polarization $L = I \cos \theta_0$ and $(\alpha, i) = (70^\circ, 30^\circ)$, at five altitudes. In landscape mode, the top row shows a pure dipole at $r \ll r_\text{LC}$. In rows 2--5, $r$ ranges from $r = 0.05\,r_\text{LC}$ (second row) to $r = 0.3\,r_\text{LC}$ (bottom row). From left to right in landscape mode, the columns show: (1) the $I$-$Q$ phase portrait, (2) the $I$-$U$ phase portrait, (3) the $Q$-$U$ phase portrait, (4) the PA swing (data points with $L \geq 10^{-2}$ plotted only), and (5) the locus of $\hat{\mathbf{x}}_0(t)$. The intensity map (greyscale) follows Figure \ref{dipoleemission}.}
\label{altitude1}
\end{figure*}

In this section, we investigate how the emission altitude changes the Stokes phase portraits and PA swings. For the sake of definiteness, we focus on the orientation $(\alpha, i) = (70^\circ, 30^\circ)$ for a filled core beam with polarization $L = I \cos \theta_0$. In Figure \ref{altitude1}, we present the Stokes phase portraits, PA swing, and locus of $\hat{\mathbf{x}}_0(t)$ for this orientation at four altitudes ranging from $r_\text{min} = 0.05\,r_\text{LC}$ to $r_\text{max} = 0.3\,r_\text{LC}$. The pure dipole case at $r \ll r_\text{LC}$ is presented at the top of the figure as a comparison. $I$ is normalised by its peak value.

We begin with an obvious yet fundamental observation: in a current-modified magnetic dipole field, the phase portraits evolve with altitude. The PA swing evolves also but to a lesser degree.

We note the following trends. (i) The peak value of $L/I$ increases from 0.90 at $r_\text{min}$ to 0.92 at $r_\text{max}$. (ii) In the $I$-$Q$ plane, at $r = r_\text{min}$, the hockey stick straightens into a line (major axis has $dQ/dI > 0$), ranging over $0 < Q < 0.8$. As $r$ increases, the slope of the line becomes less steep. At $r = r_\text{max}$, the line broadens into a tilted banana shape ranging over $-0.07 < Q < 0.15$. (iii) In the $I$-$U$ plane, the balloon narrows and tilts upwards (major axis has $dU/dI > 0$). (iv) In the $Q$-$U$ plane, the heart shape narrows into a straight line, then broadens again as $r$ increases, rotating counter-clockwise about $(Q, U) = (0,0)$. (v) The shape of the PA swing changes as $r$ increases, and the pulse longitude where the phase wrapping occurs shifts gradually from $\approx 0.65$ phase units at $r = r_\text{min}$ to $\approx 0.54$ phase units at $r = r_\text{max}$. 

To explain the trends, we look at the locus of $\hat{\mathbf{x}}_0(t)$ in the body frame. From $r = r_\text{min}$ to $r = r_\text{max}$, $\hat{\mathbf{x}}_0(t)$ traces out an oval containing a secondary loop, which is not present for $r << r_\text{LC}$. The oval tilts and becomes egg-shaped with increasing $r$. The range of $\theta_0(t)$ increases from 0.4 rad $\lesssim \theta_0 \lesssim$ 1.3 rad at $r = r_\text{min}$ to 0.4\,rad $\lesssim \theta_0 \lesssim$ 1.4\,rad at $r = r_\text{max}$. $\phi_0(t)$ shifts from 0.8 rad $\lesssim \phi_0 \lesssim$ 2.2 rad at $r = r_\text{min}$ to 0.6\,rad $\lesssim \phi_0 \lesssim$ 2.1\,rad at $r = r_\text{max}$. As $I$ and $L$ do not depend on $\phi_0$, the change in the shapes must be caused by the tilt and asymmetry of the locus. For a large tilt, the path traced from $l = 0$ to $l = 0.5$ traverses different values of $\theta_0$ than the path traced from $l = 0.5$ to $l = 1$, resulting in different ranges of $\lvert Q \rvert$ and $\lvert U \rvert$ in the first and second halves of the pulse. This causes the asymmetry about the $U$-axis and the broadening and narrowing of the shapes in the $I$-$Q$, $I$-$U$, and $Q$-$U$ planes.

Figure \ref{altitude1} refers to a particular orientation and polarization model, but the trends it depicts are fairly generic. We find that overall, from $r_\text{min}$ to $r_\text{max}$, the phase portraits change modestly with emission altitude at a fixed $(\alpha, i)$. Their sizes, orientation, and detailed substructure (e.g. extra twisting of the secondary loops) change gradually, but there are no cases of the phase portraits changing drastically, for example from a hockey stick to a trefoil. Therefore, when comparing the look-up tables to real data, we do not have to worry about very degenerate matches, i.e. equally good fits for very different combinations of $(\alpha, i)$ at two or more very different altitudes. We do, however, have to be aware that similar shapes like the hockey stick and banana can interchange as the altitude varies, leading to mild $(r, \alpha, i)$ degeneracies.

\section{Conclusion}
\label{sec:conclusion}
In this paper, we introduce Stokes phase portraits as an observational probe of a radio pulsar's magnetic geometry, orientation, and emission altitude. Our aim is to supply the reader with a set of easy-to-use tools for analysing pulsar polarization data in a physically instructive way. We present handy look-up tables of pulse and polarization profiles, Stokes phase portraits, and PA swings as functions of magnetic inclination angle $\alpha$ and line-of-sight inclination angle $i$, for a pure and current-modified dipole, emission from a filled core and hollow cone, and two simple models of the degree of linear polarization. For the current-modified dipole, we also investigate how the phase portraits change with emission altitude $r$. In all the models, relativistic aberration is incorporated fully to order $\mathcal{O} (r/r_\text{LC})$. To assist the reader, we summarise in  Table \ref{tab:list} where to find the look-up tables corresponding to each model.

\begin{table*}
 \centering
\caption{Catalog of Stokes phase portraits and PA swing look-up tables in Sections \ref{sec:dipole} and \ref{sec:toroidal}.}
\label{tab:list}
\begin{tabular}{|l|l|l|l|c|}
\hline
Magnetic geometry & Emission altitude &Beam pattern & $L$ pattern & Figures \\
\hline
Pure dipole & $r \ll r_\text{LC}$ & Filled core & $I \cos \theta_0$ & \ref{m0w10lcostheta_dqvsi}--\ref{m0w10lcostheta_dpa}\\
& & & $I \sin \theta_0$ & \ref{m0w10lsintheta_dqvsi}--\ref{m0w10lsintheta_dpa}\\
& & Hollow cone & $I \cos \theta_0$ & \ref{m25w10lcostheta_dqvsi}--\ref{m25w10lcostheta_dpa}\\
& & & $I \sin \theta_0$ & \ref{m25w10lsintheta_dqvsi}--\ref{m25w10lsintheta_dpa}\\
Pure dipole & $r = 0.1 r_\text{LC}$ & Filled core & $I \cos \theta_0$ & \ref{m0w10lcostheta_dabqvsi}--\ref{m0w10lcostheta_dabuvsq}\\
& & Hollow cone & $I \cos \theta_0$ & \ref{m25w10lcostheta_dabqvsi}--\ref{m25w10lcostheta_dabuvsq}\\
Current-modified dipole & $r = 0.1 r_\text{LC}$ & Filled core & $I \cos \theta_0$ & \ref{m0w10lcostheta_tqvsir25}--\ref{m0w10lcostheta_tpar25}\\
& & & $I \sin \theta_0$ & \ref{m0w10lsintheta_tqvsir25}--\ref{m0w10lsintheta_tpar25}\\
& & Hollow cone & $I \cos \theta_0$ & \ref{m25w10lcostheta_tqvsir25}--\ref{m25w10lcostheta_tpar25}\\
& & & $I \sin \theta_0$ & \ref{m25w10lsintheta_tqvsir25}--\ref{m25w10lsintheta_tpar25}\\
\hline
\end{tabular}
\end{table*}

For a pure dipole at low emission altitudes ($r \ll r_\text{LC}$), the Stokes phase portraits display relatively simple shapes, summarised in Table \ref{tab:dipole}. The shapes are classified according to colloquial descriptors in the Appendix. They are symmetric about $U = 0$, because the effects of aberration are negligible. The curvature in the shapes is determined by the path traced by $\hat{\mathbf{x}}_0(t)$; for example, as $\lvert \alpha - i \rvert$ increases, the curvature of the hockey stick in the $I$-$Q$ plane decreases. For a filled core, the phase portraits produced by the two polarization models, $L = I \cos \theta_0$ and $L = I \sin \theta_0$ are generally similar, except for $\alpha = i$, where $L$ is double-peaked. When the filled core is replaced by a hollow cone, the pulse profiles for $\lvert \alpha - i \rvert \leq 30^\circ$ become double-peaked, and the phase portraits transform into more complicated shapes. 

For a pure dipole at $r = 0.1 r_\text{LC}$, aberration distorts the Stokes phase portraits. The symmetry about $U = 0$ is broken, causing the shapes in the $I$-$U$ and $Q$-$U$ planes to tilt and rotate. In the $I$-$Q$ plane, the shapes broaden noticeably. A phase shift between the pulse centroid and PA swing inflection point of $\approx 4 r/r_\text{LC}$ is also observed \citep{blaskiewicz91, hibschman01, dyks08}.

\begin{table*}
 \centering
\caption{Summary of shapes seen in Stokes phase portraits for a pure dipole (Section \ref{sec:dipole}). Examples of each shape appear in Figure \ref{glossary}.}
\label{tab:dipole}
\begin{tabular}{|l|l|l|}
\hline
&$L = I \cos \theta_0$ & $L = I \sin \theta_0$ \\
\hline
\multirow{3}{*}{Filled core} & hockey stick, straight line ($I$-$Q$) & hockey stick ($I$-$Q$)\\
& balloon, figure-eight ($I$-$U$) & balloon, figure-eight ($I$-$U$)\\
& balloon, heart ($Q$-$U$) & balloon, heart ($Q$-$U$)\\
\hline
\multirow{3}{*}{Hollow cone} & $\gamma$ shape ($I$-$Q$) & $\gamma$ shape ($I$-$Q$) \\
& trefoil, mosquito, twisted triangle ($I$-$U$) & trefoil, mosquito, twisted triangle ($I$-$U$)\\
& balloon, heart, mosquito, interlocking ovals ($Q$-$U$) & balloon, heart, mosquito, interlocking ovals ($Q$-$U$) \\
\hline
\end{tabular}
\end{table*}

For a current-modified dipole at $r = 0.1 r_\text{LC}$, the phase portraits are also asymmetric about $Q = 0$ and $U = 0$. The toroidal field can either partially cancel out or enhance the effects of aberration, depending on its orientation. For $B_\phi = -\sin \theta \cos \alpha B_p r/r_\text{LC}$, the shapes of the phase portraits are summarised in Table \ref{tab:toroidal} and in the Appendix. For a hollow cone, at those orientations where the pulse is double-peaked, the shapes are more complicated that those of a filled core. The tilt angles of the shapes are roughly proportional to the emission altitude and hence $B_\phi$. The heart shapes and balloons seen in the $Q$-$U$ plane also rotate about $(U, Q) = (0, 0)$ as the orientation and/or emission altitude changes.

\begin{table*}
 \centering
\caption{Summary of shapes seen in Stokes phase portraits for a current-modified dipole (Section \ref{sec:toroidal}). Examples of each shape appear in Figure \ref{glossary}.}
\label{tab:toroidal}
\begin{tabular}{|l|l|l|}
\hline
&$L = I \cos \theta_0$ & $L = I \sin \theta_0$ \\
\hline
\multirow{3}{*}{Filled core} & Balloon, banana ($I$-$Q$) & Balloon, banana ($I$-$Q$)\\
& Balloon, banana ($I$-$U$) & Balloon, figure-eight ($I$-$U$)\\
& Balloon ($Q$-$U$) & Balloon ($Q$-$U$)\\
\hline
\multirow{3}{*}{Hollow cone} & Balloon, mosquito, twisted triangle ($I$-$Q$) & balloon, mosquito, figure-eight, trefoil ($I$-$Q$) \\
& Balloon, banana, trefoil, mosquito ($I$-$U$) & Balloon, mosquito, trefoil ($I$-$U$)\\
& balloon, heart, figure-eight ($Q$-$U$) & Balloon, heart, figure-eight ($Q$-$U$) \\
\hline
\end{tabular}
\end{table*}

We close by summarising our main conclusions. 
\begin{enumerate}
\item Stokes phase portraits contain additional information regarding a pulsar's orientation $(\alpha, i)$ and magnetic geometry when used in conjunction with PA swings. 

\item The observed phase portraits of $\approx 60$\% of the 24 pulsars studied in this paper are asymmetric (Section \ref{sec:compdipole}). There are two possible causes for this: (1) the phase portraits are distorted by relativistic aberration and are hence incompatible with a pure magnetic dipole at low emission altitudes, or (2) the axis of beam symmetry is tilted with respect to the magnetic axis, mimicing the effect of aberration even if the emission originates from a low altitude. 

\item If the former is true, the radio emission region in these objects is found to lie at $r \gtrsim 0.1 r_\text{LC}$, where the effects of relativistic aberration and a toroidal field are important. In some cases, one must have $r \gtrsim 0.2 r_\text{LC}$, i.e. the emission comes from the outer magnetosphere .

\item The idealised model of a pure or current-modified dipole at $r \gtrsim 0.1 r_\text{LC}$ with a filled core beam and $L = I \cos \theta_0$, or a hollow cone beam and $L = I \sin \theta_0$, manages to account for most gross features in the polarization data from all 26 pulsars studied in this paper, whether the pulse profiles are single-peaked or double-peaked. Small longitudinal variations in $I$ and $L$ are needed to fit certain minor details (e.g. relative peak heights) in the observations.

\end{enumerate}

This paper is the first in a series. Future papers will examine the Stokes tomography of a force-free rotator, millisecond pulsars, and the circularly polarized component of pulsar radio emission.

\section*{Acknowledgements}
We thank the anonymous referee for his/her meticulous reading of the manuscript and many suggestions that improved the paper, including alerting us to the issues raised in Sections \ref{sec:orientationxy} and \ref{sec:displaced}. Part of this research has made use of the data base of published pulse profiles and Stokes parameters maintained by the European Pulsar Network, available at: http://www.mpifr-bonn.mpg.de/pulsar/data/. CC acknowledges the support of an Australian Postgraduate Award and the Albert Shimmins Memorial Fund.

\section*{Appendix: Glossary of phase portrait descriptors}
\begin{figure*}
\includegraphics[scale=0.7]{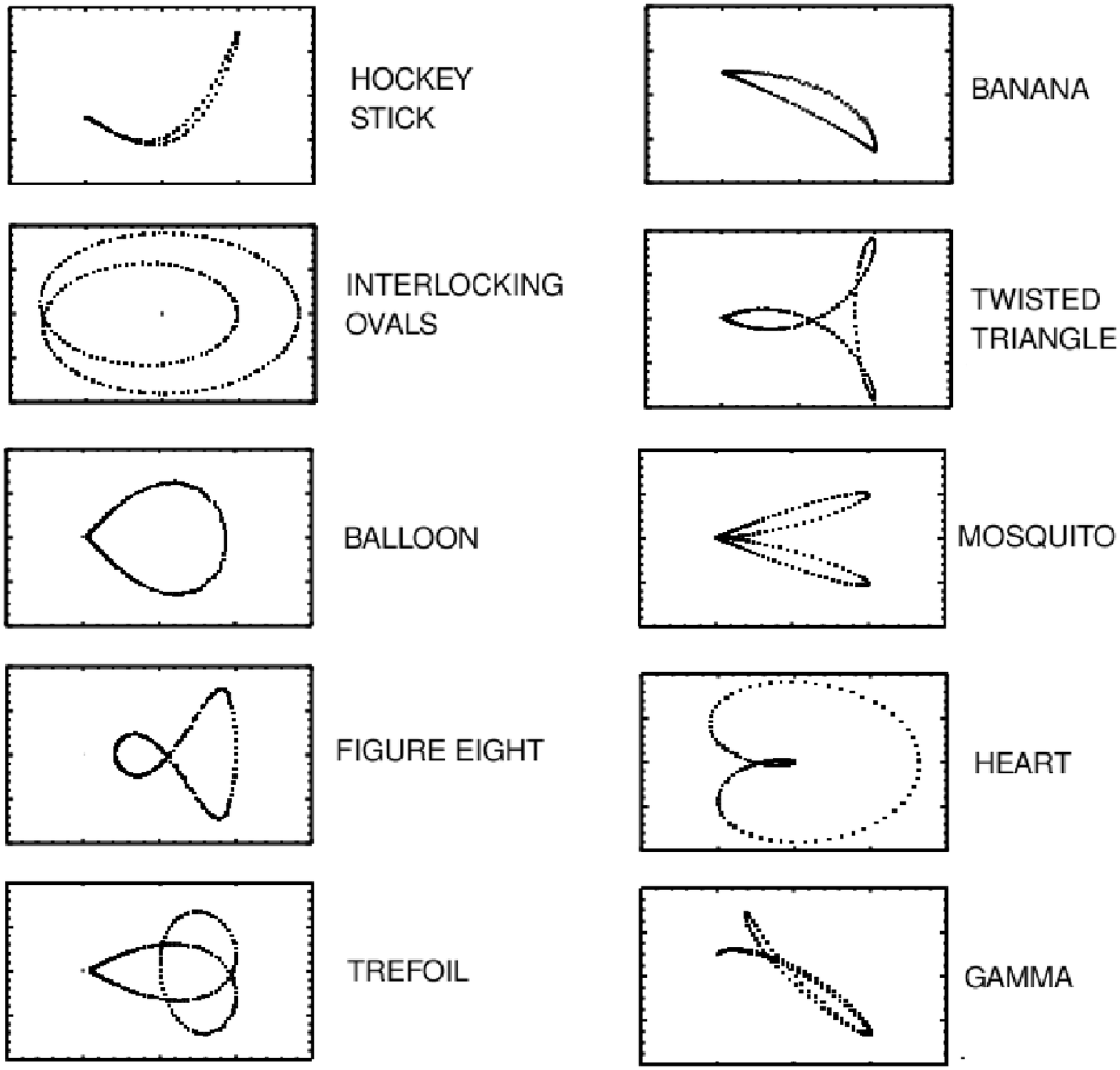}
\caption{Colloquial descriptors of phase portraits encountered in this paper. \textit{Left column, top to bottom}: (i) $I$-$Q$ for pure dipole, filled core beam, $(\alpha, i) = (60^\circ, 40^\circ)$; (ii) $U$-$Q$ for pure dipole, filled core beam, $(\alpha, i) = (20^\circ, 10^\circ)$; (iii) $I$-$U$ for pure dipole, filled core beam, $(\alpha, i) = (30^\circ, 70^\circ)$; (iv) $I$-$U$ for pure dipole, hollow cone, $(\alpha, i) = (60^\circ, 10^\circ)$; (v) $I$-$U$ for pure dipole, hollow cone, $(\alpha, i) = (50^\circ, 30^\circ)$. \textit{Right column, top to bottom}: (i) $I$-$U$ for current-modified dipole, hollow cone, $(\alpha, i) = (80^\circ, 20^\circ)$; (ii) $I$-$U$ for pure dipole, hollow cone, $(\alpha, i) = (60^\circ, 30^\circ)$; (iii) $I$-$U$ for pure dipole, hollow cone, $(\alpha, i) = (50^\circ, 50^\circ)$; (iv) $Q$-$U$ for pure dipole, hollow cone, $(\alpha, i) = (70^\circ, 40^\circ)$; (v) $I$-$Q$ for pure dipole, hollow cone, $(\alpha, i) = (50^\circ, 40^\circ)$. All examples share the same polarization pattern $L/I = \cos \theta_0$.}
\label{glossary}
\end{figure*}

Figure \ref{glossary} gathers together, in a handy and easy-to-read table, the colloquial descriptors used to label phase portraits in the text. The parameters for each example are stated in the caption.

\label{lastpage}
\end{document}